\documentclass[11pt,letterpaper]{article}
\pdfoutput=1

\usepackage[utf8x]{inputenc}%
\usepackage{tcolorbox,fancybox}%
\usepackage{physics}
\usepackage{cite}
\usepackage{moresize}
\usepackage{ntheorem}
\usepackage{mathtools}

\usepackage[english]{babel}
\usepackage{amsmath,amssymb,amsfonts}
\usepackage{graphicx}
\usepackage{booktabs}
\usepackage{color,xcolor}
\usepackage[numbers,sort&compress]{natbib}

\usepackage{tikz}
\usepackage[compat=1.1.0]{tikz-feynman}
\usetikzlibrary{positioning}
\usetikzlibrary{decorations.text}
\usetikzlibrary{decorations.pathmorphing}

\usetikzlibrary{calc}
\usetikzlibrary{shapes.misc}
\tikzset{
        >=latex,
    photon/.style={decorate, decoration={snake}, draw=black, thick},
    fermionnoarrow/.style={draw=black, postaction={decorate}, thick},
    scalar/.style={draw=black, postaction={decorate}, decoration={markings,mark=at position .55 with {\arrow{>}}}, thick, dashed},
    scalarnoarrow/.style={draw=black, postaction={decorate},  thick, dashed},
    fermion/.style={draw=black, postaction={decorate},decoration={markings,mark=at position .55 with {\arrow{>}}}, thick},
    gluon/.style={decorate, draw=black, decoration={coil,amplitude=4pt, segment length=5pt}, thick},
    vertex/.style={draw,shape=circle,fill=black,minimum size=3pt,inner sep=0pt},
    fillvertex/.style={draw,shape=circle,fill=black,minimum size=5pt,inner sep=0pt},
    openvertex/.style={draw,shape=circle,minimum size=5pt,inner sep=0pt},
    blob/.style={draw=red,shape=circle,fill=red,minimum size=6pt,inner sep=0pt},
    redvertex/.style={draw=red,shape=circle,fill=red,minimum size=3pt,inner sep=0pt},
    cross/.style={cross out, draw=black,thick, minimum size=5pt, inner sep=0pt, outer sep=0pt}
}

\usepackage{ulem}
\usepackage{physics}
\usepackage{orcidlink}
\usepackage{slashed}
\usepackage{mathrsfs}
\usepackage{enumerate}
\usepackage{mdframed}
\usepackage{url}
\usepackage[makeroom]{cancel}
\usepackage{geometry}
\usepackage{simpler-wick}

\usepackage{pifont}
\newcommand{\cmark}{\ding{51}}%
\newcommand{\xmark}{\ding{55}}%

\theoremstyle{definition}

\newtheorem{thm-non}{Theorem}

\newtheorem{define}[thm-non]{Definition}

\usepackage{hyperref} 
\hypersetup{
    colorlinks=true,       
    linkcolor=red,          
    citecolor=blue,        
    filecolor=magenta,      
    urlcolor=purple           
}
\usepackage[all]{hypcap} 

\def\beqn{\begin{eqnarray}}
\def\eeqn{\end{eqnarray}}
\def\beqs{\begin{subequations}}
\def\eeqs{\end{subequations}}
\def\beq{\begin{equation}}
\def\eeq{\end{equation}}
\def\ba{\begin{array}}
\def\ea{\end{array}}

\def\non{\nonumber\\}

\def\hf{\frac{1}{2}}
\def\[{\left[}
\def\]{\right]}
\def\({\left(}
\def\){\right)}
\newcommand\para{\paragraph{}}

\def\gSU{\rm SU}

\def\gSO{\rm SO}

\newcommand{\rep}[1]{\mathbf{#1}}
\newcommand{\repb}[1]{\mathbf{\overline{#1}}}

\def\Bc{\mathcal{B}}

\def\Dc{\mathcal{D}}
\def\Ec{\mathcal{E}}
\def\Fc{\mathcal{F}}
\def\Gc{\mathcal{G}}
\def\Hc{\mathcal{H}}

\def\Lc{\mathcal{L}}
\def\Mc{\mathcal{M}}
\def\Nc{\mathcal{N}}
\def\Oc{\mathcal{O}}
\def\Pc{\mathcal{P}}
\def\Qc{\mathcal{Q}}
\def\Rc{\mathcal{R}}

\def\Tc{\mathcal{T}}
\def\Uc{\mathcal{U}}

\def\Xc{\mathcal{X}}

\def\AG{\mathfrak{A}}  \def\aG{\mathfrak{a}}
\def\BG{\mathfrak{B}}  \def\bG{\mathfrak{b}}
\def\CG{\mathfrak{C}}  \def\cG{\mathfrak{c}}
\def\DG{\mathfrak{D}}  \def\dG{\mathfrak{d}}
\def\EG{\mathfrak{E}}  \def\eG{\mathfrak{e}}

  \def\nG{\mathfrak{n}}

\def\UG{\mathfrak{U}}  \def\uG{\mathfrak{u}}

\title{
{\bf The Standard Model quark/lepton masses and the Cabibbo-Kobayashi-Maskawa mixing in an ${\rm SU}(8)$ theory} \\
\author{\large Ning Chen$^{\,\heartsuit}$\,\orcidlink{0000-0002-0032-9012}, Ying-nan Mao$^{\,\diamondsuit}$\,\orcidlink{0000-0001-8063-8968}, Zhaolong Teng$^{\,\clubsuit}$\,\orcidlink{0000-0002-7141-2331}}
\date{\small \it
$^\heartsuit\, ^\clubsuit$School of Physics, Nankai University, Tianjin, 300071, China \\
$^\diamondsuit$ Department of Physics, School of Physics and Mechanics, Wuhan University of Technology, \\ Wuhan, 430070, Hubei, China \\
}
}

\begin{document}

\maketitle
\setlength{\parskip}{0.2ex}

\begin{abstract}
\bigskip
The observed Standard Model (SM) quark/lepton mass hierarchies and the Cabibbo-Kobayashi-Maskawa (CKM) mixing pattern are described in an ${\rm SU}(8)$ theory through its realistic symmetry breaking pattern with three intermediate stages, which rely on a set of $d=5$ gravity-induced operators that break the emergent global symmetries in the chiral fermion sector, as well as the precise identifications of all non-trivially embedded SM flavors.
\end{abstract}

\vspace{9.6cm}
{\emph{Emails:}\\  
$^{\,\heartsuit}$\url{chenning_symmetry@nankai.edu.cn},\\
$^{\,\diamondsuit}$\url{ynmao@whut.edu.cn},\\
$^{\,\clubsuit}$ \url{tengcl@mail.nankai.edu.cn}
 }

\thispagestyle{empty}  
\newpage  
\setcounter{page}{1}  

\vspace{1.0cm}
\eject
\tableofcontents

\section{Introduction}
\label{section:intro}

\para
A chiral Yang-Mill theory~\cite{Yang:1954ek} based on a simple compact Lie group, coined as the Grand Unified Theory (GUT), was proposed to unify all three fundamental symmetries described by the Standard Model (SM).
Two original versions that have been widely studied over decades are the $\gSU(5)$ Georgi-Glashow theory~\cite{Georgi:1974sy} and the $\gSO(10)$ Fritzsch-Minkowski theory~\cite{Fritzsch:1974nn}.
Elegant properties of the gauge coupling unification and the charge quantization, keep the GUT a promising framework beyond the SM (BSM).
If one recalls that several puzzling phenomena of blackbody radiation~\cite{Planck:1901oar}, the photoelectric effect~\cite{Einstein:1905tem}, and the hydrogen spectrum~\cite{Bohr:1913zba}, were completely explained within the framework of quantum mechanics in the early twentieth century, the well-known $\gSU(5)$ and/or $\gSO(10)$ theories are far from being realistic given that the observed mass hierarchies of all SM quarks/leptons as well as their electroweak mixing patterns, albeit being experimentally measured to great precisions over the past one century, were not at all theoretically interpreted without additional inputs.

\para
A crucial observation is that the flavor sector of the minimal ${\rm SU}(5)$ and ${\rm SO}(10)$ GUTs resemble the trivially repetitive structure as in the SM.
In a seminal paper~\cite{Georgi:1979md} of the extended GUTs beyond the minimal ${\rm SU}(5)$, Georgi conjectured that such trivially repetitive structure may not be necessary in a class of ${\rm SU}(N >5)$ theories.
Accordingly, he proposed the third law of the flavor sector and found a minimal ${\rm SU}(11)$ theory with $561$ chiral fermions.
By decomposing these chiral fermions in terms of the ${\rm SU}(5)$ irreducible representations (irreps), Georgi's original ${\rm SU}(11)$ theory leads to three distinctive $\repb{5_F}$'s and $\rep{10_F}$'s.
The flavor structure was later relaxed in Refs.~\cite{Frampton:1979cw,Frampton:1979fd,Barr:1979xt,Barr:2008pn,Chen:2023qxi}, since one could have distinguished three-generational SM quarks/leptons as long as there are three distinctive $\rep{10_F}$'s according to the flavor identifications.
This was recently re-stated in Ref.~\cite{Chen:2023qxi} that an ${\rm SU}(8)$ theory with $156$ chiral fermions is the minimal framework where three-generational SM quarks/leptons can transform differently, when one embeds them into distinctive chiral \underline{ir}reducible \underline{a}nomaly-\underline{f}ree \underline{f}ermion \underline{s}ets (IRAFFSs) in the UV theory.
According to the previous observation by Froggatt and Nielsen~\cite{Froggatt:1978nt} that the renormalization group (RG) equations of the Yukawa couplings can not generate large mass hierarchies without the aid of some special assignments of quantum numbers, a UV setup with intrinsically distinctive symmetry properties among three-generational SM fermions seems to be encouraging as a candidate to originate the observed SM quark/lepton mass hierarchies as well as their EW mixing pattern.
Furthermore, the emergent global symmetries can be identified straightforwardly with the concept of the chiral IRAFFS, where the Abelian components of the non-anomalous global symmetries were shown to become the global $ \widetilde {\rm U}(1)_{B-L}$ symmetry when the theory flows to the EW scale~\cite{Chen:2023qxi}.
From the experimental perspective, the only guidance to unveil the flavor puzzle (particularly for quarks and leptons) to date is the discovery~\cite{ATLAS:2012yve,CMS:2012qbp} and the measurements~\cite{CMS:2022dwd,ATLAS:2022vkf} of one single SM Higgs boson.
The LHC measurements have confirmed that the Yukawa couplings at the EW scale between the SM Higgs boson and the $(t\,,b\,,\tau\,,\mu)$ are consistent with the SM predictions.
With the non-trivial flavor structure of the ${\rm SU}(8)$ theory, one naturally asks how does this SM Higgs boson distinguish all three generations such that the hierarchical Yukawa couplings are generated.
The detailed analyses in the ${\rm SU}(8)$ framework will be devoted to prove that:
\begin{enumerate}

\item there is one single SM Higgs doublet at the EW scale, and,

\item the unique SM Higgs doublet can generate the observed SM Yukawa couplings of $y_f=\sqrt{2} m_f/v_{\rm EW}$ for all quarks/leptons as well as the observed Cabibbo-Kobayashi-Maskawa (CKM) mixing pattern~\cite{Cabibbo:1963yz,Kobayashi:1973fv}, due to the intrinsic symmetry properties of the ${\rm SU}(8)$ theory.

\end{enumerate}

\para
The rest of the paper is organized as follows.
In Sec.~\ref{section:setup}, we setup an ${\rm SU}(8)$ theory where three-generational SM fermions are non-trivially embedded into a rank-$2$ chiral IRAFFS and a rank-$3$ chiral IRAFFS.
Both non-anomalous and anomalous global symmetries will be identified, and they both play crucial roles in the SM quark/lepton masses.
Based on the possible symmetry breaking pattern, we decompose the anomaly-free chiral fermions and the minimal set of Higgs fields.
Accordingly, the SM Higgs doublet is uniquely determined from the component of the Higgs field of $\rep{70_H}$ in the spectrum based on the $\widetilde {\rm U}(1)_{B-L}$-neutrality condition~\cite{Chen:2023qxi}.
In Sec.~\ref{section:pattern}, we describe a realistic symmetry breaking pattern where all vectorlike fermions obtain their masses through the renormalizable/non-renormalizable Yukawa couplings above the EW scale.
At the EWSB stage, the SM Higgs doublet is found to give the top quark mass at the tree level with the natural Yukawa coupling of $Y_\Tc\sim \Oc(1)$.
In Sec.~\ref{section:gravity_Fmass}, we analyze the $d=5$ mass terms for lighter SM quarks/leptons, where the emergent global symmetries are generically broken by the gravitational effect~\cite{Kallosh:1995hi,Harlow:2018jwu,Harlow:2018tng}.
There can be both direct Yukawa coupling and the indirect Yukawa coupling terms through the Higgs mixing operators in the ${\rm SU}(8)$ theory.
Based on all possible fermion mass terms from these $d=5$ operators, the hierarchical SM quark/lepton masses as well as the CKM mixing matrix of the quark sector are described in Sec.~\ref{section:SMql_CKM}.
Three intermediate symmetry breaking scales are suggested in a benchmark point.
Flavors of all three-generational SM quarks/leptons are identified according to the symmetry breaking pattern.
We summarize our results and make future perspective in Sec.~\ref{section:conclusion}.
App.~\ref{section:Br} defines the gauge group indices and the decomposition rules according to the symmetry breaking pattern in the ${\rm SU}(8)$ theory.

\section{The ${\rm SU}(8)$ theory setup}
\label{section:setup}

\subsection{Overview}
\label{section:overview}

\para
Georgi's third law~\cite{Georgi:1979md} was soon reconsidered by allowing the repetition of some fermion representations as long as the gauge anomaly is cancelled~\cite{Frampton:1979cw}.
Only the anti-symmetric representations of the ${\rm SU}(N)$ fermions, denoted as $\[ N\,,k \]_{ \rep{F}}$, will be considered so that no ${\rm SU}(3)_c \otimes {\rm U}(1)_{\rm EM}$-exotic fermions will be present in the spectrum.
The chiral fermion contents rely on the concept of the chiral IRAFFS in the UV theory, which was recently proposed in Ref.~\cite{Chen:2023qxi} as follows
\begin{define}\label{def:IRAFFS}

A chiral IRAFFS is a set of left-handed anti-symmetric fermions of $\sum_\Rc m_\Rc \, \Fc_L(\Rc)$, with $m_\Rc$ being the multiplicities of a particular fermion representation of $\Rc$.
Obviously, the anomaly-free condition reads $\sum_\Rc m_\Rc \, {\rm Anom}(  \Fc_L(\Rc) ) =0$.
We also require the following conditions to be satisfied for a chiral IRAFFS:
\begin{itemize}

\item the greatest common divisor (GCD) of the $\{ m_\Rc \}$ should satisfy that ${\rm GCD} \{  m_\Rc \} =1$;

\item the fermions in a chiral IRAFFS can no longer be removed, which would otherwise bring non-vanishing gauge anomalies;

\item there should not be any singlet, self-conjugate, adjoint fermions, or vectorial fermion pairs in a chiral IRAFFS.

\end{itemize}

\end{define}
The ${\rm SU}(N)$ chiral fermions can be partitioned into several distinctive rank-$k$ chiral IRAFFSs as follows~\footnote{We use the notion of $\{ k \}^\prime$ in Eqs.~\eqref{eq:SUN_fermions_realistic} and \eqref{eq:DRS} to indicate distinctive IRAFFSs. In other words, each specific  rank-$k$ chiral IRAFFS can appear no more than once.}
\beqn\label{eq:SUN_fermions_realistic}
 \{ f_L \}_{ {\rm SU}(N)}^{n_g} &=& \bigoplus_{ \{ k \}^\prime } \Big\{  \underbrace{ \overline{ \[ N\,, 1 \]_{ \rep{F}} }^\omega \oplus   \[ N\,, k \]_{ \rep{F}} }_{ \textrm{rank-$k$ chiral IRAFFS}} \Big\} \,,\quad  \omega =1 \,,...\,, m_k \,,\quad 2 \leq k\leq \[ \frac{N}{2} \]\,,
\eeqn
with
\beqn
m_k &=& {\rm Anom} ( \[ N\,,k \]_{\rep{F}}) = \frac{ (N-2k)\, ( N-3)! }{ (N-k-1)!\, (k-1)!}  \,.
\eeqn
The chiral fermions in Eq.~\eqref{eq:SUN_fermions_realistic} lead to the global flavor symmetries~\footnote{Throughout the paper, we use the notion of $\widetilde \Gc$ to denote the groups for global symmetries.} of
\beqn\label{eq:DRS}
\widetilde{ \Gc}_{\rm flavor}&=& \bigotimes_{ \{ k \}^\prime } \Big[  \widetilde{ {\rm SU}}(m_{k } )_{ \omega_k } \otimes \widetilde{ {\rm U}}(1)_{ \omega_k } \otimes  \widetilde{ {\rm U}}(1)_{ A_k}  \Big]  \,,
\eeqn
for $m_{k } \geq 2$.
This is a generalization of the global symmetry in the rank-$2$ anti-symmetric ${\rm SU}(N)$ theory, as was pointed out by Dimopoulos, Raby, and Susskind (DRS)~\cite{Dimopoulos:1980hn}.
The emergent global flavor symmetries in chiral gauge theories were also acknowledged in Refs.~\cite{Appelquist:2013oni,Shi:2015fna}.
The global $\widetilde{ {\rm U}}(1)_{ \omega_k }$ and $\widetilde{ {\rm U}}(1)_{ A_k}$ symmetries act on the chiral fermions of $\overline{ \[ N\,,1 \] }_{ \rep{F}}$'s and $\[ N\,,k \]_{ \rep{F}}$, and they are both anomalous due to the ${\rm SU}(N)$ instanton.
One can always find a linear combination of $\widetilde{ {\rm U}}(1)_{ \omega_k }$ and $\widetilde{ {\rm U}}(1)_{ A_k}$, denoted as $\widetilde{ {\rm U}}(1)_{ T_k}$, such that the mixed anomalies of $\[ {\rm SU}(N) \]^2 \cdot \widetilde{ {\rm U}}(1)_{ T_k}=0$.
We also denote the linear combinations that are anomalous as the Peccei-Quinn symmetries~\cite{Peccei:1977hh} of $\widetilde{ {\rm U}}(1)_{ {\rm PQ}_k}$, such that the mixed anomalies of $\[ {\rm SU}(N) \]^2 \cdot \widetilde{ {\rm U}}(1)_{ {\rm PQ}_k} \neq 0$~\footnote{Automatic Peccei-Quinn symmetry in the GUT was previously pointed out in the ${\rm SU}(9)$ theory by Georgi, Hall, and Wise~\cite{Georgi:1981pu}.}.

\para
The rules~\cite{Georgi:1979md} of counting the SM fermion generations in an ${\rm SU}(N)$ theory were given by:
\begin{itemize}

\item The ${\rm SU}(N)$ fundamental irrep is decomposed under the ${\rm SU}(5)$ as $\[ N\,,1 \]_{\rep{F}}= (N-5) \times \rep{1_F} \oplus \rep{5_F}$.
The decompositions of other higher-rank irreps can be obtained by tensor products.

\item For an ${\rm SU}(N)$ theory, one can eventually decompose the set of anomaly-free fermion irreps into the ${\rm SU}(5)$ irreps of $(\rep{1_F}\,, \rep{5_F}\,, \repb{5_F}\,, \rep{10_F}\,, \repb{10_F} )$.

\item Count the multiplicity of each ${\rm SU}(5)$ irrep as $\nu_{\rep{5_F}}$ and so on, the anomaly-free condition must lead to a relation of $\nu_{\rep{5_F}} + \nu_{\rep{10_F}} = \nu_{\repb{5_F}} + \nu_{\repb{10_F}}$.

\item The SM fermion generation is determined by $n_g= \nu_{\repb{5_F}} - \nu_{\rep{5_F}} = \nu_{\rep{10_F}} - \nu_{\repb{10_F}}$.

\end{itemize}
It turns out that the multiplicity difference between $\rep{10_F}$ and $\repb{10_F}$ from a given irrep of $\[ N\,,k \]_{\rep{F}}$ can be expressed as
\beqn\label{eq:gen_rkirrep}
\nu_{\rep{10_F}} \[ N\,,k \]_{\rep{F}} - \nu_{\repb{10_F}} \[ N\,,k \]_{\rep{F}}&=&  \frac{( N-2k ) (N-5)! }{ (k-2)!\, (N -k -2)! }\,, ~ (k\geq 2\,, k\neq N-1) \,.
\eeqn
Specifically, one finds $\nu_{\rep{10_F}} \[N\,,2 \]_{\rep{F}} - \nu_{\repb{10_F}} \[ N\,,2 \]_{\rep{F}} =1$, $\nu_{\rep{10_F}} \[N\,,3 \]_{\rep{F}} - \nu_{\repb{10_F}} \[ N\,, 3 \]_{\rep{F}} =N-6$, and $\nu_{\rep{10_F}} \[N\,,4 \]_{\rep{F}} - \nu_{\repb{10_F}} \[ N\,, 4 \]_{\rep{F}} = \frac{1}{2} ( N-8 )(N-5)$.
To avoid the naive replication of one generation for multiple times, it is necessary to consider anti-symmetric irreps with rank higher than two, which in turns calls for a gauge group beyond the ${\rm SU}(5)$.
For the ${\rm SU}(6)$ case, the irrep of $\[ 6\,, 3 \]_{\rep{F}}=\rep{20_F}$ is self-conjugate~\footnote{Generally, any self-conjugate irrep of the ${\rm SU}(2N)$ GUT cannot contribute to a SM fermion generation at the EW scale.} and leads to $\nu_{\rep{10_F}} \[6\,,3\]_{\rep{F}} - \nu_{\repb{10_F}} \[6\,,3 \]_{\rep{F}}=0$ according to Eq.~\eqref{eq:gen_rkirrep}.

\para
The gauge anomalies of several $\gSU(8)$ anti-symmetric representations are listed below
\beqn\label{eq:SU8_anomalies}
&& {\rm Anom}(  \repb{8_F}  ) =-1 \,,~  {\rm Anom}( \rep{ 28_F} ) =+ 4 \,, ~  {\rm Anom}( \rep{56_F } ) = + 5 \,.
\eeqn
By decomposing the ${\rm SU}(8)$ fermions in terms of the ${\rm SU}(5)$ irreps, we have
\beqn\label{eq:SU8_decompose}
\repb{ 8_F}&=& 3\times \rep{1_F} \oplus \repb{5_F} \,,\non
\rep{ 28_F}&=&  3\times \rep{1_F} \oplus 3\times \rep{5_F}  \oplus \rep{10_F}  \,,\non
\rep{ 56_F}&=&  \rep{1_F} \oplus 3\times \rep{5_F} \oplus  3\times \rep{10_F} \oplus \repb{10_F} \,.
\eeqn
The ${\rm SU}(8)$ theory contains two following chiral IRAFFSs at the GUT scale~\footnote{The same ${\rm SU}(8)$ chiral fermions were previously obtained by Barr~\cite{Barr:2008pn}, while a partition based on the chiral IRAFFS was absent.}
\beqn\label{eq:SU8_3gen_fermions}
\{ f_L \}_{ {\rm SU}(8)}^{n_g=3}&=& \Big[ \repb{8_F}^\omega \oplus \rep{28_F} \Big] \bigoplus \Big[ \repb{8_F}^{ \dot \omega } \oplus \rep{56_F} \Big] \,,~ {\rm dim}_{ \mathbf{F}}= 156\,, \non
&& \Omega \equiv ( \omega \,, \dot \omega ) \,, ~ \omega = ( 3\,, {\rm IV}\,, {\rm V}\,, {\rm VI}) \,, ~  \dot \omega = (\dot 1\,, \dot 2\,, \dot {\rm VII}\,, \dot {\rm VIII}\,, \dot {\rm IX} ) \,,
\eeqn
with undotted/dotted indices for the $\repb{8_F}$'s in the rank-$2$ chiral IRAFFS and the rank-$3$ chiral IRAFFS, respectively.
The Roman numbers and the Arabic numbers are used for the heavy partner fermions and the SM fermions.
According to Ref.~\cite{Georgi:1979md} and Eq.~\eqref{eq:gen_rkirrep}, one identifies one generational $\rep{10_F}$ from the $\rep{ 28_F}$, and $3-1=2$ generational $\rep{10_F}$'s from the $\rep{56_F}$.
These three $\rep{10_F}$'s are transforming differently~\cite{Barr:2008pn}, while three identical $\repb{5_F}$'s can be identified from nine $\repb{8_F}$'s, which are sufficient to guarantee that three-generational SM quarks and leptons must transform differently in the UV theory.
To manifest this point, recall that one $\rep{10_F}$ is decomposed into the SM components as $\rep{10_F}=q_L \oplus {u_R}^c \oplus {e_R}^c$, with the suppressed generation indices.
The setup in Eq.~\eqref{eq:SU8_3gen_fermions} thus leads to three up-type quarks with distinctive left-handed and right-handed components, three down-type quarks with distinctive left-handed components, and three charged leptons with distinctive right-handed components.
Altogether, one finds three-generational SM fermions expressed in terms of three $\[\repb{5_F} \oplus \rep{10_F} \]$'s.
There are six vectorial pairs of $(\rep{5_F}\,,\repb{5_F})$'s by mating six $\rep{5_F}$'s from the $\rep{28_F}\oplus \rep{56_F}$ and six $\repb{5_F}$'s from the $\repb{8_F}$'s.
Another vectorial pair of $(\rep{10_F}\,,\repb{10_F})$ from the $\rep{56_F}$ indicates one generational mirror fermions~\cite{Maalampi:1988va} in the spectrum, which will obtain vectorlike masses during the intermediate symmetry breaking stages.

\para
The non-anomalous global DRS symmetries from fermions in Eq.~\eqref{eq:SU8_3gen_fermions} are
\beqn\label{eq:DRS_SU8}
\widetilde{ \Gc}_{\rm DRS} \[{\rm SU}(8)\,, n_g=3 \]&=& \Big[ \widetilde{ {\rm SU}}(4)_\omega  \otimes \widetilde{ {\rm U}}(1)_{T_2} \Big]  \bigotimes \Big[ \widetilde{ {\rm SU}}(5)_{\dot \omega } \otimes \widetilde{ {\rm U}}(1)_{T_3}   \Big]  \,,
\eeqn
and we also denote the anomalous global Peccei-Quinn symmetries~\cite{Peccei:1977hh} as
\beqn\label{eq:PQ_SU8}
\widetilde{ \Gc}_{\rm PQ} \[{\rm SU}(8)\,, n_g=3 \]&=& \widetilde{ {\rm U}}(1)_{{\rm PQ}_2} \bigotimes \widetilde{ {\rm U}}(1)_{ {\rm PQ}_3}  \,.
\eeqn
Both $\widetilde{ {\rm U}}(1)_{T_2}$ and $\widetilde{ {\rm U}}(1)_{{\rm PQ}_2}$ can be thought as the linear combinations of two global $\widetilde{ {\rm U}}(1)$'s of the $\repb{8_F}^\omega$ and the $\rep{28_F}$ in the rank-2 chiral IRAFFS, and similarly for the $\widetilde{ {\rm U}}(1)_{T_3}$ and $\widetilde{ {\rm U}}(1)_{{\rm PQ}_3}$ symmetries in the rank-3 chiral IRAFFS.
Since the global $B-L$ symmetry should be identical for all three generations, we further require a common non-anomalous $\widetilde{ {\rm U}}(1)_{T}\equiv \widetilde{ {\rm U}}(1)_{T_2} = \widetilde{ {\rm U}}(1)_{T_3}$ between two chiral IRAFFSs~\footnote{If one viewed all ${\rm SU}(8)$ chiral fermions as a whole instead of partitioning them into two chiral IRAFFSs in Eq.~\eqref{eq:SU8_3gen_fermions}, one may identify the non-anomalous global symmetry as $\widetilde{ \rm SU}(9)_\Omega \otimes \widetilde{ \rm U}(1)_T$. Correspondingly, one assigns the global $\widetilde{ \rm U}(1)_T$ charges according to the equation of $3 \Tc( \repb{8_F}^\Omega ) + 2 \Tc ( \rep{28_F}) + 5 \Tc ( \rep{56_F} )=0$, which leads to infinitely many solutions with ${\rm GCD}( \Tc( \repb{8_F}^\Omega ) \,,  \Tc ( \rep{28_F}) \,, \Tc ( \rep{56_F} ) )=1$.}.
The general $\widetilde{ {\rm U}}(1)_T$ charge assignments in a rank-$k$ chiral IRAFFS read~\cite{Appelquist:2013oni,Shi:2015fna}
\beqn\label{eq:rk_Tcharges_Ferm}
&& \Tc( \overline{ \[ N\,, 1 \]_{ \rep{F}} }^\Omega ) = - \frac{ N-2 }{d_k}  t\,, \quad  \Tc( \[ N\,, k \]_{ \rep{F}} ) = \frac{  N-2k }{ d_k} t \,, \non
&& d_k = {\rm GCD} ( N-2\,, N- 2k ) \,.
\eeqn
The other non-vanishing global anomalies in a rank-$k$ chiral IRAFFS read~\footnote{These two non-vanishing anomalies of $\[  {\rm grav} \]^2 \cdot \widetilde{ {\rm U}}(1)_{T }$ and $\[  \widetilde{ {\rm U}}(1)_{T} \]^3$ imply that the global $\widetilde{ {\rm U}}(1)_{T }$ symmetries cannot be gauged.}
\beqs
\beqn
&& \[  {\rm grav} \]^2 \cdot \widetilde{ {\rm U}}(1)_{T } = \frac{N ( N-2)! }{ d_k ( N-k)! \, k!} ( N-k -1) (N-2k ) (1-k )t \,, \\[1mm]
&& \[  \widetilde{ {\rm U}}(1)_{T} \]^3   = \Big[   \frac{ N! }{ k!\, ( N-k )! } ( N-2k)^3 - N ( N-2)^3 \Big]  \frac{ t^3 }{ d_k^3 }\,,
\eeqn
\eeqs
which can be used to count the massless sterile neutrinos in the spectrum through the `t Hooft anomaly matching~\cite{Chen:2023qxi}.
Likewise, we also assume a common anomalous $\widetilde{ {\rm U}}(1)_{\rm PQ}\equiv \widetilde{ {\rm U}}(1)_{{\rm PQ}_2} = \widetilde{ {\rm U}}(1)_{ {\rm PQ}_3}$ between two chiral IRAFFSs in Eq.~\eqref{eq:PQ_SU8}.
Accordingly, the non-anomalous global $\widetilde{ {\rm U}}(1)_T$ charges and the anomalous global $\widetilde{ {\rm U}}(1)_{\rm PQ}$ charges for fermions are assigned in Tab.~\ref{tab:U1TU1PQ}.

\begin{table}[htp]
\begin{center}
\begin{tabular}{c|cccc}
\hline\hline
 Fermions &  $\repb{8_F}^\Omega$ &  $\rep{28_F}$  &  $\rep{56_F}$  &     \\[1mm]
\hline
$\widetilde{ {\rm U}}(1)_T$ &  $-3t$  &  $+2t$  & $+t$ &      \\[1mm]
$\widetilde{ {\rm U}}(1)_{\rm PQ}$ &  $p$  &  $q_2$  & $q_3$ &      \\[1mm]
\hline
Higgs  &  $\repb{8_H}_{\,, \omega }$  & $\repb{28_H}_{\,, \dot \omega }$   & $\rep{70_H}$ &  $\rep{63_H}$    \\[1mm]
\hline
$\widetilde{ {\rm U}}(1)_T$ &  $+t$  &  $+2t$   &  $-4t$  &  $0$    \\[1mm]
$\widetilde{ {\rm U}}(1)_{\rm PQ}$ &  $-(p+q_2)$  &  $-(p+q_3 )$  & $-2q_2$ &  $0$ \\[1mm]
\hline\hline
\end{tabular}
\end{center}
\caption{The non-anomalous $\widetilde{ {\rm U}}(1)_T$ charges and the anomalous global $\widetilde{ {\rm U}}(1)_{\rm PQ}$ charges for the $\gSU(8)$ fermions and Higgs fields.}
\label{tab:U1TU1PQ}
\end{table}%

\para
The most general gauge-invariant Yukawa couplings at least include the following renormalizable and non-renormalizable terms~\footnote{The term of $\rep{56_F} \rep{56_F} \rep{28_H}  + H.c.$ vanishes due to the anti-symmetric property~\cite{Barr:2008pn}. Instead, only a $d=5$ non-renormalizable term of $\frac{1}{ M_{\rm pl} } \rep{56_F}  \rep{56_F}  \repb{28_{H}}_{\,,\dot \omega }^\dag  \rep{63_{H}} $ is possible to generate masses for vectorlike fermions in the $\rep{56_F}$. Since it transforms as an $\widetilde{ {\rm SU}}(5)_{\dot \omega }$ vector and carries non-vanishing $\widetilde {\rm U}(1)_{\rm PQ}$ charge of $p+3q_3 \neq 0$ from Eq.~\eqref{eq:PQcharges_SU8}, it is only possible due to the gravitational effect.}
\beqn\label{eq:Yukawa_SU8}
-\Lc_Y&=& Y_\Bc  \repb{8_F}^\omega  \rep{28_F}  \repb{8_{H}}_{\,,\omega }  +  Y_\Tc \rep{28_F} \rep{28_F} \rep{70_H} \non
&+&  Y_\Dc \repb{8_F}^{\dot \omega  } \rep{56_F}  \repb{28_{H}}_{\,,\dot \omega }   + \frac{ c_4 }{ M_{\rm pl} } \rep{56_F}  \rep{56_F}  \repb{28_{H}}_{\,,\dot \omega }^\dag  \rep{63_{H}} + H.c.\,.
\eeqn
All renormalizable Yukawa couplings are assumed to be $(Y_\Bc \,, Y_\Tc\,, Y_\Dc)\sim\Oc(1)$, and the Wilson coefficient $c_4$ of the non-renormalizable term is assigned in accordance with our conventions in Sec.~\ref{section:direct_Yukawa}.
Altogether, we collect the ${\rm SU}(8)$ Higgs fields as follows
\beqn\label{eq:SU8_Higgs}
 \{ H \}_{ {\rm SU}(8)}^{n_g=3} &=& \repb{8_H}_{ \,, \omega}  \oplus \repb{28_H}_{ \,, \dot \omega}  \oplus \rep{70_H}  \oplus \underline{ \rep{63_H} } \,,~ {\rm dim}_{ \mathbf{H}}= 547 \,,
\eeqn
where the adjoint Higgs field of $\rep{63_H}$ is real while all others are complex.
The adjoint Higgs field of $\rep{63_H}$ will maximally break the GUT symmetry as ${\rm SU}(8)\to \Gc_{441}$ through its VEV of
\beqn\label{eq:63H_VEV}
\langle  \rep{63_H}\rangle &=&\frac{1}{ 4 } {\rm diag}(- \mathbb{I}_{4\times 4} \,, +\mathbb{I}_{4\times 4} ) v_U \,.
\eeqn
We assign the non-anomalous global $\widetilde{ {\rm U}}(1)_T$ charges and the anomalous global $\widetilde{ {\rm U}}(1)_{\rm PQ}$ charges for Higgs fields in Tab.~\ref{tab:U1TU1PQ} according to the renormalizable Yukawa couplings in Eq.~\eqref{eq:Yukawa_SU8}.
The anomalous global $\widetilde{ {\rm U}}(1)_{\rm PQ}$ charges are assigned such that
\beqn\label{eq:PQcharges_SU8}
&&  p : q_2  \neq -3 : +2  \,, \quad p  : q_3 \neq -3  : +1  \,.
\eeqn
Obviously, the only non-renormalizable term in Eq.~\eqref{eq:Yukawa_SU8} is both $\widetilde{ {\rm SU}}(5)_{\dot \omega}$-violating and $\widetilde{ {\rm U}}(1)_{\rm PQ}$-charged according to Tab.~\ref{tab:U1TU1PQ} and Eq.~\eqref{eq:PQcharges_SU8}.
Thus, it can be explicitly broken by the gravitational effect.

\subsection{The symmetry breaking pattern}
\label{section:patterns}

\para
It was first pointed out in Ref.~\cite{Li:1973mq} that the Higgs representations are responsible for the gauge symmetry breaking patterns, as well as the proper choices of the Higgs self couplings.
For this purpose, we tabulate the patterns of the symmetry breaking for the ${\rm SU}(N)$ groups with various Higgs representations in Tab.~\ref{tab:HiggsPatterns}.

\begin{table}[htp]
\begin{center}
\begin{tabular}{c|cc}
\hline\hline
 Higgs irrep&  dimension & pattern   \\
\hline
fundamental  &  $N$  &  $ {\rm SU}(N-1)$  \\[1mm]
rank-$2$ symmetric  &  $\frac{1}{2}N(N+1)$  & ${\rm SO}(N)$   \\
  &  & or ${\rm SU} (N-1)$ \\[1mm]
rank-$2$ anti-symmetric  &  $\frac{1}{2}N(N-1)$  & ${\rm SO}(2k+1)\,, k=[ \frac{N}{2}]$  \\
  &  & or ${\rm SU} (N-2)$ \\[1mm]
adjoint  &  $N^2 -1$  &  ${\rm SU}(N-k) \otimes {\rm SU}(k)\otimes {\rm U}(1)\,, k=[ \frac{N}{2}]$ \\
 &   &  or $ {\rm SU}(N-1)$  \\
\hline\hline
\end{tabular}
\end{center}
\caption{The Higgs representations and the corresponding symmetry breaking patterns for the ${\rm SU}(N)$ group.}
\label{tab:HiggsPatterns}
\end{table}%

\para
We consider the following symmetry breaking pattern of the ${\rm SU}(8)$ theory
\beqn\label{eq:Pattern}
&& {\rm SU}(8) \xrightarrow{ v_U } \Gc_{441} \xrightarrow{ v_{441} } \Gc_{341} \xrightarrow{v_{341} } \Gc_{331} \xrightarrow{ v_{331} } \Gc_{\rm SM} \xrightarrow{ v_{\rm EW} } {\rm SU}(3)_{c}  \otimes  {\rm U}(1)_{\rm EM} \,, \non
&&\Gc_{441} \equiv {\rm SU}(4)_{s} \otimes {\rm SU}(4)_W \otimes  {\rm U}(1)_{X_0 } \,, ~ \Gc_{341} \equiv {\rm SU}(3)_{c} \otimes {\rm SU}(4)_W \otimes  {\rm U}(1)_{X_1 } \,,\non
&&\Gc_{331} \equiv {\rm SU}(3)_{c} \otimes {\rm SU}(3)_W \otimes  {\rm U}(1)_{X_2 } \,,~ \Gc_{\rm SM} \equiv  {\rm SU}(3)_{c} \otimes {\rm SU}(2)_W \otimes  {\rm U}(1)_{Y } \,,\non
&&  \textrm{with}~  v_U\gg v_{441}  \gg v_{341} \gg v_{331} \gg v_{\rm EW} \,.
\eeqn
At the GUT scale, the maximal symmetry breaking pattern of $ {\rm SU}(8) \xrightarrow{v_U  } \Gc_{441}$ is favored~\cite{Li:1973mq}, and gauge symmetries of both strong and weak sectors are extended.
This symmetry breaking pattern has three intermediate scales between the GUT scale of $v_U$ and the EW scale of $v_{\rm EW}$.
The GUT scale symmetry breaking will be due to the VEV of the ${\rm SU}(8)$ adjoint Higgs field.
All intermediate symmetry breaking stages will be due to the VEVs of the Higgs fields from the gauge-invariant Yukawa couplings of the ${\rm SU}(8)$ theory.
After the first stage of symmetry breaking pattern of $\Gc_{441} \to \Gc_{341}$ in Eq.~\eqref{eq:Pattern}, the sequential symmetry breaking stages are uniquely determined.
Along this symmetry breaking pattern, the non-anomalous global $\widetilde{ {\rm U}}(1)_T$ symmetry becomes the global $\widetilde{ {\rm U}}(1)_{B-L}$ at the EW scale according to the following sequence~\cite{Chen:2023qxi}
\beqn\label{eq:U1T_def}
&& \Gc_{441}~:~ \Tc^\prime \equiv \Tc - 4t \Xc_0 \,, \quad  \Gc_{341}~:~ \Tc^{ \prime \prime} \equiv \Tc^\prime + 8 t \Xc_1 \,, \non
&& \Gc_{331}~:~   \Tc^{ \prime \prime \prime} \equiv  \Tc^{ \prime \prime} \,, \quad  \Gc_{\rm SM}~:~  \Bc- \Lc \equiv  \Tc^{ \prime \prime \prime} \,,
\eeqn
where the charges of $\Xc_0$ and $\Xc_1$ are defined by Eq.~\eqref{eq:X0charge} and \eqref{eq:X1charge_4sfund}.
With these definitions, we have also checked that all Higgs components that can develop the VEVs at each symmetry breaking stage are neutral under the corresponding global $\widetilde{ {\rm U}}(1)_{T}$ symmetries~\cite{Chen:2023qxi}.

\subsection{Decompositions of the ${\rm SU}(8)$ fermions}
\label{section:SU8_fermions}

\begin{table}[htp] {\small
\begin{center}
\begin{tabular}{c|c|c|c|c}
\hline \hline
   $\gSU(8)$   &  $\Gc_{441}$  & $\Gc_{341}$  & $\Gc_{331}$  &  $\Gc_{\rm SM}$  \\
\hline \hline
 $\repb{ 8_F}^\Omega $   & $( \repb{4} \,, \rep{1}\,,  +\frac{1}{4} )_{ \mathbf{F} }^\Omega$  & $(\repb{3} \,, \rep{1} \,, +\frac{1}{3} )_{ \mathbf{F} }^\Omega $  & $(\repb{3} \,, \rep{1} \,, +\frac{1}{3} )_{ \mathbf{F} }^\Omega $  &  $( \repb{3} \,, \rep{ 1}  \,, +\frac{1}{3} )_{ \mathbf{F} }^{\Omega }~:~ { \Dc_R^\Omega }^c$  \\[1mm]
 &  &  $( \rep{1} \,, \rep{1} \,, 0)_{ \mathbf{F} }^{\Omega }$  &  $( \rep{1} \,, \rep{1} \,, 0)_{ \mathbf{F} }^{\Omega }$ &  $( \rep{1} \,, \rep{1} \,, 0)_{ \mathbf{F} }^{\Omega } ~:~ \check \Nc_L^{\Omega }$  \\[1.5mm]
 & $(\rep{1}\,, \repb{4}  \,,  -\frac{1}{4})_{ \mathbf{F} }^\Omega $  &  $(\rep{1}\,, \repb{4}  \,,  -\frac{1}{4})_{ \mathbf{F} }^\Omega$  &  $( \rep{1} \,, \repb{3} \,,  -\frac{1}{3})_{ \mathbf{F} }^{\Omega }$  &  $( \rep{1} \,, \repb{2} \,,  -\frac{1}{2})_{ \mathbf{F} }^{\Omega } ~:~\Lc_L^\Omega =( \Ec_L^\Omega \,, - \Nc_L^\Omega )^T$   \\[1mm]
 &   &   &   &  $( \rep{1} \,, \rep{1} \,,  0)_{ \mathbf{F} }^{\Omega^\prime} ~:~ \check \Nc_L^{\Omega^\prime }$  \\[1mm]
  &   &  &   $( \rep{1} \,, \rep{1} \,, 0)_{ \mathbf{F} }^{\Omega^{\prime\prime} }$ &   $( \rep{1} \,, \rep{1} \,, 0)_{ \mathbf{F} }^{\Omega^{\prime\prime} } ~:~ \check \Nc_L^{\Omega^{\prime \prime} }$   \\[1mm]   
\hline\hline
\end{tabular}
\caption{
The $\gSU(8)$ fermion representation of $\repb{8_F}^\Omega$ under the $\Gc_{441}\,,\Gc_{341}\,, \Gc_{331}\,, \Gc_{\rm SM}$ subgroups for the three-generational ${\rm SU}(8)$ theory, with $\Omega\equiv (\omega \,, \dot \omega )$.
Here, we denote $\underline{ {\Dc_R^\Omega}^c={d_R^\Omega}^c}$ for the SM right-handed down-type quarks, and ${\Dc_R^\Omega}^c={\DG_R^\Omega}^c$ for the right-handed down-type heavy partner quarks.
Similarly, we denote $\underline{ \Lc_L^\Omega = ( \ell_L^\Omega \,, - \nu_L^\Omega)^T}$ for the left-handed SM lepton doublets, and $\Lc_L^\Omega =( \eG_L^\Omega \,, - \nG_L^\Omega )^T$ for the left-handed heavy lepton doublets.
All left-handed neutrinos of $\check \Nc_L$ are sterile neutrinos, which are $\Gc_{\rm SM}$-singlets and do not couple to the EW gauge bosons.
}
\label{tab:SU8_8barferm}
\end{center}
}
\end{table}%

\begin{table}[htp] {\small
\begin{center}
\begin{tabular}{c|c|c|c|c}
\hline \hline
   $\gSU(8)$   &  $\Gc_{441}$  & $\Gc_{341}$  & $\Gc_{331}$  &  $\Gc_{\rm SM}$  \\
\hline \hline
 $\rep{28_F}$   & $( \rep{6}\,, \rep{ 1} \,, - \frac{1}{2})_{ \mathbf{F}}$ &  $ ( \rep{3}\,, \rep{ 1} \,, - \frac{1}{3})_{ \mathbf{F}}$   & $( \rep{3}\,, \rep{ 1} \,, - \frac{1}{3})_{ \mathbf{F}}$  & $( \rep{3}\,, \rep{ 1} \,, - \frac{1}{3})_{ \mathbf{F}} ~:~\DG_L$  \\[1mm]
                        &   & $( \repb{3}\,, \rep{ 1} \,, - \frac{2}{3})_{ \mathbf{F}}$  & $( \repb{3}\,, \rep{ 1} \,, - \frac{2}{3})_{ \mathbf{F}}$  & $\underline{( \repb{3}\,, \rep{ 1} \,, - \frac{2}{3})_{ \mathbf{F}}~:~ {t_R }^c }$   \\[1.5mm]
                        & $( \rep{1}\,, \rep{ 6} \,, +\frac{1}{2})_{ \mathbf{F}}$ & $( \rep{1}\,, \rep{ 6} \,, +\frac{1}{2})_{ \mathbf{F}}$   &  $( \rep{1}\,, \rep{ 3} \,, +\frac{1}{3})_{ \mathbf{F}}$ & $( \rep{1}\,, \rep{2} \,, +\frac{1}{2})_{ \mathbf{F}} ~:~( {\eG_R }^c \,, { \nG_R }^c)^T$  \\[1mm]
                       &   &   &   & $( \rep{1}\,, \rep{1} \,, 0 )_{ \mathbf{F}} ~:~ \check \nG_R^c $ \\[1mm]
                       &   &   & $( \rep{1}\,, \repb{ 3} \,, +\frac{2}{3})_{ \mathbf{F}}$  & $( \rep{1}\,, \repb{2} \,, +\frac{1}{2})_{ \mathbf{F}}^\prime ~:~( { \nG_R^{\prime} }^c\,, - {\eG_R^{\prime} }^c  )^T$   \\[1mm]
                       &   &   &   & $\underline{ ( \rep{1}\,, \rep{1} \,, +1 )_{ \mathbf{F}} ~:~ {\tau_R}^c}$ \\[1.5mm]
                        & $( \rep{4}\,, \rep{4} \,,  0)_{ \mathbf{F}}$ &  $( \rep{3}\,, \rep{4} \,,  -\frac{1}{12})_{ \mathbf{F}}$   & $( \rep{3}\,, \rep{3} \,,  0)_{ \mathbf{F}}$  & $\underline{ ( \rep{3}\,, \rep{2} \,,  +\frac{1}{6})_{ \mathbf{F}}~:~ (t_L\,, b_L)^T}$  \\[1mm]
                        &   &   &   & $( \rep{3}\,, \rep{1} \,,  -\frac{1}{3})_{ \mathbf{F}}^{\prime} ~:~\DG_L^\prime$  \\[1mm]
                        &   &   & $( \rep{3}\,, \rep{1} \,,  -\frac{1}{3})_{ \mathbf{F}}^{\prime\prime}$  & $( \rep{3}\,, \rep{1} \,,  -\frac{1}{3})_{ \mathbf{F}}^{\prime\prime} ~:~\DG_L^{\prime \prime}$ \\[1mm]
                        &   & $ ( \rep{1}\,, \rep{4} \,,  +\frac{1}{4} )_{ \mathbf{F}}$  & $( \rep{1}\,, \rep{3} \,,  +\frac{1}{3} )_{ \mathbf{F}}^{\prime\prime}$  & $( \rep{1}\,, \rep{2} \,,  +\frac{1}{2} )_{ \mathbf{F}}^{\prime\prime} ~:~( {\eG_R^{\prime\prime} }^c \,, { \nG_R^{\prime\prime}}^c )^T$  \\[1mm]
                        &   &   &   & $( \rep{1}\,, \rep{1}\,, 0)_{ \mathbf{F}}^{\prime} ~:~ \check \nG_R^{\prime\,c}$ \\[1mm]  
                        &   &   & $( \rep{1}\,, \rep{1}\,, 0)_{ \mathbf{F}}^{\prime\prime}$ & $( \rep{1}\,, \rep{1}\,, 0)_{ \mathbf{F}}^{\prime\prime} ~:~\check \nG_R^{\prime \prime \,c}$ \\[1mm]  
\hline\hline
\end{tabular}
\caption{
The $\gSU(8)$ fermion representation of $\rep{28_F}$ under the $\Gc_{441}\,,\Gc_{341}\,, \Gc_{331}\,, \Gc_{\rm SM}$ subgroups for the three-generational ${\rm SU}(8)$ theory.
All SM fermions are marked with underlines.}
\label{tab:SU8_28ferm}
\end{center}
}
\end{table}%

\begin{table}[htp] {\small
\begin{center}
\begin{tabular}{c|c|c|c|c}
\hline \hline
   $\gSU(8)$   &  $\Gc_{441}$  & $\Gc_{341}$  & $\Gc_{331}$  &  $\Gc_{\rm SM}$  \\
\hline \hline
     $\rep{56_F}$   & $( \rep{ 1}\,, \repb{4} \,, +\frac{3}{4})_{ \mathbf{F}}$  &  $( \rep{ 1}\,, \repb{4} \,, +\frac{3}{4})_{ \mathbf{F}}$ & $( \rep{ 1}\,, \repb{3} \,, +\frac{2}{3})_{ \mathbf{F}}^\prime$   &  $( \rep{ 1}\,, \repb{2} \,, +\frac{1}{2})_{ \mathbf{F}}^{\prime\prime\prime} ~:~( {\nG_R^{\prime\prime\prime }}^c \,, -{\eG_R^{\prime\prime\prime } }^c )^T$  \\[1mm]
                            &   &   &   & $\underline{( \rep{ 1}\,, \rep{1} \,, +1)_{ \mathbf{F}}^{\prime} ~:~ {\mu_R}^c}$ \\[1mm]
                            &   &   & $( \rep{ 1}\,, \rep{1} \,, +1)_{ \mathbf{F}}^{\prime\prime}$  & $( \rep{ 1}\,, \rep{1} \,, +1)_{ \mathbf{F}}^{\prime \prime} ~:~{\EG_R}^c$   \\[1.5mm]
                       & $( \repb{ 4}\,, \rep{1} \,, -\frac{3}{4})_{ \mathbf{F}}$  &  $( \repb{3}\,, \rep{1} \,, -\frac{2}{3})_{ \mathbf{F}}^{\prime}$ & $( \repb{3}\,, \rep{1} \,, -\frac{2}{3})_{ \mathbf{F}}^\prime$  & $\underline{ ( \repb{3}\,, \rep{1} \,, -\frac{2}{3})_{ \mathbf{F}}^{\prime} ~:~{u_R}^c }$ \\[1mm]
                       &   &  $( \rep{1}\,, \rep{1} \,, -1)_{ \mathbf{F}}$ & $( \rep{1}\,, \rep{1} \,, -1)_{ \mathbf{F}}$  &  $( \rep{1}\,, \rep{1} \,, -1)_{ \mathbf{F}} ~:~\EG_L$  \\[1.5mm]
                       & $( \rep{ 4}\,, \rep{6} \,, +\frac{1}{4})_{ \mathbf{F}}$  &  $( \rep{3}\,, \rep{6} \,, +\frac{1}{6})_{ \mathbf{F}}$ & $( \rep{3}\,, \rep{3} \,, 0 )_{ \mathbf{F}}^\prime$ & $\underline{ ( \rep{3}\,, \rep{2} \,, +\frac{1}{6} )_{ \mathbf{F}}^{\prime} ~:~ ( c_L \,, s_L)^T} $  \\[1mm]
                       &   &   &   & $( \rep{3}\,, \rep{1} \,, -\frac{1}{3})_{ \mathbf{F}}^{\prime\prime \prime } ~:~\DG_L^{\prime \prime \prime}$ \\[1mm]
                       &   &   & $( \rep{3}\,, \repb{3} \,, +\frac{1}{3})_{ \mathbf{F}}$ & $( \rep{3}\,, \repb{2} \,, +\frac{1}{6})_{ \mathbf{F}}^{\prime\prime} ~:~ (\dG_L \,, - \uG_L )^T$   \\[1mm]
                       &   &   &   & $( \rep{3}\,, \rep{1} \,, +\frac{2}{3})_{ \mathbf{F}} ~:~\UG_L$  \\[1mm]
                       &   & $( \rep{1}\,, \rep{6} \,, +\frac{1}{2})_{ \mathbf{F}}^\prime$ & $( \rep{1}\,, \rep{3} \,, +\frac{1}{3})_{ \mathbf{F}}^\prime $ & $( \rep{1}\,, \rep{2} \,, +\frac{1}{2})_{ \mathbf{F}}^{\prime\prime \prime \prime} ~:~ ( {\eG_R^{\prime\prime\prime\prime }}^c \,, {\nG_R^{\prime\prime\prime\prime } }^c )^T$ \\[1mm]
                       &   &   &   & $( \rep{1}\,, \rep{1} \,, 0 )_{ \mathbf{F}}^{\prime\prime \prime} ~:~ {\check \nG_R}^{\prime \prime\prime \,c}$ \\[1mm]
                       &   &   & $( \rep{1}\,, \repb{3} \,, +\frac{2}{3})_{ \mathbf{F}}^{\prime\prime}$  & $( \rep{1}\,, \repb{2} \,, +\frac{1}{2})_{ \mathbf{F}}^{\prime\prime \prime \prime \prime} ~:~( {\nG_R^{\prime\prime\prime\prime\prime }}^c \,, -{\eG_R^{\prime\prime\prime\prime\prime } }^c )^T$  \\[1mm]
                       &   &   &   & $\underline{ ( \rep{1}\,, \rep{1} \,, +1 )_{ \mathbf{F}}^{\prime\prime \prime } ~:~ {e_R}^c }$ \\[1.5mm]
                       & $( \rep{ 6}\,, \rep{4} \,, -\frac{1}{4})_{ \mathbf{F}}$  & $( \rep{3}\,, \rep{4} \,, -\frac{1}{12})_{ \mathbf{F}}^\prime$ & $( \rep{3}\,, \rep{3} \,, 0)_{ \mathbf{F}}^{\prime\prime}$ & $\underline{ ( \rep{3}\,, \rep{2} \,, +\frac{1}{6})_{ \mathbf{F}}^{\prime\prime \prime } ~:~  (u_L \,, d_L)^T} $ \\[1mm]
                       &   &   &   &  $( \rep{3}\,, \rep{1} \,, -\frac{1}{3})_{ \mathbf{F}}^{\prime \prime \prime \prime} ~:~\DG_L^{\prime \prime \prime\prime}$ \\[1mm]
                       &   &   &  $( \rep{3}\,, \rep{1} \,, -\frac{1}{3})_{ \mathbf{F}}^{\prime \prime \prime\prime \prime}$ & $( \rep{3}\,, \rep{1} \,, -\frac{1}{3})_{ \mathbf{F}}^{\prime \prime \prime \prime \prime} ~:~ \DG_L^{\prime \prime \prime\prime \prime}$ \\[1mm]
                       &   & $( \repb{3}\,, \rep{4} \,, -\frac{5}{12})_{ \mathbf{F}}$ & $( \repb{3}\,, \rep{3} \,, -\frac{1}{3})_{ \mathbf{F}}$ & $( \repb{3}\,, \rep{2} \,, -\frac{1}{6})_{ \mathbf{F}} ~:~ ( {\dG_R}^c \,,{\uG_R}^c )^T$  \\[1mm]
                       &   &   &   & $( \repb{3}\,, \rep{1} \,, -\frac{2}{3})_{ \mathbf{F}}^{\prime \prime} ~:~{\UG_R}^c$  \\[1mm]
                       &   &   & $( \repb{3}\,, \rep{1} \,, -\frac{2}{3})_{ \mathbf{F}}^{\prime \prime \prime}$ & $\underline{ ( \repb{3}\,, \rep{1} \,, -\frac{2}{3})_{ \mathbf{F}}^{\prime \prime \prime} ~:~{c_R}^c }$  \\[1mm]
\hline\hline
\end{tabular}
\caption{
The $\gSU(8)$ fermion representation of $\rep{56_F}$ under the $\Gc_{441}\,,\Gc_{341}\,, \Gc_{331}\,, \Gc_{\rm SM}$ subgroups for the three-generational ${\rm SU}(8)$ theory.
All SM fermions are marked with underlines.
}
\label{tab:SU8_56ferm}
\end{center}
}
\end{table}%

\para
By following the symmetry breaking pattern in Eq.~\eqref{eq:Pattern}, we tabulate the fermion representations at various stages of the ${\rm SU}(8)$ theory in Tabs.~\ref{tab:SU8_8barferm}, \ref{tab:SU8_28ferm}, and \ref{tab:SU8_56ferm}.
All ${\rm U}(1)$ charges at different stages are obtained according to Eqs.~\eqref{eq:X1charge_4sfund}, \eqref{eq:X2charge_4Wfund}, \eqref{eq:Ycharge_4Wfund}, and~\eqref{eq:Qcharge_4Wfund}.
All chiral fermions are named in terms of their $\Gc_{\rm SM}$ irreps.
For the right-handed down-type quarks of ${\Dc_R^\Omega}^c$, they are named as follows
\beqn\label{eq:DR_names}
&& {\Dc_R^{ \dot 1} }^c \equiv {d_R}^c \,, ~ {\Dc_R^{  \dot 2} }^c \equiv {s_R}^c \,,~ {\Dc_R^{\dot {\rm VII} } }^c \equiv {\DG_R^{\prime\prime\prime\prime \prime }}^c \,, ~  {\Dc_R^{\dot {\rm VIII} } }^c \equiv {\DG_R^{\prime\prime \prime }}^c  \,,~ {\Dc_R^{  \dot {\rm IX} } }^c \equiv {\DG_R^{\prime\prime\prime \prime }}^c \,, \non
&&{\Dc_R^{3} }^c \equiv {b_R}^c \,,~  {\Dc_R^{\rm IV } }^c \equiv {\DG_R}^c \,, ~  {\Dc_R^{\rm V } }^c \equiv {\DG_R^{\prime\prime }}^c  \,,~ {\Dc_R^{\rm VI } }^c \equiv {\DG_R^{\prime }}^c  \,.
\eeqn
For the left-handed ${\rm SU}(2)_W$ lepton doublets of $(\Ec_L^\Omega \,, - \Nc_L^\Omega )$, they are named as follows
\beqn\label{eq:ELNL_names}
&&   ( \Ec_L^{ \dot 1} \,,  - \Nc_L^{\dot 1})  \equiv (e_L\,, - \nu_{e\,L} ) \,, ~( \Ec_L^{  \dot 2} \,,   - \Nc_L^{\dot 2})  \equiv( \mu_L \,, - \nu_{\mu\,L} )  \,, \non
&&  ( \Ec_L^{ \dot {\rm VII} } \,,  - \Nc_L^{ \dot {\rm VII} })  \equiv ( \eG_L^{ \prime\prime \prime \prime} \,, - \nG_L^{\prime\prime \prime \prime } )  \,, ~ ( \Ec_L^{ \dot {\rm VIII} } \,, - \Nc_L^{ \dot {\rm VIII} }  )  \equiv ( \eG_L^{ \prime\prime  \prime} \,, - \nG_L^{\prime\prime \prime} ) \,,~  ( \Ec_L^{\dot {\rm IX} } \,,  - \Nc_L^{ \dot {\rm IX} } ) \equiv ( \eG_L^{ \prime\prime \prime \prime \prime } \,,  - \nG_L^{\prime\prime \prime\prime \prime} )  \,,\non
&&  ( \Ec_L^{ 3} \,, - \Nc_L^{3}) \equiv ( \tau_L \,, - \nu_{\tau\,L})\,,~  ( \Ec_L^{\rm IV } \,, - \Nc_L^{\rm IV }) \equiv ( \eG_L^{\prime\prime} \,, - \nG_L^{\prime\prime} ) \,, \non
&& ( \Ec_L^{\rm V } \,, -  \Nc_L^{\rm V }) \equiv ( \eG_L \,, - \nG_L )  \,,~ ( \Ec_L^{\rm VI } \,, - \Nc_L^{\rm VI } ) \equiv  ( \eG_L^\prime\,, - \nG_L^\prime ) \,.
\eeqn
Through the analysis in Sec.~\ref{section:pattern}, we shall see that all heavy $(\Dc^\Omega\,, \Ec^\Omega\,, \Nc^\Omega)$ (with $\Omega={\rm IV}\,, \ldots \,,\dot {\rm IX}$) acquire vectorlike masses during the intermediate symmetry breaking stages.
For the remaining left-handed sterile neutrinos of $(  \check \Nc_L^\Omega \,, \check \Nc_L^{\Omega^\prime } \,, \check \Nc_L^{\Omega^{ \prime\prime} } )$, several of them are massive and they are named as follows
\beqn\label{eq:stNL_names}
&& \check \Nc_L^{{\rm IV}^\prime }  \equiv \check \nG_L^\prime \,,~   \check \Nc_L^{{\rm IV}^{\prime \prime} }  \equiv \check \nG_L^{\prime \prime} \,, ~ \check \Nc_L^{{\rm V}^{\prime } }  \equiv \check \nG_L \,,~  \check \Nc_L^{\dot {\rm VII}^{\prime} }  \equiv \check \nG_L^{\prime \prime \prime } \,.
\eeqn
%
%
%
%
%
%
The flavor indices in Eqs.~\eqref{eq:DR_names} and \eqref{eq:ELNL_names} are chosen in accordance with the symmetry breaking pattern depicted in Sec.~\ref{section:pattern}.
The SM flavor names in Tabs.~\ref{tab:SU8_8barferm}, \ref{tab:SU8_28ferm}, and~\ref{tab:SU8_56ferm} will be justified according to the inter-generational mass hierarchies and the CKM mixing pattern in Sec.~\ref{section:SMql_CKM}.
Nevertheless, the decompositions of the ${\rm SU}(8)$ fermions have already confirmed Georgi's counting rule~\cite{Georgi:1979md}, which is obviously independent of the symmetry breaking pattern.

\para
Through Tabs.~\ref{tab:SU8_8barferm}, \ref{tab:SU8_28ferm}, and \ref{tab:SU8_56ferm}, it is straightforward to find three-generational SM fermions transform differently in the UV theories.
For example, three left-handed quark doublets belong to $(\rep{3}\,, \rep{2}\,, +\frac{1}{6} )_{\mathbf{F}}\subset (\rep{4}\,, \rep{4}\,, 0 )_{\mathbf{F}}$, $(\rep{3}\,, \rep{2}\,, +\frac{1}{6} )_{\mathbf{F}}^\prime \subset (\rep{4}\,, \rep{6}\,, +\frac{1}{4} )_{\mathbf{F}}$, and $(\rep{3}\,, \rep{2}\,, +\frac{1}{6} )_{\mathbf{F}}^{\prime\prime } \subset (\rep{6}\,, \rep{4}\,, -\frac{1}{4} )_{\mathbf{F}}$, when one traces back their irreps in the $\Gc_{441}$ theory.
Likewise, three $(\repb{3}\,, \rep{1}\,, -\frac{2}{3} )_{\mathbf{F}}$'s and three $(\rep{1}\,, \rep{1}\,, +1 )_{\mathbf{F}}$'s all stem from distinctive $\Gc_{441}$ irreps.
Thus, the flavor non-universality under the flavor-conserving neutral currents of the $\Gc_{441}$ theory can be expected~\footnote{Some of the earlier and recent references addressing the flavor non-universality scenario with product gauge groups include~\cite{Li:1981nk,Ma:1987ds,Li:1992fi,Craig:2011yk,Bordone:2017bld,Fuentes-Martin:2020pww,Davighi:2022fer,Fuentes-Martin:2022xnb,Davighi:2022bqf,FernandezNavarro:2023rhv,FernandezNavarro:2023hrf}.}.
When the theory flows to the $\Gc_{331}$ along the symmetry breaking pattern in Eq.~\eqref{eq:Pattern}, all three-generational SM fermions together with their heavy partner fermions, have already transformed the same.
Therefore, the flavor universality under the flavor-conserving neutral currents of the $\Gc_{331}$ theory can be expected.

\subsection{Decompositions of the ${\rm SU}(8)$ Higgs fields}
\label{section:SU8_Higgs}

\para
We decompose the Higgs fields in Eq.~\eqref{eq:Yukawa_SU8} into components that can be responsible for the sequential symmetry breaking pattern in Eq.~\eqref{eq:Pattern}.
For Higgs fields of $\repb{8_H}_{\,,\omega }$, they read
\beqn\label{eq:SU8_Higgs_Br01}
\repb{8_H}_{\,,\omega }  &\supset&  \langle  ( \repb{4} \,, \rep{1} \,, +\frac{1}{4} )_{\mathbf{H}\,, \omega }  \rangle \oplus  \underline{ ( \rep{1} \,, \repb{4} \,, -\frac{1}{4} )_{\mathbf{H}\,, \omega } }\non
&\supset&  \langle  ( \rep{1} \,, \repb{4} \,, -\frac{1}{4} )_{\mathbf{H}\,, \omega }  \rangle  \supset  \langle ( \rep{1} \,, \repb{3} \,, -\frac{1}{3} )_{\mathbf{H}\,, \omega } \rangle \supset \langle ( \rep{1} \,, \repb{2} \,, -\frac{1}{2} )_{\mathbf{H}\,, \omega } \rangle  \,.
\eeqn
For Higgs fields of $\repb{28_H}_{\,,\dot \omega } $, they read
\beqn\label{eq:SU8_Higgs_Br02}
\repb{28_H}_{\,,\dot \omega } &\supset& ( \rep{6} \,, \rep{1} \,,  +\frac{1}{2} )_{\mathbf{H}\,, \dot\omega }  \oplus  \underline{ ( \rep{1} \,, \rep{6} \,, -\frac{1}{2} )_{\mathbf{H}\,, \dot\omega } } \oplus \underline{ ( \repb{4} \,, \repb{4} \,, 0 )_{\mathbf{H}\,, \dot\omega } }  \non
&\supset & \underline{  ( \rep{1} \,, \rep{6} \,, -\frac{1}{2} )_{\mathbf{H}\,, \dot\omega }  } \oplus \langle ( \rep{1} \,, \repb{4} \,, -\frac{1}{4} )_{\mathbf{H}\,, \dot\omega }  \rangle  \non
&\supset& \[ \langle  ( \rep{1} \,, \repb{3} \,, -\frac{1}{3} )_{\mathbf{H}\,, \dot\omega }^\prime \rangle  \oplus \underline{ ( \rep{1} \,, \rep{3} \,, -\frac{2}{3} )_{\mathbf{H}\,, \dot\omega }}  \] \oplus \langle  ( \rep{1} \,, \repb{3} \,, -\frac{1}{3} )_{\mathbf{H}\,, \dot\omega } \rangle   \non
&\supset& \[ \langle ( \rep{1} \,, \repb{2} \,, -\frac{1}{2} )_{\mathbf{H}\,, \dot\omega }^\prime \rangle \oplus \langle (  \rep{1} \,, \rep{2} \,, -\frac{1}{2}  )_{\mathbf{H}\,, \dot\omega }  \rangle  \] \oplus \langle ( \rep{1} \,, \repb{2} \,, -\frac{1}{2} )_{\mathbf{H}\,, \dot\omega } \rangle  \,.
\eeqn
%
%
%
For Higgs field of $\rep{70_H}$, they read
\beqn\label{eq:SU8_Higgs_Br05}
\rep{70_H} &\supset& ( \rep{1} \,, \rep{1 } \,, -1 )_{\mathbf{H}}^{ \prime \prime } \oplus ( \rep{1} \,, \rep{1 } \,, +1 )_{\mathbf{H}}^{ \prime \prime \prime \prime } \oplus \underline{ ( \rep{4} \,, \repb{4} \,, +\frac{1}{2} )_{\mathbf{H}} } \oplus  ( \repb{4} \,, \rep{4} \,, -\frac{1}{2} )_{\mathbf{H}}   \oplus ( \rep{6 } \,, \rep{6 } \,, 0 )_{\mathbf{H}}  \non
&\supset& \underline{ ( \rep{1} \,, \repb{4} \,, +\frac{3}{4} )_{\mathbf{H}}^\prime } \supset  \underline{ ( \rep{1} \,, \repb{3} \,, +\frac{2}{3} )_{\mathbf{H}}^{\prime\prime\prime} }  \supset  \langle ( \rep{1} \,, \repb{2} \,, +\frac{1}{2} )_{\mathbf{H}}^{\prime \prime\prime } \rangle \,.
\eeqn
%
%

\begin{table}[htp]
\begin{center}
\begin{tabular}{c|cccc}
\hline \hline
Higgs & $\Gc_{441}\to \Gc_{341}$ & $\Gc_{341}\to \Gc_{331}$  &  $\Gc_{331}\to \Gc_{\rm SM}$ & $\Gc_{\rm SM} \to {\rm SU}(3)_{c}  \otimes  {\rm U}(1)_{\rm EM}$  \\
\hline
$\repb{8_H}_{\,,\omega }$ & \cmark$\,(\omega ={\rm IV})$  & \cmark$\,(\omega ={\rm V})$  &  \cmark$\,(\omega =3\,, {\rm VI} )$  & \cmark  \\
$\repb{28_H}_{\,,\dot \omega }$ &  \xmark  & \cmark$\,(\dot \omega = \dot 1\,, \dot {\rm VII} )$  &  \cmark$\,(\dot \omega =\dot 2 \,, \dot {\rm VIII}\,, \dot{\rm IX})$  & \cmark  \\
$\rep{70_H}$ & \xmark  & \xmark  & \xmark  & \cmark \\
\hline\hline
\end{tabular}
\end{center}
\caption{The Higgs fields and their symmetry breaking directions in the ${\rm SU}(8)$ theory.
The \cmark and \xmark represent possible and impossible symmetry breaking directions for a given Higgs field.
The flavor indices for the Higgs fields that develop VEVs at different stages are also specified in the parentheses.
}
\label{tab:SU8Higgs_directions}
\end{table}%

\para
All Higgs components that are likely to develop VEVs for the sequential symmetry breaking stages are labelled by $\langle ... \rangle$, and other ${\rm SU}(3)_c\otimes {\rm U}(1)_{\rm EM}$-exact components are neglected.
Accordingly, we mark the allowed/disallowed symmetry breaking directions for all Higgs fields in Tab.~\ref{tab:SU8Higgs_directions}.
At this point, these components are determined by whether they contain the singlets in the gauge groups of the broken symmetries.
Given that the $\rep{70_H}$ is self-conjugate, one may speculate two EWSB components of $( \rep{1} \,, \repb{2} \,, +\frac{1}{2} )_{\mathbf{H}}^{\prime \prime\prime }\subset( \rep{4} \,, \repb{4} \,, +\frac{1}{2} )_{\mathbf{H}}$ and $( \rep{1} \,, \rep{2} \,, -\frac{1}{2} )_{\mathbf{H}}^{\prime \prime\prime }\subset( \repb{4} \,, \rep{4} \,, -\frac{1}{2} )_{\mathbf{H}}$ from the $\rep{70_H}$.
In Ref.~\cite{Chen:2023qxi}, we have proved that only one component of $( \rep{1} \,, \repb{2} \,, +\frac{1}{2} )_{\mathbf{H}}^{\prime \prime\prime }\subset \rep{70_H}$ is $\widetilde{ {\rm U}}(1)_{B-L}$-neutral according to the definition in Eq.~\eqref{eq:U1T_def}, and thus can develop the EWSB VEV.
Furthermore, this component also gives the top quark mass through the $\Oc(1)$ Yukawa coupling of $Y_\Tc \rep{28_F} \rep{28_F} \rep{70_H} +H.c.$, according to Eq.~\eqref{eq:top_Yukawa} below~\footnote{The naturalness of the top quark mass at the EW scale was previously pointed our in Ref.~\cite{Haba:2000be}.}.
An immediate result~\footnote{This was pointed out in Ref.~\cite{Barr:2008pn} previously.} is that the third-generational SM $\rep{10_F}$ must reside in the $\rep{28_F}$, while the first- and second-generational SM $\rep{10_F}$'s must reside in the $\rep{56_F}$.

\para
Schematically, we assign the Higgs VEVs according to the symmetry breaking pattern in Eq.~\eqref{eq:Pattern} as follows
\beqs\label{eqs:SU8_Higgs_VEVs_mini}
\beqn
\Gc_{441} \to \Gc_{341} ~&:&~ \langle ( \repb{4} \,, \rep{1} \,, +\frac{1}{4} )_{\mathbf{H}\,, {\rm IV}} \rangle \equiv \frac{1 }{ \sqrt{2} } W_{ \repb{4}\,, {\rm IV}}\,, \label{eq:SU8_Higgs_VEVs_mini01}\\[1mm]
\Gc_{341} \to \Gc_{331} ~&:&~ \langle ( \rep{1} \,, \repb{4} \,, -\frac{1}{4} )_{\mathbf{H}\,, {\rm V} } \rangle \equiv \frac{1 }{ \sqrt{2} } w_{\repb{4}\,, {\rm V} }\,, ~   \langle ( \rep{1} \,, \repb{4} \,, -\frac{1}{4} )_{\mathbf{H}\,,  \dot 1\,, \dot {\rm VII} } \rangle \equiv \frac{1 }{ \sqrt{2} } w_{\repb{4}\,,  \dot 1\,, \dot {\rm VII} }  \,, \label{eq:SU8_Higgs_VEVs_mini02}\\[1mm]
\Gc_{331} \to \Gc_{\rm SM} ~&:&~  \langle ( \rep{1} \,, \repb{3} \,, -\frac{1}{3} )_{\mathbf{H}\,, 3\,, {\rm VI}} \rangle \equiv \frac{1 }{ \sqrt{2} } V_{ \repb{3}\,, 3\,, {\rm VI} } \non
&&   \langle ( \rep{1} \,, \repb{3} \,, -\frac{1}{3} )_{\mathbf{H}\,, \dot {\rm IX} } \rangle \equiv \frac{1 }{ \sqrt{2} }V_{ \repb{3}\,, \dot {\rm IX} }  \,, ~  \langle ( \rep{1} \,, \repb{3} \,, -\frac{1}{3} )_{\mathbf{H}\,, \dot 2\,, \dot {\rm VIII}  }^\prime \rangle \equiv \frac{1 }{ \sqrt{2} } V_{ \repb{3}\,, \dot 2\,, \dot {\rm VIII} }^\prime  \,, \label{eq:SU8_Higgs_VEVs_mini03} \\[1mm]
 {\rm EWSB} ~&:&~   \langle ( \rep{1} \,, \repb{2} \,, +\frac{1}{2} )_{\mathbf{H} }^{ \prime\prime \prime} \rangle \equiv \frac{1 }{ \sqrt{2} } v_{\rm EW}   \,.\label{eq:SU8_Higgs_VEVs_mini05}
\eeqn
\eeqs
For our later convenience, we parametrize different symmetry breaking VEVs
\beqn\label{eq:scale_params}
&& \zeta_0 \equiv \frac{ v_U }{ M_{\rm pl} } \,, \quad \zeta_1 \equiv \frac{ W_{ \repb{4}\,, {\rm IV} } }{ M_{\rm pl} } \,, \quad \zeta_2 \equiv \frac{ w_{ \repb{4}\,, {\rm V} }  }{ M_{\rm pl} } \,,  \quad \dot \zeta_2 \equiv \frac{ w_{ \repb{4}\,, \dot 1\,, \dot {\rm VII} }  }{ M_{\rm pl} }  \,,   \non
&&  \zeta_3 \equiv \frac{ V_{ \repb{3}\,, 3\,, {\rm VI} } }{ M_{\rm pl} } \,, \quad \dot \zeta_3^\prime \equiv \frac{ V_{ \repb{3}\,, \dot 2\,, \dot {\rm VIII} }^\prime  }{ M_{\rm pl} }   \,, \quad \dot \zeta_3 \equiv \frac{ V_{ \repb{3}\,, \dot {\rm IX} } }{ M_{\rm pl} }     \,, \non
&& \zeta_0 \gg \zeta_1 \gg  \zeta_2 \sim \dot \zeta_2  \gg \zeta_3 \sim \dot \zeta_3^\prime \sim  \dot \zeta_3  \,, \quad \zeta_{i j} \equiv \frac{ \zeta_j }{ \zeta_i } \,, \quad ( i < j ) \,,
\eeqn
in terms of dimensionless quantities.

\para
Several remarks on the Higgs VEV assignments in Eqs.~\eqref{eqs:SU8_Higgs_VEVs_mini} above are necessary.
\begin{enumerate}

\item If one eliminated the Higgs VEV components of $\omega=3$ and $\dot \omega=(\dot 1\,, \dot 2)$ in Eqs.~\eqref{eq:SU8_Higgs_VEVs_mini02} and \eqref{eq:SU8_Higgs_VEVs_mini03}, it is already sufficient for all electrically charged heavy partner fermions to obtain their vectorlike masses through the intermediate symmetry breaking stages with the remaining Higgs VEVs, which has been previously derived in Ref.~\cite{Chen:2023qxi} and will be revisited in Sec.~\ref{section:pattern}.

\item Through the decompositions in Eqs.~\eqref{eq:SU8_Higgs_Br01} and \eqref{eq:SU8_Higgs_Br02}, both the $\repb{8_H}_{\,,\omega } $ and the $\repb{28_H}_{\,,\dot \omega }$ contain the EWSB components.
According to the $\widetilde{ \rm U}(1)_{B-L}$ charges defined in Eq.~\eqref{eq:U1T_def}, these EWSB components are also $\widetilde{ \rm U}(1)_{B-L}$-neutral.
If these components from the $\repb{8_H}_{\,,\omega=3 } $ and the $\repb{28_H}_{\,,\dot \omega = \dot 1\,, \dot 2 }$ developed the EWSB VEVs as were indicated in Tab.~\ref{tab:SU8Higgs_directions}, one expects a total of four Higgs doublets at the EW scale together with the $( \rep{1} \,, \repb{2} \,, +\frac{1}{2} )_{\mathbf{H} }^{ \prime\prime \prime}\subset \rep{70_H}$, which then lead to thirteen physical Higgs bosons instead of one $125$ GeV SM Higgs boson that has been discovered at the LHC~\cite{ATLAS:2012yve,CMS:2012qbp}.
Obviously, this scenario is experimentally challenging with the current LHC results.

\item Alternatively, it will be more reasonable for relevant components from the $\repb{8_H}_{\,,\omega=3 } $ and the $\repb{28_H}_{\,,\dot \omega = \dot 1\,, \dot 2 }$ to develop VEVs at certain intermediate stages above the EW scale.
The VEV assignments for specific flavor indices in Eqs.~\eqref{eq:SU8_Higgs_VEVs_mini02} and \eqref{eq:SU8_Higgs_VEVs_mini03} are made such that the inter-generational down-type quark and charge lepton masses can become hierarchical.
The corresponding results will be demonstrated explicitly in Sec.~\ref{section:gravity_Fmass}.
According to the minimal survival hypothesis~\cite{Georgi:1979md,Glashow:1979nm,Barbieri:1979ag,Barbieri:1980vc,Barbieri:1980tz,delAguila:1980qag}, one expect all components of Higgs fields of $\repb{8_H}_{\,,\omega } $ and $\repb{28_H}_{\,,\dot \omega  }$ to obtain their masses no less than their VEVs in Eqs.~\eqref{eqs:SU8_Higgs_VEVs_mini}.

\item By denoting the EWSB VEV as $\langle ( \rep{1} \,, \repb{2} \,, +\frac{1}{2} )_{\mathbf{H} }^{ \prime\prime\prime } \rangle \equiv \frac{1 }{ \sqrt{2}} v_{\rm EW}$, we conjecture that the $( \rep{1} \,, \repb{2} \,, +\frac{1}{2} )_{\mathbf{H}}^{\prime \prime\prime }\subset \rep{70_H}$ is the only SM Higgs doublet which contains the $125$ GeV Higgs boson discovered at the LHC.
This conjecture is also supported from the natural tree-level top quark Yukawa coupling, as will be shown in Eq.~\eqref{eq:top_Yukawa} below.

\item All VEVs in Eqs.~\eqref{eqs:SU8_Higgs_VEVs_mini} are assumed to be real for simplicity.

\end{enumerate}

\section{The symmetry breaking pattern and vectorlike fermions in the ${\rm SU}(8)$}
\label{section:pattern}

\para
In this section, we describe the intermediate symmetry breaking stages between the GUT scale and the EW scale, where we focus on the Yukawa couplings in Eq.~\eqref{eq:Yukawa_SU8} that lead to the vectorlike fermion masses.
According to the discussions in Ref.~\cite{Chen:2023qxi}, the Higgs fields that can develop the VEVs for the specific symmetry breaking stage of $\Gc_1\to \Gc_2$ should be: (i) the gauge singlets of $\Gc_2$, (ii) neutral under the non-anomalous global $\widetilde{\rm U}(1)_T$ symmetries defined with the $\Gc_2$.

\subsection{The first stage}

\begin{table}[htp]
\begin{center}
\begin{tabular}{c|cccc}
\hline\hline
$\repb{8_F}^\Omega$ & $( \repb{4}\,, \rep{1}\,, +\frac{1}{4} )_{\rep{F}}^\Omega$  &  $( \rep{1}\,, \repb{4}\,,  -\frac{1}{4} )_{\rep{F}}^\Omega$  &   &     \\[1mm]
\hline
$\Tc^\prime$  &  $ -4t$  &  $-2t$  &   &     \\[1mm]
\hline
$\rep{28_F}$ &  $( \rep{6}\,, \rep{1}\,,  -\frac{1}{2} )_{\rep{F}}$ &  $( \rep{1}\,, \rep{ 6}\,,  + \frac{1}{2} )_{\rep{F}}$ &  $( \rep{4 }\,, \rep{4}\,,  0 )_{\rep{F}}$   &    \\[1mm]
\hline
$\Tc^\prime$  &  $4t$  & $0$ & $2t$   &    \\[1mm]
\hline
 $\rep{56_F}$  &  $( \rep{1}\,, \repb{4}\,,  +\frac{3}{4} )_{\rep{F}}$ &  $( \repb{4}\,,  \rep{1}\,,  -\frac{3}{4} )_{\rep{F}}$ & $( \rep{4}\,,  \rep{6}\,,  +\frac{1}{4} )_{\rep{F}}$  & $( \rep{6}\,,  \rep{4}\,,  -\frac{1}{4} )_{\rep{F}}$   \\[1mm]
 \hline
 $\Tc^\prime$  & $-2t$  &  $4t$  & $0$  & $2 t$  \\[1mm]
 \hline\hline
$\repb{8_H}_{\,, \omega }$  & $( \repb{4}\,, \rep{1}\,, +\frac{1}{4} )_{\rep{H}\,,\omega}$  &  $( \rep{1}\,, \repb{4}\,,  -\frac{1}{4} )_{\rep{H}\,,\omega }$  &   &      \\[1mm]
\hline
$\Tc^\prime$  &  $0$  &  $2t$  &   &     \\[1mm]
\hline
 $\repb{28_H}_{\,,\dot \omega}$ & $( \rep{1}\,, \rep{6}\,,  -\frac{1}{2} )_{\rep{H}\,, \dot \omega}$  &  $( \repb{4}\,, \repb{4}\,,  0 )_{\rep{H}\,, \dot \omega}$  &   &    \\[1mm]
  \hline
  $\Tc^\prime$  &  $4t$  &  $2t$  &      & $$  \\[1mm]
  \hline
$\rep{70_H}$  & $( \rep{4}\,, \repb{4}\,,  +\frac{1}{2} )_{\rep{H}}$   &  $( \repb{4}\,, \rep{4}\,,  - \frac{1}{2} )_{\rep{H}}$  &   &       \\[1mm]
\hline
$\Tc^\prime$  & $-6t$  & $-2t$ &   &      \\[1mm]
 \hline\hline
\end{tabular}
\end{center}
\caption{The $\widetilde{ {\rm U}}(1)_{T^\prime } $ charges for massless fermions and possible symmetry breaking Higgs components in the $\Gc_{441}$ theory.}
\label{tab:G441_Tcharges}
\end{table}%

\para
The field contents at this stage come from the GUT scale symmetry breaking of ${\rm SU}(8) \to \Gc_{441}$.
The non-anomalous $\widetilde {\rm U}(1)_{T^\prime}$ charges are tabulated for all $\Gc_{441}$ fields in Tab.~\ref{tab:G441_Tcharges}.
Accordingly, the first-stage symmetry breaking of $\Gc_{441} \to \Gc_{341}$ can only be achieved by $( \repb{4} \,, \rep{1} \,, +\frac{1}{4} )_{\mathbf{H}\,,\omega } \subset \repb{8_H}_{\,,\omega }$ in the rank-2 sector, given that it is both $\Gc_{341}$-invariant as well as $\widetilde {\rm U}(1)_{T^\prime}$-neutral.
The Yukawa coupling between $\repb{8_F}^\omega$ and $\rep{28_F}$ and the corresponding vectorlike fermion masses can be expressed as
\beqn\label{eq:Yukawa_441_01}
&& Y_\Bc  \repb{8_F}^\omega \rep{28_F} \repb{8_H}_{\,,\omega } + H.c. \non
&\supset& Y_\Bc  \Big[ ( \rep{1}\,, \repb{4}\,, -\frac{1}{4})_{ \mathbf{F}}^\omega \otimes ( \rep{4}\,, \rep{4}\,, 0 )_{ \mathbf{F}} \oplus ( \repb{4}\,, \rep{1}\,, +\frac{1}{4} )_{ \mathbf{F}}^\omega \otimes ( \rep{6}\,, \rep{1}\,, -\frac{1}{2} )_{ \mathbf{F}}   \Big] \otimes \langle ( \repb{4}\,, \rep{1}\,, +\frac{1}{4})_{ \mathbf{H}\,,\omega } \rangle + H.c. \non
%
%
&\Rightarrow& \frac{1}{\sqrt{2} }  Y_\Bc \Big( \eG_L^{\prime\prime }  {\eG_R^{\prime\prime}}^c - \nG_L^{\prime\prime}  {\nG_R^{\prime\prime}}^c +  \check \Nc_L^{ {\rm IV}^\prime}  \check \nG_R^{\prime\, c} + \check \Nc_L^{ {\rm IV}^{\prime\prime} }  \check \nG_R^{\prime\prime\, c} +  \DG_L  {\DG_R}^c  \Big) W_{ \repb{4}\,,{\rm IV} }  + H.c.  \,.
\eeqn
Without loss of generality, we choose the massive anti-fundamental fermion to be $\omega ={\rm IV}$ at this stage.
Thus, we can identify that $ ( \rep{1}\,, \repb{2}\,, -\frac{1}{2} )_{ \mathbf{F}}^{\rm IV } \equiv ( \eG_L^{\prime\prime} \,, - \nG_L^{\prime \prime})^T$, $(\check \Nc_L^{ {\rm IV}^\prime}\,, \check \Nc_L^{ {\rm IV}^{\prime\prime} })\equiv ( \check \nG_L^\prime\,, \check \nG_L^{ \prime \prime} )$, and $( \repb{3}\,, \rep{1}\,, +\frac{1}{3} )_{ \mathbf{F}}^{\rm IV} \equiv {\DG_R}^c$.

\para
After this stage, the remaining massless fermions expressed in terms of the $\Gc_{341}$ irreps are the following
\beqn\label{eq:341_fermions}
&& \Big[ ( \repb{3}\,, \rep{1}\,, +\frac{1}{3} )_{ \mathbf{F}}^\Omega \oplus ( \rep{1}\,, \rep{1}\,, 0 )_{ \mathbf{F}}^\Omega \Big]  \oplus ( \rep{1}\,, \repb{4}\,, -\frac{1}{4} )_{ \mathbf{F}}^\Omega  \subset \repb{8_F}^\Omega \,, \non
&& \Omega = ( \omega \,, \dot \omega ) \,,\quad \omega = (3\,, {\rm V}\,, {\rm VI} ) \,,\quad \dot \omega = ( \dot 1\,, \dot 2\,, \dot {\rm VII} \,,\dot {\rm VIII} \,, \dot {\rm IX}) \,,\non
&& ( \rep{1}\,, \rep{1}\,, 0 )_{ \mathbf{F}}^{{\rm IV} } \subset \repb{8_F}^{ {\rm IV}} \,,\non
&& \Big[ \cancel{ ( \rep{3}\,, \rep{1}\,, -\frac{1}{3} )_{ \mathbf{F}} } \oplus ( \repb{3}\,, \rep{1}\,, -\frac{2}{3} )_{ \mathbf{F}} \Big] \oplus ( \rep{1}\,, \rep{6}\,, +\frac{1}{2} )_{ \mathbf{F}}  \oplus \Big[ ( \rep{3}\,, \rep{4}\,, -\frac{1}{12} )_{ \mathbf{F}} \oplus \cancel{ ( \rep{1}\,, \rep{4}\,, +\frac{1}{4} )_{ \mathbf{F}} } \Big] \subset \rep{28_F}\,, \non
&& ( \rep{1}\,, \repb{4}\,, +\frac{3}{4} )_{ \mathbf{F}} \oplus \Big[ ( \repb{3}\,, \rep{1}\,, -\frac{2}{3} )_{ \mathbf{F}}^\prime \oplus ( \rep{1}\,, \rep{1}\,, -1 )_{ \mathbf{F}} \Big] \oplus  \Big[ ( \rep{3}\,, \rep{6}\,, +\frac{1}{6} )_{ \mathbf{F}} \oplus ( \rep{1}\,, \rep{6}\,, +\frac{1}{2})_{ \mathbf{F}}^\prime \Big] \non
&\oplus& \Big[ ( \rep{3}\,, \rep{4}\,, -\frac{1}{12} )_{ \mathbf{F}}^\prime \oplus ( \repb{3}\,, \rep{4}\,, -\frac{5}{12})_{ \mathbf{F}} \Big]   \subset \rep{56_F}\,.
\eeqn
Fermions that become massive at this stage are crossed out by slashes.
From the Yukawa couplings in Eq.~\eqref{eq:Yukawa_441_01}, no components from the $\rep{56_F}$ obtain their masses at this stage.
Loosely speaking, we find that only one of the massive $\repb{8_F}^\Omega$ is integrated out from the anomaly-free conditions of $[ {\rm SU}(3)_c]^2 \cdot {\rm U}(1)_{X_1}=0$, $\[ {\rm SU}(4)_W \]^2 \cdot {\rm U}(1)_{X_1}=0$, and $\[ {\rm U}(1)_{X_1} \]^3=0$, except for one left-handed sterile neutrino of $\check \Nc_L^{\rm IV}\equiv ( \rep{1}\,, \rep{1}\,, 0 )_{ \mathbf{F}}^{\rm IV} \subset \repb{8_F}^{\rm IV}$.

\subsection{The second stage}

\para
The second symmetry breaking stage of $\Gc_{341} \to \Gc_{331}$ can be achieved by $( \rep{1} \,, \repb{4} \,, -\frac{1}{4} )_{\mathbf{H}\,,\omega } \subset \repb{8_H}_{\,,\omega }$ in the rank-2 sector, and $( \rep{1} \,, \repb{4} \,, -\frac{1}{4} )_{\mathbf{H}\,, \dot\omega } \subset \repb{28_H}_{\,,\dot \omega }$ in the rank-3 sector, according to their $\Gc_{331}$-invariant and $\widetilde{ {\rm U}}(1)_{T^{\prime \prime} }$-neutral components in Tab.~\ref{tab:G341_Tcharges}.

\begin{table}[htp]
\begin{center}
\begin{tabular}{c|cccc}
\hline\hline
$\repb{8_F}^\Omega $ & $( \repb{3}\,, \rep{1}\,, +\frac{1}{3} )_{\rep{F}}^\Omega $ & $( \rep{1}\,, \rep{1}\,,  0 )_{\rep{F}}^\Omega $  &  $( \rep{1}\,, \repb{4}\,,  -\frac{1}{4} )_{\rep{F}}^\Omega$   &   \\[1mm]
\hline
$\Tc^{\prime\prime}$  &  $ - \frac{4}{3} t$ & $ -4t$  &  $ - 4 t$   &    \\[1mm]
\hline
 $\rep{28_F}$  &  $( \repb{3}\,, \rep{1}\,,  -\frac{2}{3 } )_{\rep{F}}$ &  $( \rep{1}\,, \rep{ 6}\,,  + \frac{1}{2} )_{\rep{F}}$ & $( \rep{3  }\,, \rep{4}\,, - \frac{1 }{12 } )_{\rep{F}}$  &    \\[1mm]
\hline
$\Tc^{\prime\prime}$   & $- \frac{4}{3} t$  &  $ 4 t$  &  $\frac{4 }{3 } t$  &    \\[1mm]
\hline
$\rep{56_F}$ & $( \rep{1}\,, \repb{4}\,,  + \frac{3 }{4} )_{\rep{F}}$ & $( \repb{3}\,,  \rep{1}\,,  - \frac{2 }{3} )_{\rep{F}}^\prime$ &  $( \rep{1}\,,  \rep{1}\,,  - 1 )_{\rep{F}}$  &     \\[1mm]
 \hline
 $\Tc^{\prime\prime}$  &  $4 t$  &  $- \frac{4 }{3} t$  & $-4 t$  &     \\[1mm]
    \hline
   & $( \rep{3}\,,  \rep{6}\,,  +\frac{1}{6} )_{\rep{F}}$ & $( \rep{1}\,,  \rep{6}\,,  +\frac{1}{2} )_{\rep{F}}^\prime$ & $( \rep{3}\,,  \rep{4}\,,  -\frac{1}{12 } )_{\rep{F}}^\prime$  &   $( \repb{3}\,,  \rep{4}\,,  -\frac{5}{12 } )_{\rep{F}}^\prime$   \\[1mm]
   \hline
  $\Tc^{\prime\prime}$ & $ \frac{4}{3 } t$ & $ 4 t$    & $ \frac{4}{3 } t$ &   $ -\frac{4}{3 } t$ \\[1mm]
    \hline\hline
$\repb{8_H}_{\,, \omega}$  &  $( \rep{1}\,, \repb{4}\,,  -\frac{1}{4} )_{\rep{H}\,,\omega}$  &   &  &    \\[1mm]
\hline
$\Tc^{\prime\prime}$  &  $0$  &   &   &     \\[1mm]
\hline
 $\repb{28_H}_{\,, \dot \omega }$  & $( \rep{1}\,, \rep{6}\,,  -\frac{1}{2} )_{\rep{H}\,,\dot \omega }$  &  $( \rep{1}\,, \repb{4}\,,  -\frac{1}{4} )_{\rep{H}\,,\dot \omega }$  &   &     \\[1mm]
\hline
$\Tc^{\prime\prime}$  &  $ 0$  &  $0 $  &  &     \\[1mm]
\hline
$\rep{70_H}$ &  $( \rep{1 }\,, \repb{4}\,,  +\frac{3}{4} )_{\rep{H}}^\prime$  & $( \rep{1}\,, \rep{4}\,,  - \frac{3 }{4} )_{\rep{H}}^\prime$   &  &    \\[1mm]
\hline
$\Tc^{\prime\prime}$  & $0$  &  $ -8 t$  &   &     \\[1mm]
 \hline\hline
\end{tabular}
\end{center}
\caption{The $\widetilde{ {\rm U}}(1)_{T^{\prime \prime} }$ charges for massless fermions and possible symmetry breaking Higgs components in the $\Gc_{341}$ theory.
}
\label{tab:G341_Tcharges}
\end{table}%

\para
The Yukawa coupling between $\repb{8_F}^\omega$ and $\rep{28_F}$ and the corresponding vectorlike fermion masses can be expressed as
\beqn\label{eq:Yukawa_341_01}
&& Y_\Bc  \repb{8_F}^\omega \rep{28_F} \repb{8_H}_{\,,\omega } + H.c. \non
&\supset& Y_\Bc  \Big[ ( \rep{1}\,, \repb{4}\,, -\frac{1}{4})_{ \mathbf{F}}^\omega \otimes ( \rep{1}\,, \rep{6}\,, +\frac{1}{2} )_{ \mathbf{F}} \oplus ( \repb{4}\,, \rep{1}\,, +\frac{1}{4})_{ \mathbf{F}}^\omega \otimes ( \rep{4}\,, \rep{4}\,, 0 )_{ \mathbf{F}} \Big] \otimes  ( \rep{1}\,, \repb{4}\,, -\frac{1}{4})_{ \mathbf{H}\,,\omega } + H.c. \non
&\supset& Y_\Bc  \Big[ ( \rep{1}\,, \repb{4}\,, -\frac{1}{4})_{ \mathbf{F}}^\omega \otimes ( \rep{1}\,, \rep{6}\,, +\frac{1}{2} )_{ \mathbf{F}} \oplus ( \rep{1 }\,, \rep{1}\,, 0 )_{ \mathbf{F}}^\omega \otimes ( \rep{1}\,, \rep{4}\,, +\frac{1 }{ 4} )_{ \mathbf{F}} \oplus ( \repb{3}\,, \rep{1}\,, +\frac{1}{3})_{ \mathbf{F}}^\omega \otimes ( \rep{3 }\,, \rep{4}\,, - \frac{1 }{12 } )_{ \mathbf{F}}  \Big] \non
&\otimes&  \langle ( \rep{1}\,, \repb{4}\,, -\frac{1}{4})_{ \mathbf{H}\,,\omega } \rangle + H.c. \non
%
%
&\Rightarrow& \frac{1}{ \sqrt{2} } Y_\Bc  \Big(  \eG_L {\eG_R}^c -   \nG_L {\nG_R }^c +  \check \Nc_L^{{\rm V}^\prime}  \check \nG_R^c  + \check \Nc_L^{\rm V}  \check \nG_R^{\prime\prime\,c} + \DG_L^{\prime\prime}  {\DG_R^{\prime\prime}}^c   \Big) w_{ \repb{4}\,, {\rm V}} + H.c.  \,.
\eeqn
Without loss of generality, we choose $\omega = {\rm V}$ at this stage.
Thus, we can identify that $ ( \rep{1}\,, \repb{2}\,, -\frac{1}{2} )_{ \mathbf{F}}^{\rm V} \equiv ( \eG_L \,, - \nG_L )^T$, $\check \Nc_L^{{\rm V}^\prime} \equiv \check \nG_L$, and $( \repb{3}\,, \rep{1}\,, +\frac{1}{3} )_{ \mathbf{F}}^{\rm V} \equiv {\DG_R^{\prime\prime}}^c$.
Notice that the term of $\check \Nc_L^{\rm V}  \check \nG_R^{\prime\prime\,c}+H.c.$ is a mass mixing term, and the sterile neutrino of $( \rep{1} \,, \rep{1} \,, 0 )_{ \rep{F}}^{\rm V}= \check \Nc_L^{\rm V} $ remains massless.

\para
The Yukawa coupling between $\repb{8_F}^{\dot \omega }$ and $\rep{56_F}$ and the corresponding vectorlike fermion masses can be expressed as
\beqn\label{eq:Yukawa_341_02}
&& Y_\Dc \repb{8_F}^{\dot \omega } \rep{56_F} \repb{28_H}_{\,,\dot \omega } + H.c. \non
&\supset& Y_\Dc  \Big[  ( \rep{1}\,, \repb{4}\,, -\frac{1}{4})_{ \mathbf{F}}^{\dot \omega  } \otimes  ( \rep{4 }\,, \rep{6}\,, +\frac{1}{4 } )_{ \mathbf{F}} \oplus ( \repb{4}\,, \rep{1}\,, +\frac{1}{4} )_{ \mathbf{F}}^{\dot \omega  } \otimes   ( \rep{6 }\,, \rep{4 }\,, - \frac{1}{4 } )_{ \mathbf{F}} \Big] \otimes ( \repb{4 }\,, \repb{4}\,, 0 )_{ \mathbf{H}\,, \dot \omega } + H.c. \non
&\supset& Y_\Dc  \Big[ ( \rep{1}\,, \repb{4}\,, -\frac{1}{4})_{ \mathbf{F}}^{\dot \omega  } \otimes ( \rep{1}\,, \rep{6}\,, +\frac{1}{2} )_{ \mathbf{F}}^\prime  \oplus ( \repb{3}\,, \rep{1}\,, +\frac{1}{3})_{ \mathbf{F}}^{\dot \omega  } \otimes ( \rep{3}\,, \rep{4}\,, -\frac{1}{12} )_{ \mathbf{F}}^\prime \Big]  \otimes  \langle ( \rep{1}\,, \repb{4}\,, -\frac{1}{4})_{ \mathbf{H}\,, \dot \omega }  \rangle + H.c. \non
%
%
&\Rightarrow&  \frac{1}{ \sqrt{2}} Y_\Dc  \Big(  \eG_L^{\prime \prime  \prime\prime }  {\eG_R^{\prime \prime  \prime\prime}}^c -   \nG_L^{\prime \prime  \prime\prime }  {\nG_R^{\prime \prime  \prime\prime }}^c +  \check \Nc_L^{ \dot {\rm VII }^\prime }  \check \nG_R^{\prime\prime\prime\,c } + \DG_L^{\prime\prime\prime\prime \prime }  {\DG_R^{\prime\prime\prime\prime \prime } }^c  \Big) w_{\repb{4}\,, \dot {\rm VII}} \non
&+& \frac{1}{ \sqrt{2}} Y_\Dc  \Big( e_L  {\eG_R^{\prime \prime  \prime\prime}}^c -   \nu_{e \, L}  {\nG_R^{\prime \prime  \prime\prime }}^c +  \check \Nc_L^{ \dot  1^\prime }  \check \nG_R^{\prime\prime\prime\,c } + \DG_L^{\prime\prime\prime\prime \prime }  {d_R}^c  \Big) w_{\repb{4}\,, \dot 1  } + H.c.   \,.
\eeqn
Without loss of generality, we choose $\dot \omega = \dot 1\,, \dot {\rm VII}$ according to the VEV assignment in Eq.~\eqref{eq:SU8_Higgs_VEVs_mini02} at this stage.
Thus, we can identify that $ ( \rep{1}\,, \repb{2}\,, -\frac{1}{2} )_{ \mathbf{F}}^{\dot {\rm VII} } \equiv ( \eG_L^{ \prime\prime \prime \prime } \,, - \nG_L^{ \prime\prime \prime \prime } )^T$, $\check \Nc_L^{ \dot {\rm VII }^\prime } \equiv \check \nG_L^{ \prime \prime \prime }$, and $( \repb{3}\,, \rep{1}\,, +\frac{1}{3} )_{ \mathbf{F}}^{\dot {\rm VII}} \equiv {\DG_R^{\prime\prime\prime \prime \prime }}^c$.
The first-generational SM fermions of $(e\,, \nu_e \,, d)$ only form mass mixing terms with their heavy partner fermions of $(  \eG^{ \prime\prime \prime \prime } \,,   \nG^{ \prime\prime \prime \prime } \,,  \DG^{\prime\prime\prime \prime \prime } )$, and remain massless at this stage.

\para
The Yukawa coupling between two $\rep{56_F}$ cannot be realized at the renormalizable level due to the anti-symmetric property.
It was suggested in Ref.~\cite{Barr:2008pn} that these vectorlike fermion masses are due to the following $d=5$ operator
\beqn\label{eq:Yukawa_341_03}
&&  \frac{ c_4 }{ M_{\rm pl} } \rep{56_F} \rep{56_F} \langle \rep{63_H} \rangle \repb{28_H}_{\,, \dot \omega }^\dag + H.c. \non
&\supset&  c_4 \zeta_0  \Big[ ( \rep{1 } \,, \repb{4 } \,,  + \frac{3}{4} )_{ \rep{F}} \otimes ( \repb{4 } \,, \rep{1 } \,, - \frac{3}{4} )_{ \rep{F}}  \oplus  ( \rep{4 } \,, \rep{6 } \,, + \frac{1 }{4} )_{ \rep{F}} \otimes  ( \rep{6 } \,, \rep{4 } \,, -  \frac{1 }{4} )_{ \rep{F}} \Big] \otimes ( \repb{4 } \,, \repb{4 } \,, 0 )_{ \rep{H}\,, \dot \omega }^\dag + H.c. \non
&\supset& c_4 \zeta_0 \Big[ ( \rep{1 } \,, \repb{4 } \,,  + \frac{3}{4} )_{ \rep{F}} \otimes ( \rep{1 } \,, \rep{1 } \,, - 1 )_{ \rep{F}}  \oplus  ( \rep{3 } \,, \rep{6 } \,, + \frac{1 }{6 } )_{ \rep{F}} \otimes  ( \repb{3 } \,, \rep{4 } \,, -  \frac{5 }{12 } )_{ \rep{F}}  \Big]  \otimes \langle ( \rep{1 } \,, \repb{4 } \,, -\frac{1 }{4 } )_{ \rep{H}\,, \dot \omega }^\dag \rangle + H.c. \non
%
%
&\Rightarrow& \frac{ c_4  }{ \sqrt{2}}  \zeta_0 \Big(  \EG_L { \EG_R}^c + \dG_L {\dG_R }^c - \uG_L  {\uG_R}^c +  \UG_L {\UG_R }^c  \Big) w_{ \repb{4}\,, \dot 1\,, \dot {\rm VII} } + H.c. \,.
\eeqn
This $d=5$ operator breaks the global $\widetilde{ {\rm SU}}(5)_{\dot \omega }$ and $\widetilde{\rm U}(1)_{\rm PQ}$ symmetries explicitly, hence it can only be suppressed by $1/M_{\rm pl}$.

\para
After integrating out the massive fermions in Eqs.~\eqref{eq:Yukawa_341_01}, \eqref{eq:Yukawa_341_02}, and \eqref{eq:Yukawa_341_03}, the remaining massless fermions expressed in terms of the $\Gc_{331}$ irreps are the following
\beqn\label{eq:331_fermions}
&& \Big[ ( \repb{3}\,, \rep{1}\,, +\frac{1}{3} )_{ \mathbf{F}}^\Omega \oplus ( \rep{1}\,, \rep{1}\,, 0 )_{ \mathbf{F}}^\Omega \Big]  \oplus \Big[ ( \rep{1}\,, \repb{3}\,, -\frac{1}{3} )_{ \mathbf{F}}^\Omega \oplus ( \rep{1}\,, \rep{1}\,, 0 )_{ \mathbf{F}}^{\Omega^{\prime \prime }}  \Big]  \subset \repb{8_F}^\Omega \,, \non
&& \Omega = ( \omega \,, \dot \omega ) \,, \quad \omega = (3\,, {\rm VI})\,,  \quad \dot \omega = (\dot 1\,, \dot 2\,, \dot {\rm VIII}\,, \dot {\rm IX} )\,,\non
&& ( \rep{1}\,, \rep{1}\,, 0)_{ \mathbf{F}}^{{\rm IV} } \subset \repb{8_F}^{{\rm IV} } \,, \quad ( \rep{1}\,, \rep{1}\,, 0)_{ \mathbf{F}}^{{\rm V} \,, {\rm V}^{\prime\prime} } \subset \repb{8_F}^{{\rm V} } \,, \non
&&  ( \rep{1}\,, \rep{1}\,, 0)_{ \mathbf{F}}^{\dot {\rm VII} } \oplus ( \rep{1}\,, \rep{1}\,, 0)_{ \mathbf{F}}^{\dot {\rm VII}^{ \prime \prime} } \subset \repb{8_F}^{\dot {\rm VII} } \,,  \non
&&\Big[ \cancel{ ( \rep{3}\,, \rep{1}\,, -\frac{1}{3} )_{ \mathbf{F}} } \oplus ( \repb{3}\,, \rep{1}\,, -\frac{2}{3} )_{ \mathbf{F}} \Big] \oplus \Big[ \bcancel{ ( \rep{1}\,, \rep{3}\,, +\frac{1}{3} )_{ \mathbf{F}} } \oplus ( \rep{1}\,, \repb{3}\,, +\frac{2}{3} )_{ \mathbf{F}} \Big] \non
&\oplus& \Big[ ( \rep{3}\,, \rep{3}\,, 0 )_{ \mathbf{F}} \oplus \bcancel{ ( \rep{3}\,, \rep{1}\,, -\frac{1}{3} )_{ \mathbf{F}}^{\prime\prime} } \Big]  \oplus \cancel{ \Big[ ( \rep{1}\,, \rep{3}\,, +\frac{1}{3} )_{ \mathbf{F}} \oplus  ( \rep{1}\,, \rep{1}\,, 0 )_{ \mathbf{F}}^{\prime\prime}  \Big] }  \subset \rep{28_F}\,, \non
&&\Big[ ( \rep{1}\,, \repb{3}\,, +\frac{2}{3} )_{ \mathbf{F}}^\prime \oplus \bcancel{ ( \rep{1}\,, \rep{1}\,, +1 )_{ \mathbf{F}}^{\prime\prime} } \Big] \oplus \Big[  ( \repb{3}\,, \rep{1}\,, -\frac{2}{3} )_{ \mathbf{F}}^\prime \oplus  \bcancel{ ( \rep{1}\,, \rep{1}\,, -1 )_{ \mathbf{F}} }  \Big] \non
&\oplus&  \Big[ ( \rep{3}\,, \rep{3}\,, 0 )_{ \mathbf{F}}^\prime \oplus \bcancel{( \rep{3}\,, \repb{3}\,, +\frac{1}{3} )_{ \mathbf{F}} } \oplus \bcancel{  ( \rep{1}\,, \rep{3}\,, +\frac{1}{3} )_{ \mathbf{F}}^\prime } \oplus ( \rep{1}\,, \repb{3}\,, +\frac{2}{3})_{ \mathbf{F}}^{\prime \prime} \Big] \non
&\oplus& \Big[ ( \rep{3}\,, \rep{3}\,, 0)_{ \mathbf{F}}^{\prime\prime} \oplus \bcancel{ ( \rep{3}\,, \rep{1}\,, -\frac{1}{3})_{ \mathbf{F}}^{\prime\prime\prime\prime\prime}  } \oplus \bcancel{ ( \repb{3}\,, \rep{3}\,, -\frac{1}{3})_{ \mathbf{F}} }  \oplus ( \repb{3}\,, \rep{1}\,, -\frac{2}{3})_{ \mathbf{F}}^{\prime\prime\prime} \Big]   \subset \rep{56_F}\,.
\eeqn
We use the slashes and the back slashes to cross out massive fermions at the first and the second stages, respectively.
From the anomaly-free conditions of $[ {\rm SU}(3)_c]^2 \cdot {\rm U}(1)_{X_2}=0$, $\[ {\rm SU}(3)_W \]^2 \cdot {\rm U}(1)_{X_2}=0$, and $\[ {\rm U}(1)_{X_2} \]^3=0$, we find that one of the $\repb{8_F}^\omega$ and one of the $\repb{8_F}^{\dot \omega}$ are integrated out.
Without loss of generality, we choose the massive anti-fundamental fermions to be $\omega={\rm V}$ and $\dot \omega=\dot {\rm VII}$ at this stage.

\subsection{The third stage}

\para
The third-stage symmetry breaking of $\Gc_{331} \to \Gc_{\rm SM}$ can be achieved by Higgs fields of $( \rep{1} \,, \repb{3} \,, -\frac{1}{3} )_{\mathbf{H}\,,\omega } \subset ( \rep{1} \,, \repb{4} \,, -\frac{1}{4} )_{\mathbf{H}\,,\omega }  \subset \repb{8_H}_{\,,\omega }$, and $\[ ( \rep{1} \,, \repb{3} \,, -\frac{1}{3} )_{\mathbf{H}\,,\dot \omega }^\prime \oplus ( \rep{1} \,, \repb{3} \,, -\frac{1}{3} )_{\mathbf{H}\,,\dot \omega }  \]  \subset ( \rep{1} \,, \rep{6} \,, -\frac{1}{2} )_{\mathbf{H}\,,\dot \omega }  \oplus  ( \rep{1} \,, \repb{4} \,, -\frac{1}{4} )_{\mathbf{H}\,,\dot \omega}  \subset ( \rep{1} \,, \rep{6} \,, -\frac{1}{2} )_{\mathbf{H}\,,\dot \omega } \oplus  ( \repb{4} \,, \repb{4} \,, 0 )_{\mathbf{H}\,, \dot \omega }  \subset \repb{28_H}_{\,,\dot\omega }$, according to the decompositions in Eqs.~\eqref{eq:SU8_Higgs_Br01} and \eqref{eq:SU8_Higgs_Br02}, as well as their $\widetilde{\rm U}(1)_{T^{ \prime\prime \prime} }$ charges in Tab.~\ref{tab:G331_Tcharges}.

\begin{table}[htp]
\begin{center}
\begin{tabular}{c|cccc}
\hline\hline
 $\repb{8_F}^\Omega$ & $( \repb{3}\,, \rep{1}\,, +\frac{1}{3} )_{\rep{F}}^\Omega$ & $( \rep{1}\,, \rep{1}\,,  0 )_{\rep{F}}^\Omega$  &  $( \rep{1}\,, \repb{3}\,,  -\frac{1}{3} )_{\rep{F}}^\Omega$  &  $( \rep{1}\,, \rep{1}\,,  0 )_{\rep{F}}^{\Omega^{ \prime\prime} }$   \\[1mm]
\hline
$\Tc^{ \prime \prime \prime }$  &  $- \frac{4}{3} t$ & $- 4 t$  &  $- 4 t$  & $ -4 t$     \\[1mm]
\hline
$\rep{28_F}$ & $( \repb{3}\,, \rep{1}\,, - \frac{2}{3} )_{\rep{F}}$  &  $( \rep{1}\,, \repb{3}\,,  + \frac{2}{3} )_{\rep{F}}$  &  $( \rep{3}\,, \rep{3 }\,,  0 )_{\rep{F}}$   & $$   \\[1mm]
 \hline
 $\Tc^{ \prime \prime \prime }$  & $ - \frac{4}{3} t$  &  $ 4 t$  &  $   \frac{4}{3} t$ & $$   \\[1mm]
\hline
$\rep{56_F}$  & $( \rep{1}\,,  \repb{3}\,, +\frac{2}{3} )_{\rep{F}}^\prime$ &  $( \repb{3 }\,, \rep{1}\,,  -\frac{2}{3 } )_{\rep{F}}^\prime$  &    &     \\[1mm]
 \hline
 $\Tc^{ \prime \prime \prime }$  &  $ 4 t$  &  $  -\frac{4}{3 } t$ &   &   \\[1mm]
 \hline
   $$ & $( \rep{3}\,,  \rep{3}\,, 0 )_{\rep{F}}^\prime$ & $(  \rep{1}\,, \repb{3 }\,,  +\frac{2}{3 } )_{\rep{F}}^{\prime\prime} $ &  $( \rep{3}\,,  \rep{3}\,, 0 )_{\rep{F}}^{\prime \prime} $  & $( \repb{3 }\,,   \rep{1}\,,  -\frac{2}{3 } )_{\rep{F}}^{\prime\prime \prime}$  \\[1mm]
 \hline
 $\Tc^{ \prime \prime \prime }$  &  $\frac{4}{3} t$  &  $  4 t$  &  $\frac{4}{3} t$  &  $ -\frac{4}{3} t$  \\[1mm]
 \hline\hline
$\repb{8_H}_{ \,, \omega }$   &  $( \rep{1}\,, \repb{3}\,,  - \frac{1}{3} )_{\rep{H}\,,\omega }$  &   &   & $$   \\[1mm]
\hline
$\Tc^{ \prime \prime \prime }$  &  $0$  &   &  & $$  \\[1mm]
\hline
  $\repb{28_H}_{ \,, \dot \omega }$  &  $( \rep{1}\,,  \repb{3}\,, -\frac{1}{ 3} )_{\rep{H}\,, \dot \omega }^\prime$  & $( \rep{1}\,,  \rep{3}\,, -\frac{2 }{ 3} )_{\rep{H}\,, \dot \omega }$   &  $( \rep{1}\,,  \repb{3}\,, -\frac{1 }{ 3} )_{\rep{H}\,, \dot \omega }$  & $$   \\[1mm]
 \hline
 $\Tc^{ \prime \prime \prime }$  & $0$  & $0$  & $0$ &     \\[1mm]
 \hline
 $\rep{70_H}$  &  $( \rep{1}\,, \repb{3}\,,  + \frac{2}{3} )_{\rep{H}}^{ \prime\prime \prime}$  &  $( \rep{1}\,, \rep{3}\,,  - \frac{2}{3} )_{\rep{H}}^{ \prime\prime \prime}$  &  & $$   \\[1mm]
\hline
$\Tc^{ \prime \prime \prime }$  &   $0$ &   $-8 t$ &  & $$  \\[1mm]
 \hline\hline
\end{tabular}
\end{center}
\caption{The $\widetilde{ {\rm U}}(1)_{T^{\prime \prime \prime } }$ charges for massless fermions and possible symmetry breaking Higgs components in the $\Gc_{331}$ theory.
}
\label{tab:G331_Tcharges}
\end{table}

\para
The Yukawa couplings between the $\repb{8_F}^\omega$ and the $\rep{28_F}$ and the corresponding vectorlike fermion masses can be expressed as
\beqn\label{eq:Yukawa_331_01}
&& Y_\Bc \repb{8_F}^\omega \rep{28_F} \repb{8_H}_{\,,\omega } + H.c. \non
&\supset& Y_\Bc  \Big[ ( \rep{1}\,, \repb{4}\,, -\frac{1}{4} )_{ \mathbf{F}}^\omega \otimes ( \rep{1}\,, \rep{6}\,, + \frac{1}{2} )_{ \mathbf{F}}  \oplus ( \repb{4}\,, \rep{1}\,, +\frac{1}{4} )_{ \mathbf{F}}^\omega \otimes ( \rep{4}\,, \rep{4}\,, 0 )_{ \mathbf{F}}   \Big]  \otimes ( \rep{1}\,, \repb{4}\,, -\frac{1}{4} )_{ \mathbf{H}\,, \omega } + H.c. \non
&\supset& Y_\Bc  \Big[ ( \rep{1}\,, \repb{4}\,, -\frac{1}{4} )_{ \mathbf{F}}^\omega \otimes ( \rep{1}\,, \rep{6}\,, + \frac{1}{2} )_{ \mathbf{F}}  \oplus ( \rep{1}\,, \rep{1}\,, 0 )_{ \mathbf{F}}^\omega \otimes ( \rep{1}\,, \rep{4}\,, +\frac{1}{4} )_{ \mathbf{F}} \non
&\oplus& ( \repb{3}\,, \rep{1}\,, +\frac{1}{3} )_{ \mathbf{F}}^\omega \otimes ( \rep{3}\,, \rep{4}\,, -\frac{1}{12} )_{ \mathbf{F}}   \Big] \otimes ( \rep{1}\,, \repb{4}\,, -\frac{1}{4} )_{ \mathbf{H}\,, \omega } + H.c. \non
&\supset& Y_\Bc  \Big[  ( \rep{1}\,, \repb{3}\,, -\frac{1}{3} )_{ \mathbf{F}}^\omega \otimes ( \rep{1}\,, \repb{3}\,, +\frac{2}{3} )_{ \mathbf{F}} \oplus ( \rep{1}\,, \rep{1}\,, 0 )_{ \mathbf{F}}^{\omega^{ \prime \prime } } \otimes ( \rep{1}\,, \rep{3}\,, +\frac{1}{3} )_{ \mathbf{F}}  \oplus ( \rep{1}\,, \rep{1}\,, 0 )_{ \mathbf{F}}^\omega \otimes ( \rep{1}\,, \rep{3}\,, +\frac{1}{3} )_{ \mathbf{F}}^{ \prime \prime } \non
&\oplus&  (  \repb{3}\,, \rep{1}\,, +\frac{1}{3} )_{ \mathbf{F}}^\omega \otimes ( \rep{3}\,, \rep{3}\,, 0 )_{ \mathbf{F}}  \Big] \otimes \langle ( \rep{1}\,, \repb{3}\,, -\frac{1}{3} )_{ \mathbf{H}\,, \omega }  \rangle + H.c. \non
%
%
&\Rightarrow& \frac{1}{ \sqrt{2} }  Y_\Bc   \Big(  \nG_L^\prime \nG_R^{\prime\,c }- \eG_L^\prime  \eG_R^{\prime\,c}  + \check \Nc_L^{{\rm VI}^{ \prime \prime }  }\check \nG_R^{ c} + \check \Nc_L^{\rm VI}  \check \nG_R^{\prime\, c} + \DG_L^\prime  \DG_R^{\prime\,c}  \Big)  V_{ \repb{3}\,, {\rm VI} }  \non
&+& \frac{1}{ \sqrt{2} }  Y_\Bc   \Big(  \nu_{\tau\, L} \nG_R^{\prime\,c }-  \tau_L \eG_R^{\prime\,c}  + \check \Nc_L^{ 3^{ \prime \prime }  }\check \nG_R^{ c} + \check \Nc_L^{3}  \check \nG_R^{\prime\, c} + \DG_L^\prime  {b_R}^{c}  \Big)  V_{ \repb{3}\,, 3 } + H.c.  \,,
\eeqn
where we choose $\omega =3\,, {\rm VI}$ according to the VEV assignment in Eq.~\eqref{eq:SU8_Higgs_VEVs_mini03} at this stage.
Thus, we can identify that $( \rep{1}\,, \repb{2}\,, -\frac{1}{2} )_{ \mathbf{F}}^{\rm VI}  \equiv ( \eG_L^\prime\,, - \nG_L^\prime)$, and $(  \repb{3}\,, \rep{1}\,, +\frac{1}{3} )_{ \mathbf{F}}^{\rm VI} \equiv \DG_R^{\prime\,c}$.
The third-generational SM fermions of $(\tau\,, \nu_\tau\,, b)$ only form mass mixing terms with their heavy partner fermions of $(  \eG^{ \prime } \,,   \nG^{ \prime } \,, \DG^{\prime } )$, and remain massless at this stage.

\para
The Yukawa couplings between the $\repb{8_F}^{\dot \omega}$ and the $\rep{56_F}$ and the corresponding vectorlike fermion masses can be expressed as
\beqs\label{eqs:Yukawa_331_02}
\beqn
&& Y_\Dc \repb{8_F}^{\dot \omega } \rep{56_F} \repb{28_H}_{\,,\dot \omega } + H.c. \non
&\supset& Y_\Dc  \Big[ ( \rep{1}\,, \repb{4}\,, -\frac{1}{4})_{ \mathbf{F}}^{\dot \omega } \otimes ( \rep{1}\,, \repb{4}\,, +\frac{3}{4})_{ \mathbf{F}}  \oplus   ( \repb{4}\,, \rep{1}\,, +\frac{1}{4})_{ \mathbf{F}}^{\dot \omega } \otimes  ( \rep{4 }\,, \rep{6}\,, +\frac{1}{4 })_{ \mathbf{F}} \Big] \otimes  ( \rep{1}\,, \rep{6}\,, -\frac{1}{2})_{ \mathbf{H}\,, \dot \omega }  + H.c. \non
&\supset& Y_\Dc \Big[ ( \rep{1}\,, \repb{4}\,, -\frac{1}{4})_{ \mathbf{F}}^{\dot \omega } \otimes ( \rep{1}\,, \repb{4}\,, +\frac{3}{4})_{ \mathbf{F}}  \oplus  ( \repb{3}\,, \rep{1}\,, +\frac{1}{3})_{ \mathbf{F}}^{\dot \omega } \otimes ( \rep{3}\,, \rep{6}\,, +\frac{1}{6})_{ \mathbf{F}}  \non
&\oplus&  ( \rep{1}\,, \rep{1}\,, 0 )_{ \mathbf{F}}^{\dot \omega } \otimes ( \rep{1}\,, \rep{6}\,, +\frac{1}{2})_{ \mathbf{F}}^\prime \Big] \otimes   ( \rep{1}\,, \rep{6}\,, -\frac{1}{2})_{ \mathbf{H}\,, \dot \omega } + H.c.  \non
&\supset& Y_\Dc \Big[   ( \rep{1}\,, \repb{3}\,, -\frac{1}{3})_{ \mathbf{F}}^{\dot \omega } \otimes ( \rep{1}\,, \repb{3}\,, +\frac{2}{3})_{ \mathbf{F}}^\prime  \oplus ( \repb{3}\,, \rep{1}\,, +\frac{1}{3})_{ \mathbf{F}}^{\dot \omega } \otimes ( \rep{3}\,, \rep{3}\,, 0 )_{ \mathbf{F}}^\prime \non
& \oplus& ( \rep{1}\,, \rep{1}\,,  0 )_{ \mathbf{F}}^{\dot \omega } \otimes ( \rep{1}\,, \rep{3}\,, +\frac{1}{3} )_{ \mathbf{F}}^\prime  \Big]  \otimes  \langle ( \rep{1}\,, \repb{3}\,, -\frac{1}{3})_{ \mathbf{H}\,, \dot \omega }^\prime \rangle + H.c.  \non
%
%
&\Rightarrow& \frac{1}{\sqrt{2}}  Y_\Dc  \Big( \nG_L^{\prime \prime \prime }  \nG_R^{\prime \prime \prime\,c}   -   \eG_L^{\prime \prime \prime }  \eG_R^{\prime \prime \prime\,c} +   \check \Nc_L^{ \dot {\rm VIII} }  \check \nG_R^{ \prime\prime \prime\,c }  +  \DG_L^{\prime\prime \prime }  \DG_R^{\prime\prime \prime\,c } \Big) V_{ \repb{3}\,, {\dot {\rm VIII}} }^\prime   \non
&+&  \frac{1}{\sqrt{2}}  Y_\Dc  \Big( \nu_{\mu\, L}  \nG_R^{\prime \prime \prime\,c}  -   \mu_L  \eG_R^{\prime \prime \prime\,c} +   \check \Nc_L^{ \dot 2 }  \check \nG_R^{ \prime\prime \prime\,c }  +  \DG_L^{\prime\prime \prime } {s_R}^{c } \Big) V_{ \repb{3}\,, \dot 2  }^\prime + H.c.  \,,\label{eq:Yukawa_331_02a} \\[1mm]
&& Y_\Dc \repb{8_F}^{\dot \omega } \rep{56_F} \repb{28_H}_{\,,\dot \omega } + H.c. \non
&\supset& Y_\Dc \Big[ ( \rep{1}\,, \repb{4}\,, -\frac{1}{4})_{ \mathbf{F}}^{\dot \omega } \otimes ( \rep{4}\,, \rep{6}\,, +\frac{1}{4})_{ \mathbf{F}}  \oplus  ( \repb{4}\,, \rep{1}\,, +\frac{1}{4})_{ \mathbf{F}}^{\dot \omega } \otimes ( \rep{6}\,, \rep{4}\,, -\frac{1}{4})_{ \mathbf{F}}  \Big]  \otimes ( \repb{4}\,, \repb{4}\,, 0 )_{ \mathbf{H}\,, \dot \omega } + H.c.  \non
&\supset&  Y_\Dc  \Big[   ( \rep{1}\,, \repb{4}\,, -\frac{1}{4})_{ \mathbf{F}}^{\dot \omega } \otimes ( \rep{1}\,, \rep{6}\,, +\frac{1}{2})_{ \mathbf{F}}^\prime  \oplus  ( \repb{3}\,, \rep{1}\,, +\frac{1}{3})_{ \mathbf{F}}^{\dot \omega } \otimes ( \rep{3}\,, \rep{4}\,, -\frac{1}{12})_{ \mathbf{F}}^\prime  \Big]  \otimes ( \rep{1}\,, \repb{4}\,, -\frac{1}{4} )_{ \mathbf{H}\,, \dot \omega } + H.c. \non
&\supset& Y_\Dc \Big[  ( \rep{1}\,, \repb{3}\,, -\frac{1}{3})_{ \mathbf{F}}^{\dot \omega } \otimes ( \rep{1}\,, \repb{3}\,, +\frac{2}{3})_{ \mathbf{F}}^{ \prime\prime }   \oplus  ( \rep{1}\,, \rep{1}\,, 0 )_{ \mathbf{F}}^{{\dot \omega }^{\prime \prime}  } \otimes ( \rep{1}\,, \rep{3}\,, +\frac{1}{3})_{ \mathbf{F}}^{ \prime}   \non
& \oplus& ( \repb{3}\,, \rep{1}\,, +\frac{1}{3})_{ \mathbf{F}}^{\dot \omega } \otimes ( \rep{3}\,, \rep{3}\,, 0 )_{ \mathbf{F}}^{\prime \prime } \Big] \otimes  \langle ( \rep{1}\,, \repb{3}\,, -\frac{1}{3} )_{ \mathbf{H}\,, \dot \omega } \rangle + H.c. \non
%
%
&\Rightarrow&   \frac{1}{\sqrt{2}}  Y_\Dc  \Big( \nG_L^{\prime \prime \prime \prime \prime}  \nG_R^{\prime \prime \prime \prime \prime\,c}   -   \eG_L^{\prime \prime \prime \prime \prime}  \eG_R^{\prime \prime \prime \prime \prime\,c}  + \check \Nc_L^{ \dot {\rm IX}^{\prime\prime} }   \check \nG_R^{ \prime\prime \prime\,c } + \DG_L^{\prime\prime \prime \prime }  \DG_R^{\prime\prime \prime \prime\,c }  \Big) V_{ \repb{3}\,, {\dot {\rm IX}} } + H.c.  \,,\label{eq:Yukawa_331_02b}
\eeqn
\eeqs
where we choose $\dot \omega= \dot 2\,, \dot {\rm VIII}$ in Eq.~\eqref{eq:Yukawa_331_02a} and $\dot \omega = \dot {\rm IX}$ in Eq.~\eqref{eq:Yukawa_331_02b} according to the VEV assignment in Eq.~\eqref{eq:SU8_Higgs_VEVs_mini03}.
Thus, we can identify that $( \rep{1}\,, \repb{2}\,, -\frac{1}{2} )_{ \mathbf{F}}^{ \dot {\rm VIII} }  \equiv ( \eG_L^{\prime \prime\prime} \,, - \nG_L^{\prime\prime\prime} )$, $(  \repb{3}\,, \rep{1}\,, +\frac{1}{3} )_{ \mathbf{F}}^{\dot {\rm VIII}} \equiv \DG_R^{\prime\prime\prime\,c}$, $(  \rep{1}\,, \repb{2}\,, -\frac{1}{2} )_{ \mathbf{F}}^{\dot {\rm IX}} \equiv ( \eG_L^{\prime \prime\prime \prime \prime }\,, - \nG_L^{\prime \prime\prime \prime \prime } )$, and $(  \repb{3}\,, \rep{1}\,, +\frac{1}{3} )_{ \mathbf{F}}^{\dot {\rm IX}} \equiv \DG_R^{\prime \prime\prime \prime\,c }$.
The second-generational SM fermions of $(\mu\,, \nu_\mu \,, s)$ only form mass mixing terms with their heavy partner fermions of $(  \eG^{ \prime \prime \prime  } \,,   \nG^{ \prime \prime \prime } \,, \DG^{\prime \prime \prime } )$, and remain massless at this stage.

\para
One may wonder why two anti-fundamental fermions of $\repb{8_F}^{\dot \omega }\, ( \dot \omega = \dot {\rm VIII} \,, \dot {\rm IX})$ should obtain masses from different components of $( \rep{1}\,, \repb{3}\,, -\frac{1}{3})_{ \mathbf{H}\,, \dot \omega }^\prime  \subset ( \rep{1}\,, \rep{6}\,, -\frac{1}{2})_{ \mathbf{H}\,, \dot \omega }  \subset \repb{28_H}_{\,,\dot \omega }$ and $( \rep{1}\,, \repb{3}\,, -\frac{1}{3})_{ \mathbf{H}\,, \dot \omega }  \subset ( \repb{4}\,, \repb{4}\,, 0 )_{ \mathbf{H}\,, \dot \omega }  \subset \repb{28_H}_{\,,\dot \omega}$ at this stage.
If both $\repb{8_F}^{\dot \omega }$ coupled to the $( \rep{1}\,, \repb{3}\,, -\frac{1}{3})_{ \mathbf{H}\,, \dot \omega }^\prime$, the fermions of $( \nG^{ \prime\prime \prime \prime \prime } \,, \eG^{ \prime\prime \prime \prime \prime } \,, \DG^{ \prime \prime \prime \prime } )$ only mixed with the heavy fermions of $( \nG^{ \prime\prime \prime  } \,, \eG^{  \prime \prime \prime } \,, \DG^{ \prime \prime \prime } )$, and they remain massless.
Clearly, the massless fermion spectrum was no longer anomaly-free.
The same inconsistency is also obtained when both $\repb{8_F}^{\dot \omega }$ coupled to the $( \rep{1}\,, \repb{3}\,, -\frac{1}{3})_{ \mathbf{H}\,, \dot \omega }$.
Therefore, this situation should be prohibited.

\para
There can be further mass mixing terms from the same $d=5$ operator as in Eq.~\eqref{eq:Yukawa_341_03}, which read
\beqs\label{eqs:Yukawa_331_03}
\beqn
&& \frac{ c_4 }{ M_{\rm pl} } \rep{56_F} \rep{56_F} \langle \rep{63_H} \rangle \repb{28_H}_{\,, \dot \omega}^\dag + H.c. \non
&\supset&   \frac{  v_U }{ M_{\rm pl}} \Big[ ( \rep{1 } \,, \repb{4 } \,,  + \frac{3}{4} )_{ \rep{F}} \otimes ( \repb{4 } \,, \rep{1 } \,, - \frac{3}{4} )_{ \rep{F}}  \oplus  ( \rep{4 } \,, \rep{6 } \,, + \frac{1 }{4} )_{ \rep{F}} \otimes  ( \rep{6 } \,, \rep{4 } \,, -  \frac{1 }{4} )_{ \rep{F}} \Big] \otimes ( \repb{4 } \,, \repb{4 } \,, 0 )_{ \rep{H}\,, \dot \omega}^\dag + H.c. \non
&\supset& c_4  \frac{ v_U }{ M_{\rm pl} } \Big[ ( \rep{1 } \,, \repb{4 } \,,  + \frac{3}{4} )_{ \rep{F}} \otimes ( \rep{1 } \,, \rep{1 } \,, - 1 )_{ \rep{F}}  \oplus  ( \rep{3 } \,, \rep{6 } \,, + \frac{1 }{6 } )_{ \rep{F}} \otimes  ( \repb{3 } \,, \rep{4 } \,, -  \frac{5 }{12 } )_{ \rep{F}}  \Big]  \otimes  ( \rep{1 } \,, \repb{4 } \,, -\frac{1 }{4 } )_{ \rep{H}\,, \dot \omega}^\dag  + H.c. \non
&\supset& c_4 \zeta_0 \Big[ ( \rep{1 } \,, \repb{3 } \,,  + \frac{2  }{3} )_{ \rep{F}}^{ \prime } \otimes ( \rep{1 } \,, \rep{1 } \,, - 1 )_{ \rep{F}}  \oplus  ( \rep{3 } \,, \rep{3 } \,, 0 )_{ \rep{F}}^\prime \otimes  ( \repb{3 } \,, \rep{3 } \,, -  \frac{1 }{3 } )_{ \rep{F}} \non
&\oplus& ( \rep{3 } \,, \repb{3 } \,, +\frac{1 }{3 } )_{ \rep{F}} \otimes  ( \repb{3 } \,, \rep{1 } \,, -  \frac{2 }{3 } )_{ \rep{F}}^{ \prime \prime \prime }  \Big] \otimes \langle  ( \rep{1 } \,, \repb{3 } \,, -\frac{1 }{3 } )_{ \rep{H}\,, \dot \omega }^\dag \rangle + H.c.  \non
&\Rightarrow& \frac{1 }{ \sqrt{2} } c_4 \zeta_0 V_{ \repb{3}\,, \dot {\rm IX} } (  \EG_L {\mu_R}^c  + c_L {\uG_R}^c - s_L {\dG_R}^c + \UG_L {c_R}^c ) + H.c. \,, \label{eq:Yukawa_331_03a}\\[1mm]
%
&& \frac{  c_4  }{ M_{\rm pl} } \rep{56_F} \rep{56_F} \langle \rep{63_H} \rangle \repb{28_H}_{\,, \dot \omega }^\dag + H.c. \non
&\supset& \frac{  v_U }{ M_{\rm pl}}  ( \repb{4 } \,, \rep{1 } \,, - \frac{3}{4} )_{ \rep{F}}  \otimes ( \rep{4 } \,, \rep{6 } \,, + \frac{1}{4} )_{ \rep{F}}  \otimes  ( \rep{1 } \,, \rep{6 } \,, -\frac{1}{2} )_{ \rep{H}\,, \dot \omega }^\dag  + H.c. \non
&\supset& c_4   \frac{  v_U }{ M_{\rm pl}} \Big[  ( \repb{3 } \,, \rep{1 } \,, - \frac{2 }{3} )_{ \rep{F}}^\prime  \otimes ( \rep{3 } \,, \rep{6 } \,, + \frac{1}{6 } )_{ \rep{F}}   \oplus ( \rep{1 } \,, \rep{1 } \,, - 1 )_{ \rep{F}}  \otimes ( \rep{1 } \,, \rep{6 } \,, + \frac{1}{2} )_{ \rep{F}}^\prime    \Big]  \otimes  ( \rep{1 } \,, \rep{6 } \,, -\frac{1}{2} )_{ \rep{H}\,, \dot \omega }^\dag  + H.c. \non
&\supset& c_4  \zeta_0 \Big[  ( \repb{3 } \,, \rep{1 } \,, - \frac{2 }{3} )_{ \rep{F}}^\prime  \otimes ( \rep{3 } \,, \repb{3 } \,, + \frac{1}{3 } )_{ \rep{F}}   \oplus ( \rep{1 } \,, \rep{1 } \,, - 1 )_{ \rep{F}}  \otimes ( \rep{1 } \,, \repb{3 } \,, + \frac{2 }{3 } )_{ \rep{F}}^{\prime \prime}  \Big]  \otimes \langle ( \rep{1 } \,, \repb{3  } \,, -\frac{1}{3 } )_{ \rep{H}\,, \dot \omega }^{\prime\, \dag} \rangle + H.c.  \non
&\Rightarrow& \frac{1 }{ \sqrt{2} }   c_4 \zeta_0  V_{\repb{3} \,, \dot {\rm VIII} }^{\prime} ( \UG_L {u_R}^c  + \EG_L {e_R}^c  ) + H.c.  \,. \label{eq:Yukawa_331_03b}
\eeqn
\eeqs
where we insert the Higgs VEVs of $\sim \Oc(v_{331})$.

\para
The remaining massless fermions of the $\Gc_{\rm SM}$ are listed as follows
\beqn\label{eq:SM_fermions}
&& \Big[ ( \repb{3}\,, \rep{1}\,, +\frac{1}{3} )_{ \mathbf{F}}^\Omega \oplus ( \rep{1}\,, \rep{1}\,, 0 )_{ \mathbf{F}}^\Omega \Big]  \oplus \Big[ ( \rep{1}\,, \repb{2}\,, -\frac{1}{2} )_{ \mathbf{F}}^\Omega \oplus ( \rep{1}\,, \rep{1}\,, 0 )_{ \mathbf{F}}^{\Omega^{\prime}} \oplus ( \rep{1}\,, \rep{1}\,, 0 )_{ \mathbf{F}}^{\Omega^{\prime\prime}}   \Big]  \subset \repb{8_F}^\Omega \,, \quad \Omega = ( \dot 1\,, \dot 2\,, 3 ) \,, \non
&& ( \rep{1}\,, \rep{1}\,, 0)_{ \mathbf{F}}^{{\rm IV} } \oplus ... \oplus ( \rep{1}\,, \rep{1}\,, 0)_{ \mathbf{F}}^{\dot {\rm IX} } \subset \repb{8_F}^{\Omega } \,,  \non
&& ( \rep{1}\,, \rep{1}\,, 0)_{ \mathbf{F}}^{{\rm VI}^\prime } \oplus  ( \rep{1}\,, \rep{1}\,, 0)_{ \mathbf{F}}^{\dot {\rm VIII}^\prime } \oplus ( \rep{1}\,, \rep{1}\,, 0)_{ \mathbf{F}}^{\dot {\rm IX}^\prime } \subset \repb{8_F}^{\Omega^\prime } \,,  \non
&& ( \rep{1}\,, \rep{1}\,, 0)_{ \mathbf{F}}^{{\rm V}^{ \prime \prime} } \oplus ... \oplus ( \rep{1}\,, \rep{1}\,, 0)_{ \mathbf{F}}^{\dot {\rm IX}^{ \prime \prime } } \subset \repb{8_F}^{\Omega^{ \prime\prime} } \,,  \non
&& \Big[ \cancel{ ( \rep{3}\,, \rep{1}\,, -\frac{1}{3} )_{ \mathbf{F}} } \oplus ( \repb{3}\,, \rep{1}\,, -\frac{2}{3} )_{ \mathbf{F}} \Big]  \oplus \Big[ \bcancel{ ( \rep{1}\,, \rep{2}\,, +\frac{1}{2} )_{ \mathbf{F}} \oplus ( \rep{1}\,, \rep{1}\,, 0 )_{ \mathbf{F}} } \oplus \xcancel{ ( \rep{1}\,, \repb{2}\,, +\frac{1}{2} )_{ \mathbf{F}}^\prime }  \oplus ( \rep{1}\,, \rep{1}\,, +1 )_{ \mathbf{F}} \Big] \non
&\oplus& \Big[ ( \rep{3}\,, \rep{2}\,, +\frac{1}{6} )_{ \mathbf{F}} \oplus \xcancel{ ( \rep{3}\,, \rep{1}\,, -\frac{1}{3} )_{ \mathbf{F}}^\prime }\oplus \bcancel{ ( \rep{3}\,, \rep{1}\,, -\frac{1}{3} )_{ \mathbf{F}}^{\prime \prime } } \non
&\oplus&\cancel{ ( \rep{1}\,, \rep{2}\,, +\frac{1}{2} )_{ \mathbf{F}}^{ \prime \prime} \oplus ( \rep{1}\,, \rep{1}\,, 0 )_{ \mathbf{F}}^{ \prime } \oplus ( \rep{1}\,, \rep{1}\,, 0 )_{ \mathbf{F}}^{ \prime \prime } } \Big]  \subset \rep{28_F}\,, \non
&&\Big[ \xcancel{ ( \rep{1}\,, \repb{2}\,, +\frac{1}{2} )_{ \mathbf{F}}^{\prime\prime\prime }} \oplus  ( \rep{1}\,, \rep{1}\,, +1 )_{ \mathbf{F}}^\prime \oplus  \bcancel{ ( \rep{1}\,, \rep{1}\,, +1 )_{ \mathbf{F}}^{\prime\prime} } \Big] \oplus  \Big[ ( \repb{3}\,, \rep{1}\,, -\frac{2}{3} )_{ \mathbf{F}}^\prime \oplus \bcancel{ ( \rep{1}\,, \rep{1}\,, -1 )_{ \mathbf{F}}  } \Big] \non
&\oplus&  \Big[ ( \rep{3}\,, \rep{2}\,, +\frac{1}{6} )_{ \mathbf{F}}^\prime \oplus \xcancel{ ( \rep{3}\,, \rep{1}\,, -\frac{1}{3} )_{ \mathbf{F}}^{\prime\prime \prime} } \oplus  \bcancel{  ( \rep{3}\,, \repb{2}\,, +\frac{1}{6} )_{ \mathbf{F}}^{\prime\prime} \oplus ( \rep{3}\,, \rep{1}\,, +\frac{2}{3} )_{ \mathbf{F}} } \non
&\oplus& \bcancel{ ( \rep{1}\,, \rep{2}\,, +\frac{1}{2} )_{ \mathbf{F}}^{\prime\prime\prime\prime} \oplus  ( \rep{1}\,, \rep{1}\,, 0 )_{ \mathbf{F}}^{\prime\prime \prime}  } \oplus \xcancel{ ( \rep{1}\,, \repb{2}\,, +\frac{1}{2})_{ \mathbf{F}}^{\prime\prime\prime\prime \prime} } \oplus ( \rep{1}\,, \rep{1}\,, +1)_{ \mathbf{F}}^{\prime\prime \prime} \Big] \non
&\oplus& \Big[ ( \rep{3}\,, \rep{2}\,, +\frac{1}{6} )_{ \mathbf{F}}^{\prime\prime \prime} \oplus \xcancel{ ( \rep{3}\,, \rep{1}\,, -\frac{1}{3})_{ \mathbf{F}}^{\prime\prime\prime \prime} }  \oplus \bcancel{ ( \rep{3}\,, \rep{1}\,, -\frac{1}{3})_{ \mathbf{F}}^{\prime\prime\prime \prime \prime }  } \non
&\oplus& \bcancel{ ( \repb{3}\,, \rep{2}\,, -\frac{1}{6})_{ \mathbf{F}}   \oplus ( \repb{3}\,, \rep{1}\,, -\frac{2}{3})_{ \mathbf{F}}^{\prime\prime} } \oplus ( \repb{3}\,, \rep{1}\,, -\frac{2}{3})_{ \mathbf{F}}^{\prime\prime\prime} \Big]   \subset \rep{56_F} \,.
\eeqn
The fermions that become massive at this stage are further crossed outs.
After this stage of symmetry breaking, there are three generations of the SM fermion irreps, together with twenty-three left-handed massless sterile neutrinos~\footnote{The number of residual left-handed massless sterile neutrinos have been precisely obtained through the `t Hooft anomaly matching in Ref.~\cite{Chen:2023qxi}.}.
The third-generational SM fermions are from the rank-$2$ chiral IRAFFS of $\[ \repb{8_F}^\omega \oplus \rep{28_F} \]$, while the first- and second-generational SM fermions are from the rank-$3$ chiral IRAFFS of $\[ \repb{8_F}^{\dot \omega }\oplus \rep{56_F}\]$.
The flavor indices for the massive anti-fundamental fermions at this stage are chosen to be $\omega={\rm VI}$ and $\dot \omega=( \dot {\rm VIII}\,, \dot {\rm IX})$, respectively.

\subsection{A summary of the vectorlike fermion masses}

\begin{table}[htp] {\small
\begin{center}
\begin{tabular}{c|c|c|c|c}
\hline \hline
   stages &  $Q_e=- \frac{1}{3}$  & $Q_e=+ \frac{2}{3}$   & $Q_e= -1$ & $Q_e= 0$  \\
\hline \hline
  $v_{441}$   & $\DG$   &  -  &  $( \eG^{ \prime\prime}\,, \nG^{\prime\prime} )$  &  $\{ \check \nG^{ \prime}\,, \check \nG^{\prime\prime} \}$  \\
  $\{ \Omega \}$  &  ${\rm IV}$  &    & ${\rm IV}$ &  $\{ {\rm IV}^\prime \,, {\rm IV}^{\prime\prime} \}$  \\[1mm]
    \hline
  $v_{341}$   &  $\dG\,, \{  \DG^{ \prime\prime}\,,  \DG^{\prime\prime\prime\prime\prime} \}$  & $\uG\,, \UG$   &  $\EG\,,( \eG\,, \nG )\,,( \eG^{ \prime\prime\prime\prime}\,, \nG^{\prime\prime\prime\prime} )$  & $\{ \check \nG\,, \check \nG^{\prime\prime\prime} \}$   \\
   $\{ \Omega \}$   &  $\{ {\rm V} \,, \dot {\rm VII}  \}$  &    &  $\{ {\rm V} \,, \dot {\rm VII}  \}$  &  $\{ {\rm V}^\prime \,, \dot {\rm VII}^{\prime} \}$  \\[1mm]
    \hline
   $v_{331}$   &  $\{  \DG^{ \prime}\,,  \DG^{ \prime\prime\prime}\,, \DG^{\prime\prime\prime\prime} \}$  &  -  &  $( \eG^\prime \,, \nG^\prime )\,,( \eG^{ \prime\prime\prime}\,, \nG^{\prime\prime\prime} )\,, ( \eG^{ \prime\prime\prime\prime\prime}\,, \nG^{\prime\prime\prime\prime\prime} )$  &  -  \\
    $\{ \Omega\}$  &  $\{ {\rm VI} \,, \dot {\rm VIII}\,, \dot {\rm IX} \}$  &   &  $\{ {\rm VI} \,, \dot {\rm VIII}\,, \dot {\rm IX} \}$  &    \\[1mm]
\hline\hline
\end{tabular}
\caption{
The vectorlike fermions at different intermediate symmetry breaking scales in the ${\rm SU}(8)$ theory.}
\label{tab:SU8_vectorferm}
\end{center}
}
\end{table}%

\para
Here, we summarize all vectorlike fermions that acquire masses above the EWSB scale in Tab.~\ref{tab:SU8_vectorferm}, with the conventions of flavor indices displayed explicitly.

\subsection{The EWSB stage}

\para
The Higgs sector after the third-stage symmetry breaking only contains one single SM Higgs doublet of
\beqn\label{eq:321_Higgs_massless}
&&  ( \rep{1} \,, \repb{2} \,, +\frac{1}{2} )_{\mathbf{H}}^{ \prime\prime \prime} \subset \rep{70_H} \,,
\eeqn
with the $\widetilde{\rm U}(1)_{B-L}$-neutrality condition.
By decomposing the term of $\rep{28_F} \rep{28_F} \rep{70_H} + H.c.$, we find the tree-level top quark mass of
\beqn\label{eq:top_Yukawa}
&&Y_\Tc \rep{28_F} \rep{28_F} \rep{70_H}  + H.c. \non
&\supset& Y_\Tc ( \rep{6}\,, \rep{1}\,, -\frac{1}{2} )_{ \mathbf{F}} \otimes  ( \rep{4}\,, \rep{4}\,, 0 )_{ \mathbf{F}}  \otimes ( \rep{4}\,, \repb{4}\,, +\frac{1}{2} )_{ \mathbf{H}} + H.c.  \non
&\supset& Y_\Tc ( \repb{3}\,, \rep{1}\,, -\frac{2}{3} )_{ \mathbf{F}} \otimes  ( \rep{3}\,, \rep{4}\,, -\frac{1}{12} )_{ \mathbf{F}}  \otimes ( \rep{1}\,, \repb{4}\,, +\frac{3}{4} )_{ \mathbf{H}}^\prime  + H.c.\non
&\supset& Y_\Tc ( \repb{3}\,, \rep{1}\,, -\frac{2}{3} )_{ \mathbf{F}} \otimes  ( \rep{3}\,, \rep{3}\,, 0 )_{ \mathbf{F}}  \otimes ( \rep{1}\,, \repb{3}\,, +\frac{2}{3} )_{ \mathbf{H}}^{ \prime \prime \prime }  + H.c.\non
&\supset& Y_\Tc ( \repb{3}\,, \rep{1}\,, -\frac{2}{3} )_{ \mathbf{F}} \otimes  ( \rep{3}\,, \rep{2}\,, +\frac{1}{6} )_{ \mathbf{F}}  \otimes ( \rep{1}\,, \repb{2}\,, +\frac{1}{2} )_{ \mathbf{H}}^{ \prime\prime \prime } + H.c. \non
&\Rightarrow&\frac{1}{\sqrt{2}} Y_\Tc  t_L {t_R}^c v_{\rm EW} + H.c. \,,
\eeqn
with the natural Yukawa coupling of $Y_\Tc\sim \Oc(1)$.
Furthermore, this result can be even generalized to any flavor-unified ${\rm SU}(N>5)$ theory so that a rank-$2$ chiral IRAFFS is always necessary.
With the identification of $( \rep{3}\,, \rep{2}\,, +\frac{1}{6} )_{ \mathbf{F}}\equiv (t_L \,, b_L)^T$ and $( \repb{3}\,, \rep{1}\,, -\frac{2}{3} )_{ \mathbf{F}} \equiv {t_R}^c$ within the $\rep{28_F}$, it is straightforward to infer that $( \rep{1}\,, \rep{1}\,,  +1 )_{ \mathbf{F}} \equiv { \tau_R}^c$ in Tab.~\ref{tab:SU8_28ferm}.
This explains why does the third-generational SM $\rep{10_F}$ reside in the $\rep{28_F}$, while the first- and second-generational SM $\rep{10_F}$'s reside in the $\rep{56_F}$.
The same observation and argument were previously acknowledged in Ref.~\cite{Barr:2008pn}.

\section{The gravity-induced SM quark/lepton mass terms}
\label{section:gravity_Fmass}

\para
In this section, we derive the SM quark/lepton mass terms from all possible $d=5$ non-renormalizable operators that break the global symmetries in Eqs.~\eqref{eq:DRS_SU8} and \eqref{eq:PQ_SU8} explicitly.
Such operators are possible, since gravity does not respect any underlying global symmetry~\cite{Kallosh:1995hi,Harlow:2018jwu,Harlow:2018tng}.
Thus, all $d\geq5$ non-renormalizable operators are suppressed by $1/M_{\rm pl}^{d-4}$.
There are potentially two categories of such $d=5$ operators, which will phrased as the direct Yukawa couplings and the indirect Yukawa couplings.
Historically, the gravitational effects to the SM fermion mass spectrum through the direct Yukawa couplings were discussed in various GUTs~\cite{Ellis:1979fg,Chacko:2020tbu}.

\para
We wish to highlight three major points which were absent from the pioneering work of Ref.~\cite{Barr:2008pn}, where the SM quark/lepton masses also stem from the $d\geq 5$ non-renormalizable operators, namely,
\begin{enumerate}

\item the explicit breaking of the emergent global symmetries in Eqs.~\eqref{eq:DRS_SU8} and \eqref{eq:PQ_SU8} as we have identified based on the chiral IRAFFSs in Eq.~\eqref{eq:SU8_3gen_fermions} will be the guidance to organize the non-renormalizable operators, which can only be due to the gravity~\cite{Kallosh:1995hi,Harlow:2018jwu,Harlow:2018tng};

\item a realistic symmetry breaking pattern has been given in the previous Sec.~\ref{section:pattern}, where three intermediate symmetry breaking scales will enter the effective Yukawa couplings of light SM quarks/leptons;

\item it is imminent to ask if the unique SM Higgs doublet of $( \rep{1} \,, \repb{2} \,, +\frac{1}{2} )_{\mathbf{H}}^{ \prime\prime \prime}\subset \rep{70_H}$, as we have assumed in Sec.~\ref{section:SU8_Higgs} and derived its top quark Yukawa coupling in Eq.~\eqref{eq:321_Higgs_massless}, could couple to all other light SM quarks/leptons, and how would hierarchical Yukawa couplings be generated accordingly.
Below, we will show that a set of $d=5$ operators that include the unique SM Higgs doublet is sufficient to achieve this vision~\footnote{In Ref.~\cite{Barr:2008pn}, both $d=5$ and $d=6$ operators were assumed to originate the SM fermion mass hierarchies. A unique SM Higgs doublet was not assumed therein.}.

\end{enumerate}

\subsection{The direct Yukawa couplings from the non-renormalizable bi-linear fermion operators}
\label{section:direct_Yukawa}

\begin{table}[htp]
\begin{center}
\begin{tabular}{c|c|c}
\hline \hline
$\Oc_{\Fc }$ & $\widetilde {\rm SU}(4)_\omega \otimes \widetilde {\rm SU}(5)_{\dot \omega}$ indices  & $\Pc\Qc$  \\
\hline \hline
$c_1\,\Oc_{\Fc }^{(1\,,1)} \equiv c_1\, \repb{8_F}^\omega \rep{28_F} \cdot \repb{8_H}_{\,,\kappa} \cdot  \rep{63_H}  $ & $ \omega \neq \kappa $  &  $0$  \\[1mm]
    \hline
$c_2\, \Oc_{\Fc }^{(2\,,1)} \equiv c_2\, \rep{28_F}  \rep{28_F} \cdot  \repb{28_H}_{\,, \dot \omega }^\dag \cdot \repb{28_H}_{\,, \dot \kappa}^\dag$  & $-$ & $2(p + q_2 + q_3 )$   \\[1mm]
    \hline
$c_3\, \Oc_{\Fc }^{(3\,,1)} \equiv c_3\, \repb{8_F}^{ \omega } \rep{56_F} \cdot \repb{8_H}_{\,,\kappa_1 } \cdot  \repb{8_H}_{\,,\kappa_2 } $  & $-$  & $-p - 2 q_2 + q_3$   \\[1mm]
$c_3\, \Oc_{\Fc }^{(3\,,2)} \equiv c_3\, \repb{8_F}^{ \dot \omega } \rep{56_F} \cdot \repb{28_H}_{\,, \dot \kappa }^\dag  \cdot  \rep{70_H}^\dag$  & $-$  & $2(p + q_2 + q_3 )$    \\[1mm]
$c_3\, \Oc_{\Fc }^{(3\,,3)} \equiv c_3\, \repb{8_F}^{\dot \omega } \rep{56_F} \cdot \repb{28_H}_{\,,\dot \kappa } \cdot  \rep{63_H}$  &  $ \dot \omega \neq \dot \kappa$  &  $0$  \\[1mm]
    \hline
$c_4\, \Oc_{\Fc }^{(4\,,1)}\equiv c_4\, \rep{56_F}  \rep{56_F} \cdot \repb{28_H}_{\,, \dot \omega }  \cdot  \rep{70_H}$ & $-$  &  $-p - 2 q_2 + q_3$  \\[1mm]
$c_4\, \Oc_{\Fc }^{(4\,,2)}\equiv c_4\, \rep{56_F}  \rep{56_F} \cdot  \repb{8_H}_{\,,\omega }^\dag \cdot  \repb{8_H}_{\,,\kappa }^\dag $  & $-$  &$2(p + q_2 + q_3 )$    \\[1mm]
 $c_4\, \Oc_{\Fc }^{(4\,,3)}\equiv c_4\, \rep{56_F}  \rep{56_F} \cdot  \repb{28_H}_{\,, \dot \omega }^\dag \cdot  \rep{63_H} $  & $-$  & $p+3 q_3 \neq 0$   \\[1mm]
    \hline
$c_5\, \Oc_{\Fc }^{(5\,,1)}\equiv c_5\, \rep{28_F}  \rep{56_F} \cdot  \repb{8_H}_{\,,\omega }  \cdot  \rep{70_H} $ & $-$  & $-p - 2 q_2 + q_3$  \\[1mm]
\hline\hline
\end{tabular}
\caption{
The $d= 5$ $\widetilde{\rm U}(1)_T$-neutral fermion bi-linear operators for direct Yukawa couplings in the ${\rm SU}(8)$ theory.}
\label{tab:SU8_d5direct}
\end{center}
\end{table}%

\para
We proceed to write down all possible $d= 5$ fermion bi-linear operators for the mass terms, which are suppressed by $1/M_{\rm pl}$.
Only the $\widetilde{\rm U}(1)_T$-neutral operators will be listed, which can contribute to the SM quark/lepton mass terms~\footnote{The operators with non-vanishing $\widetilde{ {\rm U}}(1)_T$-charges turn out to either break the gauge invariance with the VEV insertions or lead to the Majorana neutrino mass terms.}.
To organize these operators systematically, we only include operators that explicitly break all global symmetries in Eqs.~\eqref{eq:DRS_SU8} and \eqref{eq:PQ_SU8}.
It also turns out that the observed SM quark/lepton masses and the CKM mixing patterns are not contributed from further terms involving the  possible mixings between the $\widetilde {\rm SU}(4)_\omega$ and the $\widetilde {\rm SU}(5)_{\dot \omega}$ states.
Accordingly, we categorize all $d=5$ bi-linear fermion operators for direct Yukawa couplings in Tab.~\ref{tab:SU8_d5direct}.
For simplicity, we only assume the distinctive Wilson coefficients of $(c_1\,,...\,,c_5)$ for each type of $d=5$ bi-linear fermion operators.
Here, the operator of $\Oc_{\Fc }^{(4\,,3)}$ and the corresponding vectorlike fermion masses have been obtained in Eq.~\eqref{eq:Yukawa_341_03} during the second symmetry breaking stage.
This set of $d=5$ bi-linear fermion operators all break the global $\widetilde {\rm SU}(4)_{ \omega } \otimes \widetilde {\rm SU}(5)_{\dot \omega }$ symmetries explicitly with certain choices of the flavor indices, thus, we expect them to be suppressed by $1/M_{\rm pl}$.
Two operators of $\Oc_\Fc^{ (1\,,1)}$ and $\Oc_\Fc^{ (3\,,3)}$ are $\widetilde {\rm U}(1)_{\rm PQ}$-neutral, while the $\Oc_\Fc^{ (4\,,3)}$ is manifestly $\widetilde {\rm U}(1)_{\rm PQ}$-breaking, according to the $\widetilde {\rm U}(1)_{\rm PQ}$ charge relations in Eq.~\eqref{eq:PQcharges_SU8}.
Since we have conjectured the unique SM Higgs doublet from the $\rep{70_H}$, one should particularly focus on the mass terms from the operators of $\Oc_{\Fc }^{(3\,,2)}$, $\Oc_{\Fc }^{(4\,,1)}$, and $\Oc_{\Fc }^{(5 \,,1 )}$.
The relevant mass terms that involve the SM fermions only will be labelled with \underline{underlines}.
Below, we proceed to decompose each $d=5$ fermion bi-linear operators to find the fermion mass (mixing) terms.

\para
For the operator of $\Oc_{\Fc }^{(1\,,1)}$, it is decomposed as follows
\beqs\label{eqs:OF_11}
\beqn
&& \frac{ c_1 }{ M_{\rm pl} }  \repb{8_F}^\omega \rep{28_F} \cdot \repb{8_H}_{\,,\kappa}  \cdot  \langle \rep{63_H} \rangle + H.c. \non
&\supset& c_1 \zeta_0  \Big[ ( \repb{4 } \,, \rep{1 } \,, + \frac{1 }{4} )_{ \rep{F}}^\omega \otimes ( \rep{6 } \,, \rep{1 } \,, - \frac{1 }{2 } )_{ \rep{F}} -  ( \rep{1 } \,, \repb{4 } \,, - \frac{1 }{4} )_{ \rep{F}}^\omega \otimes ( \rep{4 } \,, \rep{4 } \,, 0 )_{ \rep{F}} \Big] \otimes \langle ( \repb{4 } \,, \rep{1 } \,, + \frac{1 }{4 } )_{ \rep{H}\,, \kappa } \rangle  + H.c. \non
&\Rightarrow& \frac{1 }{ \sqrt{2} }  c_1   \zeta_0   (  \DG_L \Dc_R^{ \omega \, c} - \Ec_L^\omega {\eG_R^{\prime\prime \, c}}  + \Nc_L^\omega  {\nG_R^{\prime\prime \, c}} - \check \Nc_L^{ \omega^\prime } \check \nG_R^{ \prime\,c } - \check \Nc_L^{ \omega^{ \prime \prime} } \check \nG_R^{ \prime \prime \,c } ) W_{ \repb{4}\,,{\rm IV} } + H.c.  \,, \quad (\omega \neq {\rm IV} ) \,, \label{eq:OF_11a}\\[1mm]
&& \frac{ c_1  }{ M_{\rm pl} }  \repb{8_F}^\omega \rep{28_F} \cdot \repb{8_H}_{\,,\kappa}  \cdot  \langle \rep{63_H} \rangle + H.c. \non
&\supset& c_1  \zeta_0  \Big[  ( \repb{4 } \,, \rep{1 } \,, + \frac{1 }{4} )_{ \rep{F}}^\omega \otimes ( \rep{4 } \,, \rep{4 } \,, 0 )_{ \rep{F}} - ( \rep{1 } \,, \repb{4 } \,, - \frac{1 }{4} )_{ \rep{F}}^\omega \otimes ( \rep{1 } \,, \rep{6 } \,, + \frac{1 }{ 2 } )_{ \rep{F}}   \Big] \otimes  \langle ( \rep{1 } \,, \repb{4 } \,, - \frac{1 }{4 } )_{ \rep{H}\,, \kappa }  \rangle + H.c. \non
&\Rightarrow& \frac{1 }{ \sqrt{2} }  c_1  \zeta_0 ( \DG_L^{ \prime\prime} \Dc_R^{ \omega \, c} + \check \Nc_L^{ \omega } \check \nG_R^{ \prime \prime \,c } - \Ec_L^\omega {\eG_R}^{ c}  + \Nc_L^\omega {\nG_R}^{c} - \check \Nc_L^{ \omega^\prime } \check \nG_R^c  ) w_{ \repb{4}\,,{\rm V} } + H.c. \,, \quad ( \omega \neq {\rm V} ) \,.\label{eq:OF_11b}
\eeqn
\eeqs
The specific flavor indices are chosen to assure both terms break the global $\widetilde{ {\rm SU}}(4 )_\omega$ explicitly.

\para
For the operator of $\Oc_{\Fc }^{(2\,,1)}$, it is decomposed as
\beqs\label{eqs:OF_21}
\beqn
&& \frac{ c_2 }{ M_{\rm pl} } \rep{28_F}  \rep{28_F} \cdot  \repb{28_H}_{\,, \dot \omega }^\dag \cdot \repb{28_H}_{\,, \dot \kappa}^\dag + H.c.  \non
&\supset& \frac{c_2 }{ M_{\rm pl} }   \[  ( \rep{1 } \,, \rep{6 } \,, +\frac{1 }{2 } )_{ \rep{F}} \otimes ( \rep{6 } \,, \rep{1 } \,, -\frac{1 }{2 } )_{ \rep{F}}   \oplus ( \rep{4  } \,, \rep{4 } \,, 0 )_{ \rep{F}}  \otimes ( \rep{4  } \,, \rep{4 } \,, 0 )_{ \rep{F}}   \]  \non
&\otimes& ( \repb{4 } \,, \repb{4 } \,, 0 )_{ \rep{H}\,, \dot \omega }^\dag  \otimes ( \repb{4 } \,, \repb{4 } \,, 0 )_{ \rep{H}\,, \dot \kappa}^\dag + H.c.  \Rightarrow \textrm{not gauge-invariant} \,, \label{eq:OF_21a} \\[1mm]
&& \frac{c_2 }{ M_{\rm pl} }  \rep{28_F}  \rep{28_F} \cdot  \repb{28_H}_{\,, \dot \omega}^\dag \cdot  \repb{28_H}_{\,, \dot \kappa}^\dag \non
&\supset& \frac{c_2 }{ M_{\rm pl} } (  \rep{6 } \,, \rep{1 } \,, - \frac{1 }{2 } )_{ \rep{F}}  \otimes ( \rep{4 } \,, \rep{4 } \,, 0 )_{ \rep{F}}  \otimes ( \repb{4 } \,, \repb{4 } \,, 0 )_{ \rep{H}\,, \dot \omega }^\dag  \otimes ( \rep{1 } \,, \rep{6 } \,, -\frac{1  }{2 } )_{ \rep{H}\,, \dot \kappa}^\dag + H.c. \non
&\supset& \frac{c_2 }{ M_{\rm pl} } (  \repb{3 } \,, \rep{1 } \,, - \frac{2 }{3 } )_{ \rep{F}}  \otimes ( \rep{3 } \,, \rep{4 } \,, - \frac{1 }{12 } )_{ \rep{F}}  \otimes \langle ( \rep{1 } \,, \repb{4 } \,, - \frac{1 }{4 } )_{ \rep{H}\,, \dot \omega }^\dag  \rangle \otimes ( \rep{1 } \,, \rep{6 } \,, -\frac{1  }{2 } )_{ \rep{H}\,, \dot \kappa}^\dag + H.c. \non
&\supset& c_2 \dot \zeta_2 (  \repb{3 } \,, \rep{1 } \,, - \frac{2 }{3 } )_{ \rep{F}}  \otimes ( \rep{3 } \,, \rep{3 } \,, 0 )_{ \rep{F}}  \otimes  ( \rep{1 } \,, \rep{3 } \,, -\frac{2  }{3 } )_{ \rep{H}\,, \dot \kappa}^{ \dag} + H.c. \,. \label{eq:OF_21b}
%
\eeqn
\eeqs
The only possible gauge-invariant mass term should be formulated by assuming a non-vanishing EWSB VEV from the $( \rep{1 } \,, \rep{2 } \,, -\frac{1  }{2 } )_{ \rep{H}\,, \dot \kappa}\subset ( \rep{1 } \,, \rep{3 } \,, -\frac{2  }{3 } )_{ \rep{H}\,, \dot \kappa}$ component.
Thus, this operator can not contribute to any mass term.

\para
For the operator of $\Oc_{\Fc }^{(3\,,1)}$, it is decomposed as
\beqs\label{eqs:OF_31}
\beqn
&& \frac{ c_3 }{ M_{\rm pl} } \repb{8_F}^{ \omega }  \rep{56_F} \cdot   \repb{8_H}_{\,, \kappa_1 } \cdot  \repb{8_H}_{\,, \kappa_2 } + H.c. \non
&\supset& \frac{c_3  }{ M_{\rm pl} } \Big[  ( \repb{4 } \,, \rep{1 } \,,  + \frac{1 }{4 } )_{ \rep{F} }^{ \omega }  \otimes ( \rep{6 } \,, \rep{4 } \,,  - \frac{1 }{4 } )_{ \rep{F} }  \oplus  ( \rep{1 } \,, \repb{4 } \,,   - \frac{1 }{4 } )_{ \rep{F} }^{ \omega }  \otimes ( \rep{4 } \,, \rep{6 } \,,  + \frac{1 }{4 } )_{ \rep{F} } \Big]  \non
&\otimes& \langle ( \repb{4 } \,, \rep{1 } \,,  + \frac{1 }{4 } )_{ \rep{H}\,, \kappa_1 }  \rangle  \otimes   ( \rep{1 } \,, \repb{4 } \,, - \frac{1 }{4 } )_{ \rep{H}\,, \kappa_2 } + H.c. \non
&\supset& c_3  \frac{ W_{ \repb{4} \,,{\rm IV} }  }{ \sqrt{2} M_{\rm pl} } \Big[   ( \repb{3 } \,, \rep{1 } \,,  + \frac{1 }{3 } )_{ \rep{F} }^{ \omega }  \otimes ( \rep{3 } \,, \rep{4 } \,,  - \frac{1 }{12 } )_{ \rep{F} }^\prime  \oplus  ( \rep{1 } \,, \repb{4 } \,,   - \frac{1 }{4 } )_{ \rep{F} }^{ \omega }  \otimes ( \rep{1 } \,, \rep{6 } \,,  + \frac{1 }{2 } )_{ \rep{F} }^\prime  \Big] \otimes  \langle ( \rep{1 } \,, \repb{4 } \,, - \frac{1 }{4 } )_{ \rep{H}\,, \kappa_2 } \rangle + H.c. \non
%
%
&\Rightarrow& \frac{c_3 }{ 2 }    \zeta_1 \Big[ ( \DG_L^{ \prime\prime \prime\prime \prime } \Dc_R^{\omega \, c}  + \Ec_L^\omega  {\eG_R^{ \prime\prime \prime\prime\, c} } - \Nc_L^\omega  {\nG_R^{ \prime\prime \prime\prime\, c} } + \check \Nc_L^{ \omega^\prime } \check \nG_R^{ \prime\prime\prime\, c}  ) w_{ \repb{4} \,,{\rm V} } \non
&+&   (  \DG_L^{ \prime\prime \prime\prime } \Dc_R^{\omega\, c}  - \Ec_L^\omega  {\eG_R^{ \prime\prime \prime\prime \prime\, c} } +  \Nc_{ L}^\omega  {\nG_R^{ \prime\prime \prime\prime \prime\, c} }  + \check \Nc_L^{ \omega^{ \prime\prime} } \check \nG_R^{ \prime\prime\prime \, c} ) V_{ \repb{3} \,,3\,, {\rm VI}  }  \Big] + H.c.   \,, \label{eq:OF_31a}\\[1mm]
&& \frac{ c_3  }{ M_{\rm pl} } \repb{8_F}^{ \omega }  \rep{56_F} \cdot   \repb{8_H}_{\,, \kappa_1 } \cdot  \repb{8_H}_{\,, \kappa_2 } + H.c. \non
&\supset& \frac{c_3  }{ M_{\rm pl} }  \Big[    ( \repb{4 } \,, \rep{1 } \,,  + \frac{1 }{4 } )_{ \rep{F} }^{ \omega }  \otimes ( \rep{4 } \,, \rep{6 } \,,  + \frac{1 }{4 } )_{ \rep{F} }  \oplus  ( \rep{1 } \,, \repb{4 } \,,   - \frac{1 }{4 } )_{ \rep{F} }^{ \omega }  \otimes ( \rep{1 } \,, \repb{4 } \,,  + \frac{3 }{4 } )_{ \rep{F} }   \Big] \non
&\otimes&  ( \rep{1 } \,, \repb{4 } \,, - \frac{1 }{4 } )_{ \rep{H}\,, \kappa_1 }  \otimes  ( \rep{1 } \,, \repb{4 } \,, - \frac{1 }{4 } )_{ \rep{H}\,, \kappa_2 } + H.c. \non
&\supset&  \frac{c_3  }{ M_{\rm pl} }  \Big[   ( \repb{3 } \,, \rep{1 } \,,  + \frac{1 }{3 } )_{ \rep{F} }^{ \omega }  \otimes ( \rep{3 } \,, \rep{6 } \,,  + \frac{1 }{6 } )_{ \rep{F} }  \oplus  ( \rep{1 } \,, \rep{1 } \,,  0 )_{ \rep{F} }^{ \omega }  \otimes ( \rep{1 } \,, \rep{6 } \,,  + \frac{1 }{2 } )_{ \rep{F} }^\prime \oplus  ( \rep{1 } \,, \repb{4 } \,,   - \frac{1 }{4 } )_{ \rep{F} }^{ \omega }  \otimes ( \rep{1 } \,, \repb{4 } \,,  + \frac{3 }{4 } )_{ \rep{F} }   \Big] \non
&\otimes& \langle ( \rep{1 } \,, \repb{4 } \,, - \frac{1 }{4 } )_{ \rep{H}\,, \kappa_1 }  \rangle \otimes ( \rep{1 } \,, \repb{4 } \,, - \frac{1 }{4 } )_{ \rep{H}\,, \kappa_2 }  + H.c. \non
&\supset&  c_3 \frac{ w_{ \repb{4}\,,{\rm V} } }{ \sqrt{2} M_{\rm pl} }  \Big[   ( \repb{3 } \,, \rep{1 } \,,  + \frac{1 }{3 } )_{ \rep{F} }^{ \omega }  \otimes ( \rep{3 } \,, \rep{3 } \,,  0 )_{ \rep{F} }^\prime  \oplus  ( \rep{1 } \,, \rep{1 } \,,  0 )_{ \rep{F} }^{ \omega }  \otimes ( \rep{1 } \,, \rep{3 } \,,  + \frac{1 }{3 } )_{ \rep{F} }^\prime  \oplus ( \rep{1 } \,, \repb{3 } \,,   - \frac{1 }{3 } )_{ \rep{F} }^{ \omega }  \otimes ( \rep{1 } \,, \repb{3 } \,,  + \frac{2 }{3 } )_{ \rep{F} }^\prime  \Big]  \non
&\otimes& \langle ( \rep{1 } \,, \repb{3 } \,, - \frac{1 }{3 } )_{ \rep{H}\,, \kappa_2 }  \rangle + H.c.  \non
&\Rightarrow&  \frac{c_3 }{ 2 }  \zeta_2  (  \DG_L^{ \prime\prime \prime } \Dc_R^{\omega \, c} + \check \Nc_L^{ \omega } \check \nG_R^{ \prime\prime\prime\, c} - \Ec_L^\omega  {\eG_R^{ \prime\prime\prime\, c}} + \Nc_{ L}^\omega {\nG_R^{ \prime\prime\prime\, c}}  ) V_{ \repb{3}\,, 3\,, {\rm VI} }  +H.c. \,.\label{eq:OF_31b}
\eeqn
\eeqs

\para
For the operator of $\Oc_{\Fc }^{(3\,,2)}$, it is decomposed as
\beqs\label{eqs:OF_32}
\beqn
&& \frac{ c_3  }{ M_{\rm pl} } \repb{8_F}^{ \dot \omega }  \rep{56_F} \cdot \repb{28_H}_{\,,\dot \kappa }^\dag \cdot  \rep{70_H}^\dag  + H.c. \non
&\supset& \frac{c_3  }{ M_{\rm pl} }  \Big[ ( \repb{4 } \,, \rep{1 } \,, + \frac{1 }{4 } )_{ \rep{F}}^{ \dot \omega } \otimes  ( \rep{4 } \,, \rep{6 } \,, + \frac{1 }{4 } )_{ \rep{F}} \oplus  ( \rep{1 } \,, \repb{4 } \,,  - \frac{1 }{4 } )_{ \rep{F}}^{ \dot \omega } \otimes ( \rep{1 } \,, \repb{4 } \,, + \frac{3 }{4 } )_{ \rep{F}}  \Big] \non
&\otimes& ( \repb{4 } \,, \repb{4 } \,, 0 )_{ \rep{H}\,, \dot \kappa }^\dag \otimes ( \rep{4 } \,, \repb{4 } \,, +\frac{1 }{2 } )_{ \rep{H} }^\dag + H.c. \non
&\supset& \frac{c_3  }{ M_{\rm pl} }  \Big[ ( \repb{3 } \,, \rep{1 } \,, + \frac{1 }{3 } )_{ \rep{F}}^{ \dot \omega } \otimes  ( \rep{3 } \,, \rep{6 } \,, + \frac{1 }{6 } )_{ \rep{F}} \oplus ( \rep{1 } \,, \rep{1 } \,, 0 )_{ \rep{F}}^{ \dot \omega } \otimes  ( \rep{1 } \,, \rep{6 } \,, + \frac{1 }{2 } )_{ \rep{F}}^\prime  \oplus  ( \rep{1 } \,, \repb{4 } \,,  - \frac{1 }{4 } )_{ \rep{F}}^{ \dot \omega } \otimes ( \rep{1 } \,, \repb{4 } \,, + \frac{3 }{4 } )_{ \rep{F}}  \Big] \non
&\otimes& \langle ( \rep{1 } \,, \repb{4 } \,, -\frac{1 }{4 } )_{ \rep{H}\,, \dot \kappa }^\dag \rangle \otimes ( \rep{1 } \,, \repb{4 } \,, +\frac{3 }{4 } )_{ \rep{H} }^{\prime\,\dag} + H.c. \non
&\supset& c_3  \frac{w_{ \repb{4}\,, \dot {\rm VII} } }{ \sqrt{2} M_{\rm pl} } \Big[  ( \repb{3 } \,, \rep{1 } \,, + \frac{1 }{3 } )_{ \rep{F}}^{ \dot \omega } \otimes  ( \rep{3 } \,, \repb{3 } \,, + \frac{1 }{3 } )_{ \rep{F}} \oplus ( \rep{1 } \,, \rep{1 } \,, 0 )_{ \rep{F}}^{ \dot \omega } \otimes  ( \rep{1 } \,, \repb{3 } \,, + \frac{2 }{3 } )_{ \rep{F}}^{\prime \prime}  \non
&\oplus& ( \rep{1 } \,, \repb{3 } \,,  - \frac{1 }{3 } )_{ \rep{F}}^{ \dot \omega } \otimes ( \rep{1 } \,, \rep{1 } \,, + 1 )_{ \rep{F}}^{ \prime\prime }  \oplus ( \rep{1 } \,, \rep{1 } \,,  0 )_{ \rep{F}}^{ \dot \omega^{ \prime\prime } } \otimes ( \rep{1 } \,, \repb{3 } \,, + \frac{2 }{3 } )_{ \rep{F}}^\prime \Big]  \otimes \langle ( \rep{1 } \,, \repb{3 } \,, +\frac{2 }{3 } )_{ \rep{H} }^{\prime \prime\prime \,\dag} \rangle + H.c. \non
&\Rightarrow& \frac{ c_3 }{2 } \dot \zeta_2   \Big( \dG_L \Dc_R^{\dot \omega\, c} +  \check \Nc_L^{ \dot \omega} {\nG_R^{ \prime\prime \prime\prime \prime\, c} } +  \Ec_L^{\dot \omega } {\EG_R}^c+  \check \Nc_L^{ \dot \omega^{ \prime\prime} }  {\nG_R^{ \prime\prime\prime\, c} }  \Big) v_{\rm EW} + H.c. \,, \label{eq:OF_32a}\\[1mm]
&& \frac{c_3  }{ M_{\rm pl} } \repb{8_F}^{ \dot \omega }  \rep{56_F} \cdot \repb{28_H}_{\,,\dot \kappa }^\dag \cdot  \rep{70_H}^\dag + H.c. \non
&\supset& c_3  \frac{ V_{ \repb{3}\,, \dot {\rm IX} } }{ \sqrt{2} M_{\rm pl} }  \Big[ ( \repb{3 } \,, \rep{1 } \,, + \frac{1 }{3 } )_{ \rep{F}}^{ \dot \omega } \otimes  ( \rep{3 } \,, \rep{3 } \,, 0 )_{ \rep{F}}^\prime \oplus ( \rep{1 } \,, \rep{1 } \,, 0 )_{ \rep{F}}^{ \dot \omega } \otimes  ( \rep{1 } \,, \rep{3 } \,, + \frac{1 }{3 } )_{ \rep{F}}^\prime  \oplus  ( \rep{1 } \,, \repb{3 } \,,  - \frac{1 }{3 } )_{ \rep{F}}^{ \dot \omega } \otimes ( \rep{1 } \,, \repb{3 } \,, + \frac{2 }{3 } )_{ \rep{F}}^\prime  \Big] \non
&\otimes&  ( \rep{1 } \,, \repb{3 } \,, +\frac{2 }{3 } )_{ \rep{H} }^{\prime \prime\prime\,\dag} + H.c.  \non
&\Rightarrow& \frac{ c_3 }{2 } \dot \zeta_3    \Big( \uline{ s_L \Dc_R^{\dot \omega\, c}  } +\check \Nc_L^{ \dot \omega } {\nG_R^{\prime\prime\prime\prime\,c }} +  \check \Nc_L^{ \dot \omega^\prime } {\nG_R^{\prime\prime\prime \,c }}  - \uline{ \Ec_L^{\dot \omega } {\mu_R}^c }  \Big) v_{\rm EW} + H.c. \,, \label{eq:OF_32b} \\[1mm]
&& \frac{ c_3  }{ M_{\rm pl} } \repb{8_F}^{ \dot \omega }  \rep{56_F} \cdot  \repb{28_H}_{\,,\dot \kappa }^\dag \cdot  \rep{70_H}^\dag + H.c. \non
&\supset& \frac{c_3  }{ M_{\rm pl} }  \Big[ ( \repb{4 } \,, \rep{1 } \,, + \frac{1 }{4 } )_{ \rep{F}}^{ \dot \omega } \otimes  ( \rep{6 } \,, \rep{4 } \,, - \frac{1 }{4 } )_{ \rep{F}} \oplus  ( \rep{1 } \,, \repb{4 } \,,  - \frac{1 }{4 } )_{ \rep{F}}^{ \dot \omega } \otimes ( \rep{4 } \,, \rep{6 } \,, + \frac{1 }{4 } )_{ \rep{F}}  \Big] \non
&\otimes& ( \rep{1 } \,, \rep{6 } \,, -\frac{1 }{2 } )_{ \rep{H}\,, \dot \kappa }^\dag \otimes   ( \rep{4 } \,, \repb{4 } \,, +\frac{1 }{2 } )_{ \rep{H} }^\dag + H.c. \non
&\supset&  \frac{c_3  }{ M_{\rm pl} }  \Big[ ( \repb{3 } \,, \rep{1 } \,, + \frac{1 }{3 } )_{ \rep{F}}^{ \dot \omega } \otimes  ( \rep{3 } \,, \rep{4 } \,, - \frac{1 }{12 } )_{ \rep{F}}^\prime \oplus  ( \rep{1 } \,, \repb{4 } \,,  - \frac{1 }{4 } )_{ \rep{F}}^{ \dot \omega } \otimes ( \rep{1 } \,, \rep{6 } \,, + \frac{1 }{2 } )_{ \rep{F}}^\prime \Big] \non
&\otimes& ( \rep{1 } \,, \rep{6 } \,, -\frac{1 }{2 } )_{ \rep{H}\,, \dot \kappa }^\dag \otimes ( \rep{1 } \,, \repb{4 } \,, +\frac{3 }{4 } )_{ \rep{H}}^{\prime\, \dag }  + H.c. \non
&\supset&   \frac{c_3  }{ M_{\rm pl} }  \Big[ ( \repb{3 } \,, \rep{1 } \,, + \frac{1 }{3 } )_{ \rep{F}}^{ \dot \omega } \otimes  ( \rep{3 } \,, \rep{3 } \,, 0 )_{ \rep{F}}^{ \prime \prime} \oplus  ( \rep{1 } \,, \repb{3 } \,,  - \frac{1 }{3 } )_{ \rep{F}}^{ \dot \omega } \otimes ( \rep{1 } \,, \repb{3 } \,, + \frac{2 }{3 } )_{ \rep{F}}^{ \prime \prime } \oplus ( \rep{1 } \,, \rep{1 } \,,  0 )_{ \rep{F}}^{ \dot \omega^{ \prime\prime} } \otimes ( \rep{1 } \,, \rep{3 } \,, + \frac{1 }{3 } )_{ \rep{F}}^\prime   \Big] \non
&\otimes& \langle ( \rep{1 } \,, \repb{3 } \,, -\frac{1 }{3 } )_{ \rep{H}\,, \dot \kappa }^{\prime\,\dag} \rangle \otimes \langle ( \rep{1 } \,, \repb{3 } \,, +\frac{2 }{3 } )_{ \rep{H}}^{\prime \prime\prime\, \dag } \rangle  + H.c. \non
%
%
&\Rightarrow& \frac{ c_3 }{2 }  \dot \zeta_3^\prime   ( \uline{ d_L \Dc_R^{ \dot \omega\, c} } - \uline{ \Ec_L^{\dot \omega } {e_R}^c } + \check \Nc_L^{ \dot \omega^\prime } {\nG_R^{ \prime\prime\prime \prime\prime\, c} }  +  \check \Nc_L^{ \dot \omega^{ \prime\prime} } {\nG_R^{ \prime\prime \prime\prime\, c} } ) v_{\rm EW}  + H.c. \,.  \label{eq:OF_32c}
\eeqn
\eeqs
By taking the possible flavor indices of $\dot \omega = \dot 1\,, \dot 2$ in mass terms from Eqs.~\eqref{eq:OF_32b} and \eqref{eq:OF_32c}, one finds the following set of mass matrices of the $(d\,, s)$ and $(e\,, \mu)$
\beqs
\beqn
&&\Big( \Mc_d \Big)_{2 \times 2}^{\rm direct} = \frac{ c_3 }{2 }  \left( \ba{cc}  
\dot \zeta_3^\prime &  \dot \zeta_3^\prime   \\
 \dot \zeta_3  & \dot \zeta_3  \\  \ea  \right)   v_{\rm EW} \,, \label{eq:ds_direct}\\[1mm]
 && \Big( \Mc_e \Big)_{2 \times 2}^{\rm direct} = - \frac{ c_3 }{2 } \left( \ba{cc}  
\dot \zeta_3^\prime &  \dot \zeta_3   \\
 \dot \zeta_3^\prime  & \dot \zeta_3  \\  \ea  \right)  v_{\rm EW}  \,,\label{eq:emu_direct}
\eeqn
\eeqs
which leave the down quark and electron massless.

\para
For the operator of $\Oc_{\Fc }^{(4\,,1)}$, it is decomposed as
\beqs\label{eqs:OF_41}
\beqn
&& \frac{c_4 }{ M_{\rm pl} } \rep{56_F}  \rep{56_F} \cdot  \repb{28_H}_{\,,\dot \omega } \cdot  \rep{70_H} + H.c. \non
&\supset& \frac{c_4 }{ M_{\rm pl} } \Big[  ( \repb{4 } \,, \rep{1 } \,, -\frac{3 }{4 } )_{ \rep{F}} \otimes  ( \rep{4 } \,, \rep{6 } \,, +\frac{1 }{4 } )_{ \rep{F}}  \oplus  \cancel{ ( \rep{6 } \,, \rep{4 } \,, - \frac{1 }{4 } )_{ \rep{F}}  \otimes ( \rep{6 } \,, \rep{4 } \,, -\frac{1 }{4 } )_{ \rep{F}} }  \Big] \otimes  ( \repb{4 } \,, \repb{4 } \,, 0 )_{ \rep{H} \,, \dot \omega }  \otimes ( \rep{4 } \,, \repb{4 } \,, +\frac{1}{2} )_{ \rep{H} }  + H.c. \non
&\supset& \frac{c_4 }{ M_{\rm pl} }  \Big[    ( \repb{3 } \,, \rep{1 } \,, -\frac{2 }{3 } )_{ \rep{F}}^\prime \otimes ( \rep{3 } \,, \rep{6 } \,, +\frac{1  }{6 } )_{ \rep{F} }     \oplus   ( \rep{1 } \,, \rep{1 } \,, - 1)_{ \rep{F} } \otimes ( \rep{1} \,, \rep{6 } \,, +\frac{1  }{2 } )_{ \rep{F}}^\prime  \oplus \cancel{ ( \rep{3 } \,, \rep{4 } \,, - \frac{1 }{12 } )_{ \rep{F}}^\prime \otimes ( \repb{3 } \,, \rep{4 } \,, -\frac{5 }{12 } )_{ \rep{F}}  }  \Big] \non
&\otimes&  \langle ( \rep{1 } \,, \repb{4 } \,, -\frac{1 }{4 })_{ \rep{H} \,, \dot \omega } \rangle \otimes ( \rep{1 } \,, \repb{4 } \,, +\frac{3 }{4} )_{ \rep{H} }^\prime + H.c.  \non
&\supset&  c_4 \frac{ w_{ \repb{4}\,, \dot {\rm VII} } }{ \sqrt{2} M_{\rm pl} }  \Big[   ( \repb{3 } \,, \rep{1 } \,, -\frac{2 }{3 } )_{ \rep{F}}^\prime \otimes ( \rep{3 } \,, \rep{3 } \,, 0 )_{ \rep{F} }^\prime  \oplus   ( \rep{1 } \,, \rep{1 } \,, - 1)_{ \rep{F} } \otimes ( \rep{1} \,, \rep{3 } \,, +\frac{1 }{3 } )_{ \rep{F}}^{\prime } \Big]  \otimes ( \rep{1 } \,, \repb{3 } \,, +\frac{2 }{3 } )_{ \rep{H} }^{\prime \prime \prime} + H.c.  \non
&\Rightarrow& \frac{ c_4 }{2} \dot \zeta_2   (  \uline{ c_L {u_R}^c  }  + \EG_L {\eG_R^{ \prime\prime \prime\prime} }^c   + \cancel{  u_L {c_R}^c  }  ) v_{\rm EW}  + H.c. \,, \label{eq:OF_41a}\\[1mm]
%
&& \frac{ c_4 }{ M_{\rm pl} } \rep{56_F}  \rep{56_F} \cdot  \repb{28_H}_{\,,\dot \omega } \cdot  \rep{70_H} + H.c. \non
&\supset&  \frac{ c_4 }{ M_{\rm pl} }  \Big[    ( \repb{3 } \,, \rep{1 } \,, -\frac{2 }{3 } )_{ \rep{F}}^\prime \otimes ( \rep{3 } \,, \repb{3 } \,, +\frac{1  }{3 } )_{ \rep{F} }     \oplus   ( \rep{1 } \,, \rep{1 } \,, - 1)_{ \rep{F} } \otimes ( \rep{1} \,, \repb{3 } \,, +\frac{2  }{3 } )_{ \rep{F}}^{ \prime \prime }  \Big]   \non
&\otimes&  \langle ( \rep{1 } \,, \repb{3 } \,, -\frac{1 }{3 })_{ \rep{H} \,, \dot \omega } \rangle \otimes ( \rep{1 } \,, \repb{3 } \,, +\frac{2 }{3 } )_{ \rep{H} }^{ \prime \prime \prime } + H.c.   \Rightarrow   \frac{ c_4}{2 } \dot \zeta_3   ( - \uG_L {u_R}^c  - \EG_L {\eG_R^{ \prime\prime \prime\prime \prime\, c}}  )  v_{\rm EW}  + H.c. \,, \label{eq:OF_41b}\\[1mm]
&& \frac{ c_4 }{ M_{\rm pl} } \rep{56_F}  \rep{56_F} \cdot  \repb{28_H}_{\,,\dot \omega } \cdot  \rep{70_H} + H.c.  \non
&\supset& \frac{c_4 }{ M_{\rm pl} } \Big[   ( \rep{1 } \,,  \repb{4 } \,, +\frac{3 }{4} )_{ \rep{F} } \otimes ( \repb{4 } \,, \rep{1 } \,, - \frac{3 }{4} )_{ \rep{F} }  \oplus  ( \rep{4 } \,,  \rep{6 } \,, +\frac{1 }{4} )_{ \rep{F} } \otimes ( \rep{6 } \,, \rep{4 } \,, - \frac{1 }{4} )_{ \rep{F}}  \Big] \non
&\otimes& ( \rep{1 } \,, \rep{6 } \,, -\frac{1}{2} )_{ \rep{H} \,, \dot \omega }  \otimes ( \rep{4 } \,, \repb{4 } \,, +\frac{1}{2} )_{ \rep{H}} + H.c. \non
&\supset& \frac{ c_4 }{ M_{\rm pl} }  \Big[    ( \rep{1 } \,,  \repb{4 } \,, +\frac{3 }{4} )_{ \rep{F} } \otimes ( \rep{1 } \,, \rep{1 } \,, - 1 )_{ \rep{F} }  \oplus  ( \rep{3 } \,,  \rep{6 } \,, +\frac{1 }{6 } )_{ \rep{F} } \otimes ( \repb{3 } \,, \rep{4 } \,, - \frac{5 }{12 } )_{ \rep{F}}  \Big] \non
&\otimes&  ( \rep{1 } \,, \rep{6 } \,, -\frac{1}{2} )_{ \rep{H} \,, \dot \omega }  \otimes ( \rep{1 } \,, \repb{4 } \,, +\frac{3 }{4 } )_{ \rep{H}}^\prime + H.c. \non
&\supset& \frac{ c_4 }{ M_{\rm pl} }  \Big[   ( \rep{1 } \,,  \repb{3 } \,, +\frac{2 }{3 } )_{ \rep{F} }^\prime \otimes ( \rep{1 } \,, \rep{1 } \,, - 1 )_{ \rep{F} } \oplus   ( \rep{3 } \,,  \rep{3 } \,, 0 )_{ \rep{F}}^\prime \otimes ( \repb{3 } \,, \rep{3 } \,, - \frac{1 }{3 } )_{ \rep{F}} \oplus  ( \rep{3 } \,,  \repb{3 } \,, +\frac{1 }{3 } )_{ \rep{F} } \otimes ( \repb{3 } \,, \rep{1 } \,, - \frac{2 }{3 } )_{ \rep{F}}^{ \prime\prime \prime }  \Big]  \non
&\otimes& \langle ( \rep{1 } \,, \repb{3 } \,, -\frac{1}{3 } )_{ \rep{H} \,, \dot \omega }^\prime  \rangle \otimes ( \rep{1 } \,, \repb{3 } \,, +\frac{2 }{3 } )_{ \rep{H}}^{\prime \prime \prime } + H.c. \non
&\Rightarrow& \frac{ c_4 }{ 2 } \dot \zeta_3^\prime  (  - \EG_L {\eG_R^{ \prime\prime\prime\, c}} + c_L {\UG_R}^c - \DG_L^{ \prime\prime\prime } {\dG_R}^c - \uG_L {c_R}^c  ) v_{\rm EW}  + H.c.  \,.\label{eq:OF_41c}
\eeqn
\eeqs

\para
We cross out the bi-linear fermion product of $( \rep{6 } \,, \rep{4 } \,, -\frac{1 }{4 } )_{ \rep{F}} \otimes ( \rep{6 } \,, \rep{4 } \,, -\frac{1 }{4 } )_{ \rep{F}} \otimes ( \repb{4 } \,, \repb{4 } \,, 0 )_{ \rep{H} \,, \dot \omega }  \otimes ( \rep{4 } \,, \repb{4 } \,, +\frac{1}{2} )_{ \rep{H} } + H.c.$, since it vanishes automatically.
To see this, let us denote relevant fields in terms of their components as $\Psi_\alpha^{ \[ \bar a \, \bar b \] \,, \bar i } \equiv ( \rep{6 } \,, \rep{4 } \,, -\frac{1 }{4 } )_{ \rep{F}}$, $\Phi_{ \bar e \,, \bar k \,, \dot \omega } \equiv ( \repb{4 } \,, \repb{4 } \,, 0 )_{ \rep{H} \,, \dot \omega }$, and ${\Phi^{\prime\, \bar e}}_{  \bar l }\equiv ( \rep{4 } \,, \repb{4 } \,, +\frac{1}{2} )_{ \rep{H} }$.
Here, all gauge group indices follow from the conventions defined in Tab.~\ref{tab:notations}, and $(\alpha\,,\beta)=1\,,2$ are the two-component Weyl spinor indices.
This gauge-invariant Yukawa coupling term can be explicitly expressed as follows
\beqn
&& \epsilon^{\alpha\beta} \epsilon_{ \bar a \bar b \bar c \bar d } { \epsilon_{ \bar i \bar j  } }^{ \bar k \bar l } \,  \Psi_\alpha^{ \[ \bar a \, \bar b \] \,, \bar i }  \Psi_\beta^{ \[ \bar c\, \bar d \] \,, \bar j } \Phi_{ \bar e \,, \bar k \,, \dot \omega }  {\Phi^{\prime\, \bar e}}_{  \bar l }  + H.c. \non
&=& \epsilon^{\beta \alpha}  \epsilon_{ \bar c \bar d \bar a \bar b }  { \epsilon_{ \bar j \bar i  } }^{ \bar k \bar l } \,  \Psi_\beta^{ \[ \bar c \, \bar d \] \,, \bar j }  \Psi_\alpha^{ \[ \bar a\, \bar b \] \,, \bar i } \Phi_{ \bar e \,, \bar k \,, \dot \omega }  {\Phi^{\prime\, \bar e}}_{  \bar l } + H.c. \non
&=& + \epsilon^{ \alpha\beta}  \epsilon_{ \bar a \bar b \bar c \bar d  } { \epsilon_{ \bar i \bar j  } }^{ \bar k \bar l } \,  \Psi_\beta^{ \[ \bar c \, \bar d \] \,, \bar j }  \Psi_\alpha^{ \[ \bar a\, \bar b \] \,, \bar i }  \Phi_{ \bar e \,, \bar k \,, \dot \omega }  {\Phi^{\prime\, \bar e}}_{  \bar l } + H.c. \non
&=& - \epsilon^{ \alpha\beta}  \epsilon_{ \bar a \bar b \bar c \bar d  }  { \epsilon_{ \bar i \bar j  } }^{ \bar k \bar l } \,  \Psi_\alpha^{ \[ \bar a\, \bar b \] \,, \bar i } \Psi_\beta^{ \[ \bar c \, \bar d \] \,, \bar j }   \Phi_{ \bar e \,, \bar k \,, \dot \omega }  {\Phi^{\prime\, \bar e}}_{  \bar l }  + H.c. \non
&\Rightarrow&  ( \rep{6}\,, \rep{4 }\,, -\frac{1}{4} )_{ \mathbf{F}}  \otimes  ( \rep{6}\,, \rep{4 }\,, -\frac{1}{4} )_{ \mathbf{F}} \otimes ( \repb{4 } \,, \repb{4 } \,, 0 )_{ \rep{H} \,, \dot \omega }  \otimes ( \rep{4 } \,, \repb{4 } \,, +\frac{1}{2} )_{ \rep{H} } + H.c. = 0 \,,
\eeqn
where we have swapped all gauge and spinor indices of two Weyl fermions in the second line.
Correspondingly, we find this operator of $\Oc_{\Fc }^{(4\,,1)}$ can only contribute to the $c_L {u_R}^c+H.c.$ mass term but not the $u_L {c_R}^c+H.c.$ mass term.

\para
For the operator of $\Oc_{\Fc }^{(4\,,2)}$, it is decomposed as
\beqs\label{eqs:OF_42}
\beqn
&& \frac{ c_4 }{ M_{\rm pl} } \rep{56_F}  \rep{56_F} \cdot  \repb{8_H}_{\,, \omega }^\dag \cdot  \repb{8_H}_{\,, \kappa }^\dag \non
&\supset& \frac{ c_4 }{ M_{\rm pl} }  \[  ( \rep{1 } \,, \repb{4 } \,,  + \frac{3}{4} )_{ \rep{F}} \otimes ( \repb{4 } \,, \rep{1 } \,, - \frac{3}{4} )_{ \rep{F}}  \oplus  ( \rep{4 } \,, \rep{6 } \,,  + \frac{1 }{4} )_{ \rep{F}} \otimes ( \rep{6 } \,, \rep{4 } \,, - \frac{1 }{4} )_{ \rep{F}} \]  \non
&\otimes&  \langle ( \repb{4 } \,, \rep{1 } \,, +\frac{1}{4} )_{ \rep{H}\,,\omega }^\dag  \rangle \otimes (  \rep{1 } \,, \repb{4 } \,, -\frac{1}{4} )_{ \rep{H}\,,\kappa }^\dag  + H.c. \non
&\supset& c_4  \frac{ W_{ \repb{4}\,, {\rm IV} } }{ \sqrt{2} M_{\rm pl} }  \[    ( \rep{1 } \,, \repb{4 } \,,  + \frac{3}{4} )_{ \rep{F}} \otimes ( \rep{1 } \,, \rep{1 } \,, - 1  )_{ \rep{F}}  \oplus  ( \rep{3 } \,, \rep{6 } \,,  + \frac{1 }{6} )_{ \rep{F}} \otimes ( \repb{3 } \,, \rep{4 } \,, - \frac{5 }{12 } )_{ \rep{F}} \]  \otimes\langle (  \rep{1 } \,, \repb{4 } \,, -\frac{1}{4} )_{ \rep{H}\,,\kappa }^\dag \rangle + H.c. \non
&\Rightarrow& \frac{c_4}{2 } \zeta_1 \Big[  (  \EG_L {\EG_R}^c  + \dG_L {\dG_R}^c - \uG_L {\uG_R}^c + \UG_L {\UG_R}^c  )  w_{ \repb{4}\,, {\rm V} } \non
&+&   (  \EG_L {\mu_R}^c +  c_L {\uG_R}^c - s_L {\dG_R}^c + \UG_L {c_R}^c  )  V_{ \repb{3}\,, 3\,, {\rm VI} } \Big] + H.c. \,,\label{eq:OF_42a}\\[1mm]
&& \frac{ c_4 }{ M_{\rm pl} } \rep{56_F}  \rep{56_F} \cdot   \repb{8_H}_{\,, \omega }^\dag \cdot  \repb{8_H}_{\,, \kappa }^\dag + H.c. \non
&\supset&  \frac{c_4 }{ M_{\rm pl} }  \[  ( \repb{3 } \,, \rep{1 } \,,  -\frac{2 }{3} )_{ \rep{F} }^\prime \otimes ( \rep{3 } \,, \rep{6 } \,,  + \frac{1 }{6} )_{ \rep{F} } \oplus  ( \rep{1 } \,, \rep{1 } \,,  -1 )_{ \rep{F} } \otimes ( \rep{1 } \,, \rep{6 } \,,  + \frac{1 }{2} )_{ \rep{F} }^\prime  \] \non
&\otimes& \langle (  \rep{1 } \,, \repb{4 } \,, -\frac{1}{4} )_{ \rep{H}\,,\omega }^\dag  \rangle \otimes (  \rep{1 } \,, \repb{4 } \,, -\frac{1}{4} )_{ \rep{H}\,,\kappa }^\dag + H.c. \non
&\supset&  c_4 \frac{ w_{ \repb{4}\,, {\rm V} } }{ \sqrt{2} M_{\rm pl} } \[   ( \repb{3 } \,, \rep{1 } \,,  -\frac{2 }{3} )_{ \rep{F} }^\prime \otimes ( \rep{3 } \,, \repb{3 } \,,  + \frac{1 }{3 } )_{ \rep{F} } \oplus  ( \rep{1 } \,, \rep{1 } \,,  -1 )_{ \rep{F} } \otimes ( \rep{1 } \,, \repb{3 } \,,  + \frac{2 }{3 } )_{ \rep{F} }^{\prime \prime }  \] \otimes \langle (  \rep{1 } \,, \repb{3 } \,, -\frac{1}{3 } )_{ \rep{H}\,,\kappa }^\dag \rangle + H.c.  \non
&\Rightarrow& \frac{ c_4}{2 } \zeta_2  ( \UG_L {u_R}^c + \EG_L {e_R}^c )  V_{ \repb{3}\,, 3\,, {\rm VI} } + H.c. \,.\label{eq:OF_42b}
\eeqn
\eeqs

\para
For the operator of $\Oc_{\Fc }^{(5\,,1)}$, it is decomposed as
\beqs\label{eqs:OF_51}
\beqn
&& \frac{ c_5 }{ M_{\rm pl} } \rep{28_F}  \rep{56_F} \cdot  \repb{8_H}_{\,, \omega }  \cdot   \rep{70_H} + H.c. \non
&\supset& \frac{c_5 }{ M_{\rm pl} } \Big[ ( \rep{6 } \,, \rep{1 } \,,  -\frac{1 }{2 } )_{ \rep{F} } \otimes ( \rep{6 } \,, \rep{4 } \,,  -\frac{1 }{4 } )_{ \rep{F} }  \oplus ( \rep{4 } \,, \rep{4 } \,,  0 )_{ \rep{F} } \otimes ( \repb{4 } \,, \rep{1 } \,,  -\frac{3 }{4 } )_{ \rep{F} }   \Big] \non
&\otimes& \langle ( \repb{4 } \,, \rep{1 } \,,  +\frac{1 }{4 } )_{ \rep{H}\,,\omega } \rangle \otimes   ( \rep{4 } \,, \repb{4 } \,,  +\frac{1 }{2 } )_{ \rep{H} } + H.c. \non
&\supset& c_5 \frac{W_{ \repb{4} \,, {\rm IV} } }{ \sqrt{2} M_{\rm pl} }  \Big[ ( \rep{3 } \,, \rep{1 } \,,  -\frac{1 }{3 } )_{ \rep{F} } \otimes ( \repb{3 } \,, \rep{4 } \,,  -\frac{5 }{12 } )_{ \rep{F} }  \oplus   ( \repb{3 } \,, \rep{1 } \,,  -\frac{2 }{3 } )_{ \rep{F} } \otimes ( \rep{3 } \,, \rep{4 } \,,  -\frac{1 }{12 } )_{ \rep{F}}^\prime \non
&\oplus& ( \rep{3 } \,, \rep{4 } \,,  -\frac{1 }{12 } )_{ \rep{F}} \otimes ( \repb{3 } \,, \rep{1 } \,,  -\frac{2 }{3 } )_{ \rep{F}}^\prime \oplus ( \rep{1 } \,, \rep{4 } \,,  +\frac{1 }{4 } )_{ \rep{F}} \otimes ( \rep{1 } \,, \rep{1 } \,,  - 1 )_{ \rep{F}}  \Big]  \otimes  ( \rep{1 } \,, \repb{4 } \,,  +\frac{3 }{4 } )_{ \rep{H} }^\prime + H.c. \non
&\Rightarrow& \frac{ c_5 }{2 } \zeta_1  (  \DG_L {\dG_R}^c  + \uline{ u_L {t_R}^c }  + \uline{ t_L {u_R}^c } + \EG_L { \eG_R^{\prime\prime\, c}}  )  v_{\rm EW} + H.c.  \,,\label{eq:OF_51a}\\[1mm]
&& \frac{ c_5 }{ M_{\rm pl} } \rep{28_F}  \rep{56_F} \cdot  \repb{8_H}_{\,, \omega }  \cdot   \rep{70_H} + H.c. \non
&\supset& \frac{ c_5 }{ M_{\rm pl} }  \Big[  ( \rep{6 } \,, \rep{1 } \,,  -\frac{1 }{2 } )_{ \rep{F} } \otimes ( \rep{4 } \,, \rep{6 } \,,  + \frac{1 }{4 } )_{ \rep{F} }  \oplus  ( \rep{1 } \,, \rep{6 } \,,  +\frac{1 }{2 } )_{ \rep{F} } \otimes  ( \repb{4 } \,, \rep{1 } \,,  -\frac{3 }{4 } )_{ \rep{F} } \oplus  ( \rep{4 } \,, \rep{4 } \,,  0 )_{ \rep{F} } \otimes ( \rep{6 } \,, \rep{4 } \,,  -\frac{1 }{4 } )_{ \rep{F} }  \Big] \non
&\otimes&  ( \rep{1 } \,, \repb{4 } \,,  - \frac{1 }{4 } )_{ \rep{H}\,,\omega } \otimes  ( \rep{4 } \,, \repb{4 } \,,  +\frac{1 }{2 } )_{ \rep{H} } + H.c. \non
&\supset& \frac{c_5 }{ M_{\rm pl} } \Big[  ( \repb{3 } \,, \rep{1 } \,,  -\frac{2 }{3 } )_{ \rep{F} } \otimes ( \rep{3 } \,, \rep{6 } \,,  + \frac{1 }{6 } )_{ \rep{F} }    \oplus  ( \rep{1 } \,, \rep{6 } \,,  +\frac{1 }{2 } )_{ \rep{F} } \otimes  ( \rep{1 } \,, \rep{1 } \,,  - 1 )_{ \rep{F} }  \oplus   ( \rep{3 } \,, \rep{4 } \,, - \frac{1 }{12 } )_{ \rep{F} } \otimes ( \repb{3 } \,, \rep{4 } \,,  -\frac{5 }{12 } )_{ \rep{F} }   \Big] \non
&\otimes& \langle ( \rep{1 } \,, \repb{4 } \,,  - \frac{1 }{4 } )_{ \rep{H}\,,\omega } \rangle \otimes  ( \rep{1 } \,, \repb{4 } \,,  +\frac{3 }{4 } )_{ \rep{H} }^\prime + H.c. \non
&\supset& c_5 \frac{ w_{ \repb{4}\,, {\rm V} } }{ \sqrt{2} M_{\rm pl} }  \Big[  ( \repb{3 } \,, \rep{1 } \,,  -\frac{2 }{3 } )_{ \rep{F} }   \otimes ( \rep{3 } \,, \rep{3 } \,, 0 )_{ \rep{F} }^\prime  \oplus  ( \rep{1 } \,, \rep{3 } \,, +\frac{1 }{3 } )_{ \rep{F} }  \otimes ( \rep{1 } \,, \rep{1 } \,,  - 1 )_{ \rep{F} }  \non
&\oplus&  ( \rep{3 } \,, \rep{3 } \,, 0 )_{ \rep{F} } \otimes ( \repb{3 } \,, \rep{1 } \,,  -\frac{2 }{3 } )_{ \rep{F}}^{ \prime\prime \prime }  \oplus  ( \rep{3 } \,, \rep{1 } \,, - \frac{1 }{3 } )_{ \rep{F}}^{ \prime\prime } \otimes ( \repb{3 } \,, \rep{3 } \,,  -\frac{1 }{3 } )_{ \rep{F} }  \Big]  \otimes   ( \rep{1 } \,, \repb{3 } \,,  +\frac{2 }{3 } )_{ \rep{H} }^{ \prime \prime \prime } + H.c. \non
&\Rightarrow& \frac{c_5}{2 } \zeta_2  (  \uline{ c_L {t_R}^c  } +   \EG_L {\eG_R}^c + \uline{ t_L {c_R}^c } + \DG_L^{ \prime\prime }  {\dG_R}^c) v_{\rm EW}   + H.c.  \,,\label{eq:OF_51b}\\[1mm]
&& \frac{ c_5 }{ M_{\rm pl} }  \rep{28_F}  \rep{56_F} \cdot  \repb{8_H}_{\,, \omega }  \cdot   \rep{70_H} + H.c. \non
&\supset& \frac{c_5 }{ M_{\rm pl} } \Big[  ( \repb{3 } \,, \rep{1 } \,,  -\frac{2 }{3 } )_{ \rep{F} } \otimes ( \rep{3 } \,, \repb{3 } \,,  + \frac{1 }{3 } )_{ \rep{F} }    \oplus  ( \rep{1 } \,, \repb{3 } \,,  +\frac{2 }{3 } )_{ \rep{F} } \otimes  ( \rep{1 } \,, \rep{1 } \,,  - 1 )_{ \rep{F} }  \oplus   ( \rep{3 } \,, \rep{3 } \,, 0 )_{ \rep{F} } \otimes ( \repb{3 } \,, \rep{3 } \,,  -\frac{1 }{3 } )_{ \rep{F} }  \Big] \non
&\otimes& \langle ( \rep{1 } \,, \repb{3 } \,,  - \frac{1 }{3 } )_{ \rep{H}\,,\omega } \rangle \otimes  ( \rep{1 } \,, \repb{3 } \,,  +\frac{2 }{3 } )_{ \rep{H} }^{ \prime \prime \prime } + H.c. \non
&\Rightarrow& \frac{c_5}{2 } \zeta_3  ( - \uG_L {t_R}^c - \EG_L {\eG_R^{ \prime\, c}} - t_L {\UG_R}^c + \DG_L^\prime {\dG_R}^c ) v_{\rm EW}  + H.c.  \,.\label{eq:OF_51c}
\eeqn
\eeqs

\para
By inspecting all $d=5$ direct Yukawa couplings, we can only find the mass mixing terms of SM up-type quarks in Eqs.~\eqref{eq:OF_41a}, \eqref{eq:OF_51a}, and~\eqref{eq:OF_51c} due to the EWSB VEV from the SM Higgs doublet of $( \rep{1 } \,, \repb{2 } \,,  +\frac{1 }{2 } )_{ \rep{H} }^{ \prime \prime \prime }\subset \rep{70_H}$, while the mass (mixing) terms of all SM down-type quarks or charged leptons were absent.
Obviously, these direct Yukawa coupling terms alone are not sufficient to describe the current LHC measurements of the Yukawa couplings between the SM Higgs boson and the $(b\,,\tau\,,\mu)$~\cite{CMS:2022dwd,ATLAS:2022vkf}.

\subsection{The indirect Yukawa couplings from the non-renormalizable Higgs mixing operators}

\begin{figure}
\begin{center}
\begin{tikzpicture}[scale=1.2]
\coordinate[] (S1);
    \coordinate [above left=1.8 and 0.4 of S1 ] (T1);
    \coordinate [above right=1.8 and 0.4 of S1 ] (T2);
    \coordinate [above left=0.6 and 1.2 of S1 ] (T3);
    \coordinate [above right=0.6 and 1.2 of S1 ] (T4);
    \draw[scalar] (T1)--node[above left=0.9 and 0.1]{$\langle\Phi_2 \rangle$} (S1);
    \draw[scalar] (T2)--node[above right=0.9 and 0.1]{$\langle\Phi_3 \rangle$} (S1);   
    \draw[scalar] (T3)--node[above left=0.2 and 0.4]{$\langle\Phi_1 \rangle$} (S1);
    \draw[scalar] (T4)--node[above right=0.2 and 0.4]{$\langle\Phi_4 \rangle$} (S1);
    \coordinate [ below=1.0 of S1 ] (S2);
    \draw[scalar] (S2)--node[ right=0.2]{$\Phi^\prime$} (S1);
    \coordinate [below left=1.0 and 1.2 of S2 ] (R1);
    \coordinate [below right=1.0 and 1.2 of S2 ] (R2);
    \draw[fermion] (R1)--node[below left=0.5]{$\Fc_{L}$} (S2);
    \draw[fermion] (R2)--node[below right=0.5]{$\Fc_{R}^{\prime\, c}$} (S2);
\end{tikzpicture}
\end{center}
\caption{The indirect Yukawa couplings through the VEV insertions to the $d=5$ Higgs mixing operators.}\label{fig:indirectY_treed5}
\end{figure}
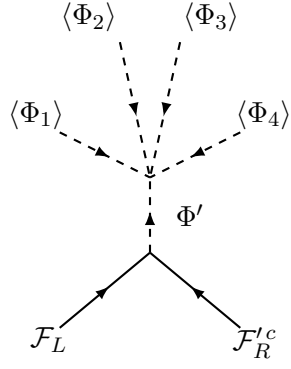

\para
Next, we list the indirect Yukawa couplings by coupling the renormalizable Yukawa coupling terms in Eq.~\eqref{eq:Yukawa_SU8} with a set of $d$-dimensional ($d\geq 5$) non-renormalizable irreducible Higgs mixing operators of $\{ \Oc_{\Hc}^{d \geq 5} \}$.
A non-renormalizable irreducible Higgs mixing operator is defined as
\begin{define}\label{def:Hmixing}

a non-renormalizable irreducible Higgs mixing operator of $\Oc_{\Hc}^{d \geq 5}$ should
\begin{itemize}

\item break the global symmetries explicitly;

\item not be further partitioned into subset of renormalizable operators, among which any of them can be allowed by both the gauge and the global symmetries.

\end{itemize}

\end{define}
In other words, we expect the $\Oc_{\Hc}^{d \geq 5}$ to break the global symmetries in Eqs.~\eqref{eq:DRS_SU8} and/or \eqref{eq:PQ_SU8} explicitly as a whole.
Schematically, we depict the indirect Yukawa couplings in Fig.~\ref{fig:indirectY_treed5} and denote the corresponding fermion mass (mixing) terms as follows
\beqn\label{eq:indirectYukawa}
&&  \Oc_{\Hc}^{d \geq 5} \equiv \Phi^{\prime\,\dag} \prod_{i=1}^{d-1}  \Phi_i \,, \quad \Fc_L \Fc_R^{\prime\, c}  \wick[offset=1.5em]{ \c \Phi^\prime \times  \frac{1 }{ M_{\rm pl}^{d-4} }  \c \Phi^{\prime\,\dag} } \prod_{i=1}^{d-1} \langle \Phi_i  \rangle + H.c. \non
&\Rightarrow& m_{\Fc_L \Fc_R^{\prime \, c}} = \frac{1}{ M_{\rm pl} }  \frac{1}{ m_{ \Phi^\prime }^2  } \prod_{i=1}^{d-1 } v_i \,,
\eeqn
where $m_{ \Phi^\prime }^2$ is the propagator mass squared.
All $d$ components of such operators can develop VEVs at various stages, where one specific component of $\Phi^{\prime\,\dag}$ acts as propagator (labeled with contraction symbol) with the $\Phi^\prime$ field in the renormalizable Yukawa coupling terms.
Below, we analyze the indirect Yukawa couplings according to two of the renormalizable Yukawa coupling terms in Eq.~\eqref{eq:Yukawa_SU8}
\beqn
&& Y_\Bc \repb{8_F}^\omega \rep{28_F}  \repb{8_{H}}_{\,,\omega } + Y_\Dc  \repb{8_F}^{\dot \omega } \rep{56_F}  \repb{28_{H}}_{\,,\dot \omega } + H.c. \,,
\eeqn
which will contribute to mass terms of all three-generational SM down-type quarks and charged leptons.

\para
Two following $d=5$ irreducible Higgs mixing operators with $\Tc=0$ will be considered
\beqs\label{eqs:d5_Hmixings}
\beqn
d_{\mathscr A}\, \Oc_{\mathscr A}^{d=5} &\equiv& d_{\mathscr A}\,  \epsilon_{ \omega_1 \omega_2 \omega_3  \omega_4 } \repb{8_{H}}_{\,, \omega_1}^\dag   \repb{8_{H}}_{\,, \omega_2}^\dag \repb{8_{H}}_{\,, \omega_3}^\dag  \repb{8_{H}}_{\,, \omega_4 }^\dag  \rep{70_H}^\dag  \,, \quad \Pc\Qc = 2 ( 2p + 3 q_2 ) \neq 0 \,, \label{eq:d5_Hmixing_A}\\[1mm]
d_{\mathscr B}\, \Oc_{\mathscr B}^{d=5} &\equiv&d_{\mathscr B}\,   ( \repb{28_H}_{\,,\dot \kappa_1 }^\dag \repb{28_H}_{\,,\dot \kappa_2 } ) \cdot  \repb{28_H}_{\,,\dot \omega_1}^\dag \repb{28_H}_{\,,\dot \omega_2}^\dag   \rep{70_H}^\dag  \,, \quad  \Pc\Qc =  2 ( p + q_2 + q_3) \,, \non
&& {\rm with} ~~ \dot \kappa_2 \neq ( \dot \kappa_1 \,, \dot \omega_1 \,, \dot \omega_2 )   \,.\label{eq:d5_Hmixing_B}
\eeqn
\eeqs
We expect both operators to be suppressed by $1/M_{\rm pl}$.
The global anomalous $\widetilde{\rm U}(1)_{\rm PQ}$ charges are listed as well.
According to Eq.~\eqref{eq:PQcharges_SU8}, the operator of $\Oc_{\mathscr A}^{d=5}$ is explicitly $\widetilde{\rm U}(1)_{\rm PQ}$-breaking, while it is manifestly $\widetilde {\rm SU}(4)_{\omega}$-preserving.
The operator of $\Oc_{\mathscr B}^{d=5}$ is explicitly $\widetilde{\rm SU}(5)_{\dot \omega}$-breaking, and may or may not be $\widetilde{\rm U}(1)_{\rm PQ}$-breaking depending on the specific $\widetilde{\rm U}(1)_{\rm PQ}$ charge assignments.
The restriction to the flavor index of $\dot \kappa_2 \neq ( \dot \kappa_1 \,, \dot \omega_1 \,, \dot \omega_2 )$ guarantees the irreducibility of the $\Oc_{\mathscr B}^{d=5}$, which would otherwise lead to a reducible operator of $| \repb{28_H}_{\,, \dot \kappa} |^2 \repb{28_H}_{\,,\dot \omega_1}^\dag \repb{28_H}_{\,,\dot \omega_2}^\dag   \rep{70_H}^\dag$.
The irreducibility in Definition \ref{def:Hmixing} further forbids operators such as $| \repb{8_H}_{\,,  \kappa} |^2 \repb{28_H}_{\,,\dot \omega_1}^\dag \repb{28_H}_{\,,\dot \omega_2}^\dag   \rep{70_H}^\dag$ and $\Tr( \rep{63_H}^2 )  \repb{28_H}_{\,,\dot \omega_1}^\dag \repb{28_H}_{\,,\dot \omega_2}^\dag   \rep{70_H}^\dag$, where both $| \repb{8_H}_{\,,  \kappa} |^2$ and $\Tr( \rep{63_H}^2 )$ are renormalizable operators allowed in the ${\rm SU}(8)$ theory.
Phenomenologically, if these subsets of operators developed their largest VEVs of $\langle | \repb{8_H}_{\,,  \kappa} |^2 \rangle \sim \Oc( v_{441}^2 )$ and $\langle \Tr( \rep{63_H}^2 ) \rangle \sim \Oc(v_U^2)$ based on the VEVs in Eqs.~\eqref{eq:63H_VEV} and \eqref{eq:SU8_Higgs_VEVs_mini01}, one would find significant enhancements to the first- and second-generational down-type quarks and charged leptons according to our analyses in Eqs.~\eqref{eq:smu_indirect} and \eqref{eq:de_indirect} below.
Manifestly, the operator of $\Oc_{\mathscr B}^{d=5}$ include two leading terms of
\beqn
 \Oc_{\mathscr B}^{d=5} &\supset& \Large( \repb{28_H}_{\,,\dot 1 }^\dag \repb{28_H}_{\,,\dot {\rm VII} }  \Large) \cdot \repb{28_H}_{\,,\dot {\rm VIII} }^\dag \repb{28_H}_{\,,\dot 1 }^\dag   \rep{70_H}^\dag \,, \non
&&  \Large( \repb{28_H}_{\,,\dot 1 }^\dag \repb{28_H}_{\,,\dot {\rm VII} }  \Large) \cdot  \repb{28_H}_{\,,\dot {\rm IX} }^\dag \repb{28_H}_{\,,\dot 2 }^\dag  \rep{70_H}^\dag  \,,
\eeqn
that will form gauge-invariant VEVs according to Eqs.~\eqref{eqs:SU8_Higgs_VEVs_mini} and lead to indirect Yukawa coupling terms.
Furthermore, one may also include the operator of
\beqn
&& ( \repb{8_H}_{\,, \kappa_1 }^\dag \repb{8_H}_{\,, \kappa_2 } ) \cdot \repb{28_H}_{\,,\dot \omega_1}^\dag \repb{28_H}_{\,,\dot \omega_2}^\dag   \rep{70_H}^\dag \,.
\eeqn
According to the VEV assignments in Eq.~\eqref{eq:SU8_Higgs_VEVs_mini03}, this subset of $( \repb{8_H}_{\,, \kappa_1 }^\dag \repb{8_H}_{\,, \kappa_2 } )$ can at most contribute to VEVs of $\langle \repb{8_H}_{\,, 3 }^\dag \repb{8_H}_{\,, {\rm VI} }  \rangle= \hf V_{ \repb{3} \,, 3} V_{ \repb{3} \,, {\rm VI} }$.
Hence its contribution is sub-leading from the operator $\Oc_{\mathscr B}^{d=5}$, and can be safely neglected.

\para
For the Yukawa coupling of $\repb{8_F}^{\omega_1} \rep{28_F}  \repb{8_{H}}_{\,, \omega_1 }$, we find the mass terms of
\beqn\label{eq:btau_indirect}
&& Y_\Bc \repb{8_F}^{\omega_1} \rep{28_F}  \repb{8_{H}}_{\,, \omega_1}  \times \frac{ d_{ \mathscr A} }{ M_{\rm pl} }  \epsilon_{ \omega_1 \omega_2 \omega_3  \omega_4 } \repb{8_{H}}_{\,, \omega_1}^\dag   \repb{8_{H}}_{\,, \omega_2}^\dag \repb{8_{H}}_{\,, \omega_3}^\dag  \repb{8_{H}}_{\,, \omega_4 }^\dag  \rep{70_{H}}^\dag + H.c. \non
&\supset& Y_\Bc \[  ( \repb{4 } \,, \rep{1 } \,,  + \frac{1 }{4 } )_{ \rep{F}}^{\omega_1 }  \otimes  ( \rep{4 } \,, \rep{4 } \,,  0 )_{ \rep{F}} \oplus ( \rep{1 } \,, \repb{4 } \,,  - \frac{1 }{4 } )_{ \rep{F}}^{\omega_1}  \otimes ( \rep{1 } \,, \rep{6 } \,,  + \frac{1 }{2 } )_{ \rep{F}} \] \otimes ( \rep{1 } \,, \repb{4 } \,,  - \frac{1 }{4 } )_{ \rep{H}\,,\omega_1 }  \non
&\times&   \frac{  d_{ \mathscr A} }{ M_{\rm pl} }  ( \rep{1 } \,, \repb{4 } \,,  - \frac{1 }{4 } )_{ \rep{H}\,,\omega_1 }^\dag  \otimes ( \rep{1 } \,, \repb{4 } \,,  - \frac{1 }{4 } )_{ \rep{H}\,,\omega_2 }^\dag \otimes ( \rep{1 } \,, \repb{4 } \,,  - \frac{1 }{4 } )_{ \rep{H}\,,\omega_3 }^\dag \otimes \langle ( \repb{4 } \,, \rep{1 } \,,  + \frac{1 }{4 } )_{ \rep{H}\,,\omega_4 }^\dag \rangle \otimes ( \rep{4 } \,, \repb{4 } \,,  + \frac{1 }{2 } )_{ \rep{H} }^\dag + H.c. \non
&\supset& Y_\Bc \Big[  ( \repb{3 } \,, \rep{1 } \,,  + \frac{1 }{3 } )_{ \rep{F}}^{\omega_1}  \otimes  ( \rep{3 } \,, \rep{4 } \,,  -\frac{1 }{12} )_{ \rep{F}} \oplus ( \rep{1 } \,, \rep{1 } \,,  0 )_{ \rep{F}}^{\omega_1 } \otimes  ( \rep{1 } \,, \rep{4 } \,,  +\frac{1 }{4 } )_{ \rep{F}}  \non
&\oplus& ( \rep{1 } \,, \repb{4 } \,,  - \frac{1 }{4 } )_{ \rep{F}}^{\omega_1 } \otimes ( \rep{1 } \,, \rep{6 } \,,  + \frac{1 }{2 } )_{ \rep{F}} \Big] \otimes ( \rep{1 } \,, \repb{4 } \,,  - \frac{1 }{4 } )_{ \rep{H}\,,\omega_1 }  \times  d_{ \mathscr A} \frac{ W_{ \repb{4}\,,{\rm IV} } }{ \sqrt{2} M_{\rm pl} } ( \rep{1 } \,, \repb{4 } \,,  - \frac{1 }{4 } )_{ \rep{H}\,,\omega_1 }^\dag \non
&\otimes& \langle ( \rep{1 } \,, \repb{4 } \,,  - \frac{1 }{4 } )_{ \rep{H}\,,\omega_2 }^\dag \rangle  \otimes ( \rep{1 } \,, \repb{4 } \,,  - \frac{1 }{4 } )_{ \rep{H}\,,\omega_3 }^\dag \otimes  ( \rep{1 } \,, \repb{4 } \,,  + \frac{3 }{4 } )_{ \rep{H} }^{\prime\,\dag}  + H.c. \non
&\supset& Y_\Bc \Big[  ( \repb{3 } \,, \rep{1 } \,,  + \frac{1 }{3 } )_{ \rep{F}}^{\omega_1}  \otimes  ( \rep{3 } \,, \rep{3 } \,,  0 )_{ \rep{F}} \oplus ( \rep{1 } \,, \rep{1 } \,,  0 )_{ \rep{F}}^{\omega_1 } \otimes  ( \rep{1 } \,, \rep{3 } \,,  +\frac{1 }{3 } )_{ \rep{F}}^{ \prime\prime }  \oplus ( \rep{1 } \,, \repb{3 } \,,  - \frac{1 }{3 } )_{ \rep{F}}^{\omega_1 } \otimes ( \rep{1 } \,, \repb{3 } \,,  + \frac{2 }{3 } )_{ \rep{F}} \non
&\oplus& ( \rep{1 } \,, \rep{1 } \,,  0 )_{ \rep{F}}^{\omega_1^{ \prime\prime} } \otimes ( \rep{1 } \,, \rep{3 } \,,  + \frac{1 }{3 } )_{ \rep{F}}  \Big] \otimes ( \rep{1 } \,, \repb{3 } \,,  - \frac{1 }{3 } )_{ \rep{H}\,,\omega_1 } \times  d_{ \mathscr A} \frac{ W_{ \repb{4}\,,{\rm IV} } w_{ \repb{4}\,,{\rm V} }  }{ 2 M_{\rm pl} }  ( \rep{1 } \,, \repb{3 } \,,  - \frac{1 }{3 } )_{ \rep{H}\,,\omega_1 }^\dag \non
&\otimes& \langle ( \rep{1 } \,, \repb{3 } \,,  - \frac{1 }{3 } )_{ \rep{H}\,,\omega_3 }^\dag \rangle \otimes \langle ( \rep{1 } \,, \repb{3 } \,,  + \frac{2 }{3 } )_{ \rep{H} }^{\prime \prime\prime \,\dag} \rangle  + H.c.  \non
&\Rightarrow& \frac{ Y_\Bc   d_{ \mathscr A}}{ 4 } \frac{ W_{ \repb{4}\,,{\rm IV} }  w_{ \repb{4}\,,{\rm V} } V_{ \repb{3}\,,{\rm VI} } }{ M_{\rm pl}  m_{( \rep{1}\,, \repb{3 }\,, - \frac{1}{3 })_{ \rep{H}\,,3} }^2}  ( \uline{ b_L {b_R}^c  } +  \uline{ \tau_L {\tau_R}^c } + \check \Nc_L^3 {\nG_R^{\prime\prime\, c}}   - \check \Nc_L^{ 3^\prime} {\nG_R^{\prime\,c}} +\check \Nc_L^{ 3^{\prime \prime}} {\nG_R^{c}}  ) v_{\rm EW}  + H.c. \,.
\eeqn
In the last line, we have fixed the flavor index of $\omega_1=3$.
Furthermore, one can only take the second component, which corresponds to the EWSB direction, from the $( \rep{1 } \,, \repb{3 } \,,  - \frac{1 }{3 } )_{ \rep{H}\,, \omega_1 = 3 }$ to form the $(b\,,\tau)$ mass terms.
Thus, one can only taken the third VEV component from the $( \rep{1 } \,, \repb{3 } \,,  - \frac{1 }{3 } )_{ \rep{H}\,, \omega_3 = {\rm VI} }$ and the first VEV component from the $( \rep{1 } \,, \repb{3 } \,,  + \frac{2 }{3 } )_{ \rep{H} }^{ \prime \prime\prime }$.
If we consider all three possibilities of the propagator masses, each of them leads to the $(b\,,\tau)$ masses of
\beqs
\beqn
m_{( \rep{1}\,, \repb{3 }\,, - \frac{1}{3 })_{ \rep{H}\,,3} } \sim \Oc(v_{441}) ~&:&~  m_b = m_\tau = \frac{ Y_\Bc  d_{ \mathscr A} }{ 4 }  \zeta_2 \zeta_{13} v_{\rm EW}  \,,\\[1mm]
m_{( \rep{1}\,, \repb{3 }\,, - \frac{1}{3 })_{ \rep{H}\,,3} } \sim \Oc(v_{341}) ~&:&~ m_b = m_\tau = \frac{ Y_\Bc  d_{ \mathscr A} }{ 4}  \zeta_1 \zeta_{23} v_{\rm EW} \,, \\[1mm]
m_{( \rep{1}\,, \repb{3 }\,, - \frac{1}{3 })_{ \rep{H}\,,3} } \sim \Oc(v_{331}) ~&:&~  m_b = m_\tau = \frac{ Y_\Bc  d_{ \mathscr A}  }{ 4 }  \frac{ \zeta_1 }{\zeta_{23} } v_{\rm EW}  \,.\label{eq:btau_mass_indirect}
\eeqn
\eeqs
The first two choices will lead to double suppression factors for the $(b\,, \tau)$ masses from the top quark mass, while only the third choice can be reasonable for the $(b\,,\tau)$ masses.
This was also the reason of our VEV assignment of $\langle ( \rep{1} \,, \repb{3} \,, -\frac{1}{3} )_{\mathbf{H}\,, 3 } \rangle \equiv \frac{1}{ \sqrt{2} } V_{ \repb{3}\,, 3 }$ in Eq.~\eqref{eq:SU8_Higgs_VEVs_mini03}.

\para
The indirect Yukawa couplings from the operator in Eq.~\eqref{eq:d5_Hmixing_B} are expected to generate the first- and second-generational down-type quark and charged lepton masses.
The gauge-invariant subset of $\Large( \repb{28_H}_{\,,\dot 1 }^\dag \repb{28_H}_{\,,\dot {\rm VII} }  \Large)$ can develop the VEV of $\langle  \repb{28_H}_{\,,\dot 1 }^\dag \repb{28_H}_{\,,\dot {\rm VII} } \rangle = \frac{1}{2} w_{\repb{4}\,,  \dot 1} w_{\repb{4}\,,  \dot {\rm VII} } \sim \Oc ( v_{341}^2 )$ according to the VEV assignments in Eq.~\eqref{eq:SU8_Higgs_VEVs_mini02}.
Similar to the indirect Yukawa couplings in Eq.~\eqref{eq:btau_indirect}, we should look for the EWSB components from the $\repb{28_{H}}_{\,, \dot 1 \,, \dot 2 }$ here.
For the Yukawa coupling of $\repb{8_F}^{\dot \omega_1} \rep{56_F}  \repb{28_{H}}_{\,, \dot \omega_1 }$, we find the mass terms of
\beqn\label{eq:smu_indirect}
&& Y_\Dc   \repb{8_F}^{\dot \omega_1} \rep{56_F}  \repb{28_{H}}_{\,, \dot \omega_1 }  \times \frac{ d_{ \mathscr B} }{ M_{\rm pl} }   \repb{28_{H}}_{\,, \dot \omega_1 }^\dag   \repb{28_{H}}_{\,, \dot \omega_2 }^\dag  \rep{70_{H}}^\dag  \Large( \repb{28_H}_{\,,\dot 1 }^\dag \repb{28_H}_{\,,\dot {\rm VII} } \Large) + H.c. \non
&\supset& Y_\Dc \[  ( \repb{4 } \,, \rep{1 } \,,  + \frac{1 }{4 } )_{ \rep{F}}^{\dot \omega_1} \otimes ( \rep{4 } \,, \rep{6 } \,,  + \frac{1 }{4 } )_{ \rep{F}} \oplus ( \rep{1 } \,, \repb{4 } \,,  - \frac{1 }{4 } )_{ \rep{F}}^{\dot \omega_1 } \otimes ( \rep{1 } \,, \repb{4 } \,,  + \frac{3 }{4 } )_{ \rep{F}} \]  \otimes ( \rep{1 } \,, \rep{6 } \,,  - \frac{1 }{2 } )_{ \rep{H}\,, \dot \omega_1} \non
&\times&  \frac{  d_{ \mathscr B} }{ M_{\rm pl} } ( \rep{1 } \,, \rep{6 } \,,  - \frac{1 }{2 } )_{ \rep{H}\,, \dot \omega_1}^\dag \otimes ( \repb{4 } \,, \repb{4 } \,,  0 )_{ \rep{H}\,, \dot \omega_2 }^\dag \otimes ( \rep{4 } \,, \repb{4 } \,,  + \frac{1 }{2 } )_{ \rep{H}}^\dag \otimes \langle  \repb{28_H}_{\,,\dot 1 }^\dag \repb{28_H}_{\,,\dot {\rm VII} }  \rangle  + H.c. \non
&\supset& Y_\Dc  d_{ \mathscr B} \frac{  w_{\repb{4}\,,  \dot 1} w_{\repb{4}\,,  \dot {\rm VII} }  }{ 2 M_{\rm pl} }  \Big[  ( \repb{3 } \,, \rep{1 } \,,  + \frac{1 }{3 } )_{ \rep{F}}^{\dot \omega_1} \otimes ( \rep{3 } \,, \rep{6 } \,,  + \frac{1 }{6 } )_{ \rep{F}}  \oplus ( \rep{1 } \,, \rep{1 } \,, 0 )_{ \rep{F}}^{\dot \omega_1} \otimes ( \rep{1 } \,, \rep{6 } \,,  + \frac{1 }{2 } )_{ \rep{F}}^\prime  \non
&\oplus& ( \rep{1 } \,, \repb{4 } \,,  - \frac{1 }{4 } )_{ \rep{F}}^{\dot \omega_1 } \otimes ( \rep{1 } \,, \repb{4 } \,,  + \frac{3 }{4 } )_{ \rep{F}} \Big] \otimes ( \rep{1 } \,, \rep{6 } \,,  - \frac{1 }{2 } )_{ \rep{H}\,, \dot \omega_1}  \non
&\times& ( \rep{1 } \,, \rep{6 } \,,  - \frac{1 }{2 } )_{ \rep{H}\,, \dot \omega_1}^\dag \otimes  ( \rep{1 } \,, \repb{4 } \,,  -\frac{1 }{4 } )_{ \rep{H}\,, \dot \omega_2 }^\dag   \otimes  ( \rep{1 } \,, \repb{4 } \,,  + \frac{3 }{4 } )_{ \rep{H}}^{\prime \, \dag }   + H.c. \non
&\supset& Y_\Dc d_{ \mathscr B}  \frac{ w_{\repb{4}\,,  \dot 1} w_{\repb{4}\,,  \dot {\rm VII} }  }{ 2 M_{\rm pl} m_{ ( \rep{1 } \,, \rep{6 } \,,  - \frac{1 }{ 2 } )_{ \rep{H}\,, \dot  \omega_1 }  }^2 }    \Big[  ( \repb{3 } \,, \rep{1 } \,,  + \frac{1 }{3 } )_{ \rep{F}}^{\dot \omega_1} \otimes ( \rep{3 } \,, \rep{3 } \,,  0 )_{ \rep{F}}^\prime \oplus ( \rep{1 } \,, \rep{1 } \,, 0 )_{ \rep{F}}^{\dot \omega_1} \otimes ( \rep{1 } \,, \rep{3 } \,,  + \frac{1  }{3 } )_{ \rep{F}}^{ \prime }  \non
&\oplus& ( \rep{1 } \,, \repb{3 } \,,  - \frac{1 }{3 } )_{ \rep{F}}^{\dot \omega_1 } \otimes ( \rep{1 } \,, \repb{3 } \,,  +\frac{2 }{3 } )_{ \rep{F}}^{\prime } \Big] \otimes   \langle ( \rep{1 } \,, \repb{3 } \,,  - \frac{1 }{3 } )_{ \rep{H}\,, \dot \omega_2 }^{ \dag} \rangle  \otimes  \langle ( \rep{1 } \,, \repb{3 } \,,  + \frac{2 }{3 } )_{ \rep{H}}^{\prime \prime \prime \,  \dag} \rangle + H.c. \non
%
%
%
&\Rightarrow& \frac{ Y_\Dc d_{ \mathscr B} }{4 }  \dot \zeta_3   \Big[ \frac{   w_{\repb{4}\,,  \dot 1} w_{\repb{4}\,,  \dot {\rm VII} } }{  m_{ ( \rep{1 } \,, \rep{6 } \,,  - \frac{1 }{ 2 })_{ \rep{H}\,, \dot  1 }  }^2 }   ( \uline{ s_L {d_R}^c } +  \uline{ e_L {\mu_R}^c  }  )  + \frac{   w_{\repb{4}\,,  \dot 1} w_{\repb{4}\,,  \dot {\rm VII} } }{  m_{ ( \rep{1 } \,, \rep{6 } \,,  - \frac{1 }{ 2 })_{ \rep{H}\,, \dot  2 }  }^2 }   ( \uline{ s_L {s_R}^c } +  \uline{ \mu_L {\mu_R}^c  }  )  \Big] v_{\rm EW}  + H.c.  \,,
\eeqn
where the SM quark/lepton components from the $\repb{8_F}^{\dot \omega_1= \dot 1}$/$\repb{8_F}^{\dot \omega_1= \dot 2}$ correspond to the $({d_R}^c \,, e_L)$ and $({s_R}^c \,, \mu_L)$, respectively.
The other mass terms read
\beqn\label{eq:de_indirect}
&& Y_\Dc  \repb{8_F}^{\dot \omega_1} \rep{56_F}  \repb{28_{H}}_{\,, \dot \omega_1 } \times \frac{ d_{ \mathscr B} }{ M_{\rm pl} }   \repb{28_{H}}_{\,, \dot \omega_1 }^\dag   \repb{28_{H}}_{\,, \dot \omega_2 }^\dag  \rep{70_{H}}^\dag \Large( \repb{28_H}_{\,,\dot 1 }^\dag \repb{28_H}_{\,,\dot {\rm VII} } \Large)  + H.c. \non
&\supset& Y_\Dc  \Big[  ( \repb{4 } \,, \rep{1 } \,,  + \frac{1 }{4 } )_{ \rep{F}}^{\dot \omega_1} \otimes ( \rep{6 } \,, \rep{4 } \,,  - \frac{1 }{4 } )_{ \rep{F}} \oplus ( \rep{1 } \,, \repb{4 } \,,  - \frac{1 }{4 } )_{ \rep{F}}^{\dot \omega_1 } \otimes ( \rep{4 } \,, \rep{6 } \,,  + \frac{1 }{4 } )_{ \rep{F}} \Big]  \otimes ( \repb{4 } \,, \repb{4 } \,,  0 )_{ \rep{H}\,, \dot \omega_1} \non
&\times& \frac{ d_{ \mathscr B} }{ M_{\rm pl} }  ( \repb{4 } \,, \repb{4 } \,,  0 )_{ \rep{H}\,, \dot \omega_1}^\dag \otimes ( \rep{1 } \,, \rep{6 } \,,  - \frac{1 }{2 } )_{ \rep{H}\,, \dot \omega_2 }^\dag  \otimes ( \rep{4 } \,, \repb{4 } \,,  + \frac{1 }{2 } )_{ \rep{H}}^\dag \otimes \langle  \repb{28_H}_{\,,\dot 1 }^\dag \repb{28_H}_{\,,\dot {\rm VII} } \rangle + H.c. \non
&\supset& Y_\Dc d_{ \mathscr B} \frac{ w_{\repb{4}\,,  \dot 1} w_{\repb{4}\,,  \dot {\rm VII} }  }{ 2 M_{\rm pl} }   \Big[  ( \repb{3 } \,, \rep{1 } \,,  + \frac{1 }{3 } )_{ \rep{F}}^{\dot \omega_1} \otimes ( \rep{3 } \,, \rep{4 } \,,  - \frac{1 }{12 } )_{ \rep{F}}^\prime \oplus ( \rep{1 } \,, \repb{4 } \,,  - \frac{1 }{4 } )_{ \rep{F}}^{\dot \omega_1 } \otimes ( \rep{1 } \,, \rep{6 } \,,  + \frac{1 }{2 } )_{ \rep{F}}^\prime \Big] \otimes  ( \rep{1 } \,, \repb{4 } \,,  - \frac{1 }{ 4 } )_{ \rep{H}\,, \dot \omega_1} \non
&\times&    ( \rep{1 } \,, \repb{4 } \,,  - \frac{1 }{ 4 } )_{ \rep{H}\,, \dot \omega_1}^\dag   \otimes   ( \rep{1 } \,, \rep{6 } \,,  - \frac{1 }{ 2} )_{ \rep{H}\,, \dot \omega_2}^\dag   \otimes  ( \rep{1 } \,, \repb{4 } \,,  + \frac{3 }{4 } )_{ \rep{H}}^{\prime\,\dag }  + H.c. \non
&\supset& Y_\Dc  d_{ \mathscr B} \frac{ w_{\repb{4}\,,  \dot 1} w_{\repb{4}\,,  \dot {\rm VII} }  }{ 2 M_{\rm pl} m_{ (\rep{1 } \,, \repb{4 } \,,   -\frac{ 1}{ 4} )_{ \rep{H}\,, \dot  \omega_1}  }^2 } \Big[  ( \repb{3 } \,, \rep{1 } \,,  + \frac{1 }{3 } )_{ \rep{F}}^{\dot \omega_1} \otimes ( \rep{3 } \,, \rep{3 } \,,  0 )_{ \rep{F}}^{ \prime \prime} \oplus  ( \rep{1 } \,, \repb{3 } \,,  - \frac{1 }{3 } )_{ \rep{F}}^{\dot \omega_1} \otimes   ( \rep{1 } \,, \repb{3 } \,,  + \frac{2 }{3 } )_{ \rep{F}}^{\prime \prime } \non
&\oplus& ( \rep{1 } \,, \rep{1 } \,,  0 )_{ \rep{F}}^{\dot \omega_1^{ \prime \prime } } \otimes   ( \rep{1 } \,, \rep{3 } \,,  + \frac{1 }{3 } )_{ \rep{F}}^{\prime }  \Big] \otimes    \langle ( \rep{1 } \,, \repb{3 } \,,  - \frac{1 }{ 3} )_{ \rep{H}\,, \dot \omega_2}^{\prime\,\dag } \rangle \otimes  ( \rep{1 } \,, \repb{3 } \,,  + \frac{2 }{3 } )_{ \rep{H}}^{\prime \prime \prime \,\dag }  + H.c.  \non
%
%
&\Rightarrow& \frac{ Y_\Dc d_{ \mathscr B} }{4 } \dot \zeta_3^\prime   \Big[ \frac{ w_{\repb{4}\,,  \dot 1} w_{\repb{4}\,,  \dot {\rm VII} }   }{  m_{ ( \rep{1 } \,, \repb{4 } \,,   -\frac{ 1}{ 4} )_{ \rep{H}\,, \dot  1}  }^2 }   ( \uline{ d_L {d_R}^c } + \uline{ e_L {e_R}^c } )  + \frac{ w_{\repb{4}\,,  \dot 1} w_{\repb{4}\,,  \dot {\rm VII} }   }{  m_{ ( \rep{1 } \,, \repb{4 } \,,   -\frac{ 1}{ 4} )_{ \rep{H}\,, \dot  2 }  }^2 }   ( \uline{ d_L {s_R}^c } + \uline{ \mu_L {e_R}^c } )   \Big] v_{\rm EW} + H.c. \,.
\eeqn
With the Higgs VEV assignments in Eqs.~\eqref{eq:SU8_Higgs_VEVs_mini02} and \eqref{eq:SU8_Higgs_VEVs_mini03}, we can obtain the following propagator masses of
\beqn
&& m_{ ( \rep{1 } \,, \repb{4 } \,,  - \frac{1 }{4} )_{ \rep{H}\,, \dot  1}  }  \sim \Oc( v_{341 } )  \,, \quad m_{ ( \rep{1 } \,, \repb{4 } \,,  - \frac{1 }{4 } )_{ \rep{H}\,, \dot  2} } \sim \Oc( v_{331 } ) \,,
\eeqn
and hypothesize two other propagator masses of
\beqn
&& m_{ ( \rep{1 } \,, \rep{6 } \,,  - \frac{1 }{2} )_{ \rep{H}\,, \dot  1}  }  \sim \Oc( v_{341 } )  \,, \quad m_{ ( \rep{1 } \,, \rep{6 } \,,  - \frac{1 }{2 } )_{ \rep{H}\,, \dot  2} } \sim \Oc( v_{331 } ) \,.
\eeqn
For convenience, we parametrize the following ratios of
\beqn
&& \Delta_{ \dot \omega } \equiv \frac{ w_{\repb{4}\,,  \dot 1} w_{\repb{4}\,,  \dot {\rm VII} }  }{ m_{ ( \rep{1 } \,, \repb{4 } \,,  - \frac{1 }{4} )_{ \rep{H}\,, \dot  \omega}  } ^2 } \,, \quad \Delta_{ \dot \omega}^\prime \equiv \frac{ w_{\repb{4}\,,  \dot 1} w_{\repb{4}\,,  \dot {\rm VII} }  }{ m_{ ( \rep{1 } \,, \rep{6 } \,,  - \frac{1 }{2} )_{ \rep{H}\,, \dot  \omega}  } ^2 } \,,
\eeqn
with $\Delta_{ \dot 1 }^\prime \simeq 1$ and $\Delta_{\dot 2} \simeq \zeta_{23}^{-2}$ in Eqs.~\eqref{eq:smu_indirect} and \eqref{eq:de_indirect}.

\section{The SM quark/lepton masses and the CKM mixing}                                                                                                                                                                                                                                                                                                                                                                                                                                                                                                                                                                                                                                                                                                                                                                                                                                                                                                                                                                                                            
\label{section:SMql_CKM}

\para
We summarize the SM quark/lepton mass matrices based on the gravity-induced mass terms in the previous section.
Based on these results, we further obtain the mass eigenvalues and the CKM mixing of the quark sector.

\subsection{The up-type quark sector}

\para
For all up-type quarks with $Q_e=+\frac{2}{3}$, we write down the following tree-level masses from both the renormalizable Yukawa couplings and the gravity-induced terms in the basis of $\Uc\equiv (u\,,c\,,t\,,\uG\,,\UG)$
\beqs\label{eqs:Uquark_masses}
\beqn
\Mc_\Uc&=& \frac{1}{\sqrt{2} }  \left( \ba{ccc|cc}  
   &    &  &  0  & 0 \\
 \multicolumn{3}{c}{ \sqrt{2} \Big( \Mc_u \Big)_{3\times 3} }  &     c_4 \zeta_0\, V_{ \repb{3}\,, \dot {\rm IX} } &    c_4 \dot \zeta_3^\prime\,  v_{\rm EW}  \\
     &     &     &  0   & -    c_5 \zeta_3\, v_{\rm EW}  \\  \hline
    -   c_4 \dot \zeta_3\, v_{\rm EW}  &  -    c_4 \dot \zeta_3^\prime\,  v_{\rm EW}&  -  c_5 \zeta_3\, v_{\rm EW} &  -  c_4  \zeta_0 \, w_{\repb{4}\,, \dot 1\,, \dot {\rm VII} } &  0 \\
     c_4 \zeta_0\, V_{ \repb{3} \,, \dot {\rm VIII} }^{\prime }   &    c_4 \zeta_0 \, V_{ \repb{3}\,, \dot {\rm IX} }  &  0   & 0   & c_4 \zeta_0 \, w_{\repb{4}\,, \dot 1\,, \dot {\rm VII} } \\  \ea  \right)     \,, \\[1mm]
\Mc_u   &=&   \frac{1}{\sqrt{2} }  \left( \ba{ccc}  
0 &  0  & c_5 \zeta_1 /\sqrt{2} \\
 c_4  \dot \zeta_2 /\sqrt{2}  & 0  & c_5 \zeta_2 / \sqrt{2}   \\
   c_5 \zeta_1 /\sqrt{2}  &  c_5 \zeta_2/\sqrt{2}  &   Y_\Tc \\  \ea  \right) v_{\rm EW}  \approx \Mc_u^{ (0)} + \Mc_u^{ (1 ) } + \Mc_u^{( 2)}   \,, \\[1mm]
\Mc_u^{ (0)}    &=& \frac{1}{\sqrt{2} }  \left( \ba{ccc}  
  0 & 0   & 0   \\
  0   &  0  & 0   \\
  0  &  0  &   Y_\Tc   \\  \ea  \right)  v_{\rm EW} \,, \label{eq:Uquark_mass00} \\[1mm]
 \Mc_u^{ (1)}    &=& \frac{1}{\sqrt{2} }  \left( \ba{ccc}  
 0&  0  &  c_5 \zeta_1 /\sqrt{2}   \\
  0   &   0 & 0    \\
  c_5 \zeta_1 /\sqrt{2} &  0 &   0   \\   \ea  \right) v_{\rm EW}   \,,\label{eq:Uquark_mass01}  \\[1mm]
\Mc_u^{ (2)}    &=&  \frac{1}{\sqrt{2} }  \left( \ba{ccc}  
  0 &  0  &  0  \\
  c_4 \dot \zeta_2  /\sqrt{2} & 0 & c_5 \zeta_2 /\sqrt{2}  \\
   0  &  c_5 \zeta_2 /\sqrt{2} &   0  \\  \ea  \right) v_{\rm EW}  \,, \label{eq:Uquark_mass02}
\eeqn
\eeqs
where we have neglected the terms of $\sim \Oc(\zeta_3\,v_{\rm EW})$ in the expansion.
One obvious feature is that the gauge eigenstates of up quark and the charm quark do not obtain tree-level masses through the $d=5$ operators with the SM Higgs doublet.
Instead, there are only off-diagonal mass mixing terms in Eqs.~\eqref{eq:Uquark_mass01} and \eqref{eq:Uquark_mass02}.
Accordingly, we find that
\beqn
{\rm det}^\prime \[ \Mc_u^{ (0)} \Mc_u^{ (0)\, \dag}   \] &=& \frac{1}{2}  Y_\Tc^2 \,  v_{\rm EW}^2  \Rightarrow m_t^2 \approx \frac{ 1 }{  2 } Y_\Tc^2 \, v_{\rm EW}^2 \,.
\eeqn
Here and below, we use the ${\rm det}^\prime$ to denote the matrix determinant that is equal to the products of all non-zero eigenvalues.
Next, we find the charm quark mass squared of
\beqn
 m_c^2 &=& {\rm det}^\prime \[ \Big( \Mc_u^{ (0)} + \Mc_u^{ (1)}  \Big) \cdot \Big( \Mc_u^{ (0)\, \dag} + \Mc_u^{ (1)\, \dag} \Big)  \]   \Big/ {\rm det}^\prime \[ \Mc_u^{ (0)} \Mc_u^{ (0)\, \dag}   \] \approx  c_5^4  \frac{  \zeta_1^4 }{8 Y_\Tc^2 } \, v_{\rm EW}^2 \,.
\eeqn
The up quark mass squared can be similarly obtained by
\beqn
 m_u^2 &=& {\rm det} \[  \Big( \Mc_u^{ (0)} + \Mc_u^{ (1)} + \Mc_u^{ (2)}  \Big) \cdot \Big( \Mc_u^{ (0)\, \dag} + \Mc_u^{ (1)\, \dag} + \Mc_u^{ (2)\, \dag} \Big)  \]  \non
&&  \Big/ {\rm det}^\prime \[ \Big( \Mc_u^{ (0)} + \Mc_u^{ (1)}  \Big) \cdot \Big( \Mc_u^{ (0)\, \dag} + \Mc_u^{ (1)\, \dag} \Big)  \]   \approx c_4^2 \frac{ \zeta_2^2 \dot \zeta_2^2  }{4 \zeta_1^2 }\, v_{\rm EW}^2 \,.
\eeqn
To summarize, all SM up-type quark masses are expressed as follows
\beqn\label{eq:SU8_SMumasses}
&&  m_u \approx c_4 \frac{ \zeta_2 \dot \zeta_2 }{ 2 \zeta_1 } v_{\rm EW} \,,\quad m_c \approx c_5^2  \frac{ \zeta_1^2 }{  2 \sqrt{2} Y_\Tc } v_{\rm EW}  \,, \quad  m_t \approx \frac{ Y_\Tc }{ \sqrt{2} } v_{\rm EW} \,.
\eeqn
%
%

\subsection{The down-type quark and charged lepton sector}

\para
For all down-type quarks with $Q_e=-\frac{1}{3}$, we find the following tree-level mass matrix in the basis of $\Dc \equiv (d\,,s\,,b\,,\dG\,,\DG\,, \DG^\prime \,, \DG^{\prime\prime}\,, \DG^{\prime\prime\prime} \,, \DG^{\prime\prime \prime \prime}\,, \DG^{\prime\prime \prime\prime \prime})$
\beqn\label{eq:Dquark_massmatrix}
\Mc_\Dc &=& \left( \ba{cccccccccc}  
   &    &  &    &    &  &  &   &   &   \\
 \multicolumn{3}{c}{ \Big( \Mc_d \Big)_{3\times 3} }    &    \multicolumn{7}{c}{ \Big(  \Mc_{d_L \DG_R }  \Big)_{3\times 7} }   \\
   &     &    &   &   &  &   &  &  &   \\
   &     &    &     &   &  &   &  &  &   \\
   &     &    &      &  &   &  &  &   \\
\multicolumn{3}{c}{ \Big(  \Mc_{\DG_L d_R} \Big)_{7\times 3} }   &   \multicolumn{7}{c}{ \Big(  \Mc_{\DG}  \Big)_{7\times 7} }   \\
   &     &    &     &   &  &   &  &  &   \\
   &     &    &    &    &  &  &   &   &   \\   \ea  \right) \,,
\eeqn
where each block except the $3\times3$ SM block is expressed as
\beqs\label{eqs:Dquark_mass_d5}
\beqn
 %
\Big(  \Mc_{d_L \DG_R }  \Big)_{3\times 7} &=&  \frac{1}{\sqrt{2} }   \left( \ba{ccccccc}  
  0 &  0 &  0  &  0   &  0  & c_3 \dot \zeta_3^\prime v_{\rm EW}   & c_3 \dot \zeta_3^\prime v_{\rm EW}   \\
  - c_4 \zeta_0 V_{ \repb{3}\,, \dot {\rm IX} }   &  0 & 0  & 0  &  c_3 \dot \zeta_3 v_{\rm EW}  &  0   & c_3 \dot \zeta_3 v_{\rm EW}  \\
  0 &  0 &   0  & 0   &  0  &  0   &  0     \\  \ea  \right) \label{eq:Dquark_massdD}  \,,  \\[1mm]
 \Big(  \Mc_{ \DG_L d_R} \Big)_{ 7 \times 3} &=&   \frac{1}{\sqrt{2} }  \left( \ba{ccc}  
c_3 \dot \zeta_2 v_{\rm EW} & c_3 \dot \zeta_2 v_{\rm EW} & 0      \\
0  &   0 &  c_1 \zeta_0 W_{ \repb{4}\,,{\rm IV} }    \\
0  & 0  & 0    \\
 0 & 0  &   c_1 \zeta_0 w_{ \repb{4}\,,{\rm V} }  \\
c_3 \zeta_0 V_{ \repb{3}\,, \dot {\rm VIII} }^\prime  & c_3 \zeta_0 V_{ \repb{3}\,, \dot {\rm VIII} }^\prime & c_3 \zeta_2 V_{ \repb{3} \,, {\rm VI} }   \\
 c_3 \zeta_0 V_{ \repb{3}\,, \dot {\rm IX}}  &  c_3 \zeta_0 V_{ \repb{3}\,, \dot {\rm IX}}  & c_3 \zeta_1 V_{ \repb{3} \,, {\rm VI} }     \\
  c_3 \zeta_0 w_{ \repb{4}\,, \dot {\rm VII}}  & c_3 \zeta_0 w_{ \repb{4}\,, \dot {\rm VII}}  & c_3 \zeta_1 w_{ \repb{4} \,, {\rm V} }  \\
 \ea  \right) \label{eq:Dquark_massDd}  \,,  \\[1mm]
\Big( \Mc_{\DG}  \Big)_{7\times 7} &\approx&  \frac{1}{\sqrt{2} }  {\rm diag} (  c_4 \zeta_0 w_{ \repb{4 }\,, \dot {\rm VII} } \,, Y_\Bc W_{ \repb{4}\,, {\rm IV} } \,, Y_\Bc V_{ \repb{3}\,, {\rm VI } }  \,, Y_\Bc  w_{ \repb{4}\,, {\rm V} } \,, \non
 && Y_\Dc V_{ \repb{3} \,, \dot {\rm VIII} }^\prime \,, Y_\Dc V_{ \repb{3} \,, \dot {\rm IX} }  \,, Y_\Dc w_{ \repb{4}\,, \dot {\rm VII} }  ) \,. \label{eq:Dquark_massDD}
\eeqn
\eeqs
Here, we have neglected the mass mixing terms in the $\Big( \Mc_{\DG}  \Big)_{7\times 7}$ for simplicity.
To obtain the SM down-type quark masses, it is sufficient to focus on the SM block of $\Big( \Mc_d \Big)_{3\times 3}$, and its entries are from the Eqs.~\eqref{eq:ds_direct}, \eqref{eq:emu_direct}, \eqref{eq:btau_indirect}, \eqref{eq:smu_indirect}, and \eqref{eq:de_indirect} as follows
\beqn\label{eq:Dquark_massdd}
&&  \Big( \Mc_d \Big)_{3\times 3}   \approx  \frac{1}{4} \left( \ba{ccc}
 ( 2 c_3 + Y_\Dc d_{\mathscr B}  )\dot \zeta_3^\prime  &  ( 2 c_3  +  Y_\Dc d_{\mathscr B}  \Delta_{ \dot 2} ) \dot \zeta_3^\prime  & 0   \\
  ( 2 c_3  +  Y_\Dc d_{\mathscr B}  \Delta_{ \dot 1}^\prime )  \dot \zeta_3  &  (  2 c_3  +  Y_\Dc d_{\mathscr B}   \zeta_{23 }^{-2 } ) \dot \zeta_3   &  0   \\
  0 & 0  &  Y_\Bc d_{\mathscr A}   \zeta_{23}^{-1} \zeta_1   \\  \ea  \right) v_{\rm EW}    \,.
\eeqn
For convenience, we parametrize all $\Gc_{331}$-breaking VEVs as follows
\beqn\label{eq:331VEVratio}
&& \zeta_3 = \dot \zeta_3 =  \dot \zeta_3^\prime  \tan \lambda \,. 
\eeqn
It is straightforward to find the following SM down-type quark masses of
\beqs\label{eqs:SU8_SMdmasses}
\beqn
 m_b &\approx&   \frac{ 1 }{ 4  }  Y_\Bc d_{\mathscr A} \zeta_{ 23 }^{-1} \zeta_1 \, v_{\rm EW} \,,  \\[1mm]
m_s  &\approx& \frac{1 }{4 } ( 2c_3 + Y_\Dc d_{\mathscr B} \zeta_{ 23 }^{-2 } ) \dot \zeta_3 \, v_{\rm EW} \,,  \\[1mm]
 m_d &\approx& \hf c_3 \dot \zeta_3  | \tan \lambda |   \, v_{\rm EW} \,,
\eeqn
\eeqs
from Eq.~\eqref{eq:Dquark_massdd}.

\para
For all charged leptons with $Q_e=-1$, their tree-level mass matrix is correlated with the down-type quark mass matrix.
In the basis of $\Lc \equiv (e\,,\mu\,, \tau\,,\EG\,,\eG\,, \eG^\prime \,, \eG^{\prime\prime}\,, \eG^{\prime\prime\prime} \,, \eG^{\prime\prime \prime \prime}\,, \eG^{\prime\prime \prime\prime \prime})$, it is expressed as follows
\beqn\label{eq:Lep_massmatrix}
\Mc_\Lc &=& \left( \ba{cccccccccc}  
   &    &  &    &    &  &  &   &   &   \\
 \multicolumn{3}{c}{ \Big( \Mc_\ell \Big)_{3\times 3} }    &    \multicolumn{7}{c}{ \Big(  \Mc_{\ell_L \eG_R }  \Big)_{3\times 7} }   \\
   &     &    &   &   &  &   &  &  &   \\
   &     &    &     &   &  &   &  &  &   \\
   &     &    &      &  &   &  &  &   \\
\multicolumn{3}{c}{ \Big(  \Mc_{\eG_L \ell_R} \Big)_{7\times 3} }   &   \multicolumn{7}{c}{ \Big(  \Mc_{\eG}  \Big)_{7\times 7} }   \\
   &     &    &     &   &  &   &  &  &   \\
   &     &    &    &    &  &  &   &   &   \\   \ea  \right) \,,
\eeqn
where the $3\times 3$ SM block is expressed as
%
%
\beqn
 %
    \Big( \Mc_\ell \Big)_{ 3\times 3} &=&   \Big( \Mc_d^T \Big)_{3\times 3} \non
    &=& \frac{1}{4} \left( \ba{ccc}
 ( 2 c_3 + Y_\Dc d_{\mathscr B}  )\dot \zeta_3^\prime  &  ( 2 c_3  +  Y_\Dc d_{\mathscr B} \Delta_{ \dot 1}^\prime ) \dot \zeta_3  & 0   \\
  ( 2 c_3  +  Y_\Dc d_{\mathscr B}   \Delta_{ \dot 2}   )  \dot \zeta_3^\prime  &  (  2 c_3  +  Y_\Dc d_{\mathscr B}   \zeta_{23 }^{-2 } ) \dot \zeta_3   &  0   \\
  0 & 0  &  Y_\Bc d_{\mathscr A}   \zeta_{23}^{-1} \zeta_1   \\  \ea  \right) v_{\rm EW}  \label{eq:Lep_massll} \,.
\eeqn
%
%
Thus, it is straightforward to find the tree-level mass relations of
\beqn
&& m_\tau = m_b \,,\quad m_\mu = m_s \,,\quad m_e = m_d  \,.
\eeqn

\subsection{The CKM matrix and the benchmark}

\para
The bi-unitary transformations of
\beqn
&& \hat \Fc_L \Mc_\Fc  \hat \Fc_R^\dag = \Mc_\Fc^{\rm diag}\,,\quad \Mc_\Fc \Mc_\Fc^\dag = \hat \Fc_L^\dag ( \Mc_\Fc^{\rm diag} )^2  \hat \Fc_L  \,,\quad \Fc = (\Uc \,, \Dc) \,,
\eeqn
diagonalize the un-hatted flavor eigenstates into their hatted mass eigenstates.
To obtain the CKM matrix of the quark sector, we derive the left-handed mixing matrices of $(\hat \Uc_L \,, \hat \Dc_L)$ of
\beqn
&&  \left( \ba{c}  \hat u_L \\  \hat c_L  \\ \hat t_L  \\  \ea  \right)  = \hat \Uc_L \cdot \left( \ba{c}  u_L \\  c_L  \\  t_L  \\  \ea  \right)  \,, \quad  \left( \ba{c}  \hat d_L \\  \hat s_L  \\ \hat b_L  \\  \ea  \right)  = \hat \Dc_L \cdot \left( \ba{c}  d_L \\  s_L  \\  b_L  \\  \ea  \right) \,,
\eeqn
through their perturbative expansions in Eqs.~\eqref{eqs:Uquark_masses} and \eqref{eqs:SU8_SMdmasses}.
Explicitly, we find that
\beqn
&& \hat \Uc_L = \hat \Uc_L^{ (12)} \cdot \hat \Uc_L^{ (13)} \cdot \hat \Uc_L^{ (23)} \cdot \hat \Uc_L^{ \rm ID} \,,  \non
&& \hat \Uc_L^{ \rm ID} =  \left( \ba{ccc}
0  &   1 &  0    \\
 -1 & 0   &   0    \\
0  & 0   &  1     \\   \ea  \right)  \,, \quad  \hat \Uc_L^{ (12)} = \left( \ba{ccc}
\cos \epsilon  &   -\sin\epsilon &  0    \\
 \sin\epsilon & \cos \epsilon  &   0    \\
0  & 0   &  1     \\   \ea  \right)  \,, \quad \sin\epsilon \simeq \frac{m_u }{ m_c } \frac{\zeta_1 }{ \zeta_2} - \frac{\zeta_2}{ \zeta_1} \sim \Oc(10^{-2})\,, \non
   &&  \hat \Uc_L^{(13)} \approx   \left( \ba{ccc}
1  &  0  &  - \frac{ c_5 \zeta_2 }{ \sqrt{2} Y_\Tc  } \\
 0 &  1  &  0   \\
\frac{ c_5 \zeta_2 }{ \sqrt{2} Y_\Tc  } &  0  &  1  \\   \ea  \right)  \,, \quad
\hat \Uc_L^{ (23)} \approx   \left( \ba{ccc}
1  & 0   & 0   \\
 0 &  1  &  \frac{ c_5 \zeta_1 }{ \sqrt{2} Y_\Tc  }    \\
 0 & - \frac{ c_5 \zeta_1 }{ \sqrt{2} Y_\Tc  } &  1   \\    \ea  \right)  \,,
\eeqn
and
\beqn
&& \hat \Dc_L= \left( \ba{ccc}
\sin \lambda  &  \cos \lambda  &       \\
- \cos \lambda &  \sin \lambda &       \\
  &    & 1   \\   \ea  \right)  \approx
  \left( \ba{ccc}
 \lambda  &  1-\lambda^2/2   &       \\
- 1 + \lambda^2/2  &  \lambda &       \\
  &    & 1   \\   \ea  \right) \,.
\eeqn
The CKM matrix can be approximated as the Wolfenstein parametrization~\cite{Wolfenstein:1983yz}
\beqn\label{eq:SU8_CKM}
\hat V_{\rm CKM} \Big|_{ {\rm SU}(8) }&=& \hat \Uc_L \hat \Dc_L^\dag  \approx \left( \ba{ccc}
1-\lambda^2/2  & \lambda   & - \frac{ c_5 \zeta_2 }{ \sqrt{2} Y_\Tc} \\
-\lambda  &  1-\lambda^2/2  &   \frac{ c_5 \zeta_1 }{ \sqrt{2} Y_\Tc } \\
 \frac{ c_5  ( \lambda \zeta_1 + \zeta_2 ) }{ \sqrt{2} Y_\Tc} &  -  \frac{ c_5 \zeta_1 }{ \sqrt{2} Y_\Tc } & 1 \\  \ea \right) \,,
\eeqn
where the mixing parameter of $\lambda = |V_{us}|$ is given by the ratio between two $\Gc_{331}$-breaking VEVs in Eq.~\eqref{eq:331VEVratio}.
The natural hierarchy of $\zeta_1 \gg \zeta_2$ in Eq.~\eqref{eq:scale_params} gives arise to the observed hierarchy of $|V_{cb}| \gg |V_{ub}|$.
Notice that, the correct relation of $|V_{cb}| \gg |V_{ub}|$ that we can obtain in Eq.~\eqref{eq:SU8_CKM} also relies on the precise flavor identifications of ${u_R}^c$ and ${c_R}^c$ in Tab.~\ref{tab:SU8_56ferm}, which would otherwise lead to the wrong relation of $|V_{cb}| \ll |V_{ub}|$ if these two flavors were swapped.
Though we have assumed all VEVs in Eqs.~\eqref{eqs:SU8_Higgs_VEVs_mini} to be real, it is also straightforward to expect a CP-violating phase that originates from the relative phase between the $\zeta_1$ and $\zeta_2$, as long as we restore these VEVs to be complex.

\begin{table}[htp]
\begin{center}
\begin{tabular}{cccccc}
\hline\hline
$\zeta_1$  &  $\zeta_2$ &  $\zeta_3$ & $Y_\Dc$  & $Y_\Bc$  &   $Y_\Tc$  \\
$ 6.0 \times10^{-2}$  &  $2.0\times10^{-3}$  & $2.0\times10^{-5}$  &  $0.5$ & $0.5$ & $0.8$  \\
\hline
  $c_3$&$c_4$&$c_5$&$d_{\mathscr A}$&  $d_{\mathscr B}$  &$\lambda$    \\
  $1.0$& $ 0.2$ & $1.0$ & $ 0.01$ & $ 0.01$ & $0.22$   \\
\hline
$m_u$  &  $m_c$ &  $m_t$  &  $m_d=m_e$  &  $m_s=m_\mu$  &  $m_b=m_\tau$       \\
$1.6 \times 10^{-3}$  &  $0.6$ &  $139.2$  &  $ 0.5\times 10^{-3}$  &  $6.4 \times 10^{-2}$  &  $1.5$      \\
\hline
$|V_{ud}|$  &  $|V_{us}|$ &  $|V_{ub}|$  &   &   &        \\
$0.98$  &  $0.22$ &  $2.1 \times10^{-3}$  &   &  &       \\
\hline
  $|V_{cd}|$  &  $|V_{cs}|$  &  $|V_{cb}|$  &   &   &  $$    \\
  $0.22$  &  $0.98$  &  $ 5.3\times 10^{-2}$  &   &   &  $$    \\
\hline
$|V_{td}|$  &  $|V_{ts}|$ &   $|V_{tb}|$  &    &   &        \\
$0.013$  &  $ 5.3 \times 10^{-2}$ &  $1$  &   &    &      \\
\hline\hline
\end{tabular}
\end{center}
\caption{The parameters of the $\gSU(8)$ benchmark point and the predicted SM quark/lepton masses (in unit of ${\rm GeV}$) as well as the CKM mixings.}
\label{tab:SU8_benchmark}
\end{table}%

\para
Based on the predicted quark masses in Eqs.~\eqref{eq:SU8_SMumasses}, \eqref{eqs:SU8_SMdmasses}, and the CKM matrix in Eq.~\eqref{eq:SU8_CKM}, we suggest a set of benchmark point of the ${\rm SU}(8)$ input parameters, as well as the predicted SM quantities in Tab.~\ref{tab:SU8_benchmark}.
With the natural renormalizable Yukawa couplings of $(Y_\Bc \,, Y_\Tc\,, Y_\Dc) \sim \Oc(1)$, non-renormalizable direct Yukawa couplings of $c_{ 3\,, 4\,, 5} \sim \Oc(0.1) - \Oc(1)$, and the relatively suppressed indirect Higgs mixing coefficients of $d_{ { \mathscr A}\,, {\mathscr B} } \sim \Oc(0.01)$, three-generational SM quark/lepton hierarchical masses and the CKM mixing patterns have been well displayed by the leading-order predictions in the ${\rm SU}(8)$ framework.
Three dimensionless parameters of $(\zeta_1\,, \zeta_2\,, \zeta_3)$ can be translated into three intermediate symmetry breaking scales in Eq.~\eqref{eq:Pattern} of
\beqn\label{eq:benchmark}
&&  v_{441}\simeq 1.4 \times 10^{17 }\,{\rm GeV}  \,, \quad v_{341} \simeq  4.8\times 10^{15} \,{\rm GeV} \,, \quad v_{331} \simeq 4.8\times 10^{13} \,{\rm GeV} \,,
\eeqn
with the reduced Planck scale of $M_{\rm pl}= ( 8 \pi G_N)^{-1/2}= 2.4 \times 10^{18}\, {\rm GeV}$.
Therefore, the highest intermediate scales of the benchmark point indicate a unification scale over $\sim \Oc(10^{17})\,{\rm GeV}$.

\section{Summary and outlook}
\label{section:conclusion}

\subsection{The main results}

\para
We explore the origin of three-generational SM quark/lepton masses and the CKM mixing pattern in an ${\rm SU}(8)$ theory, where three SM generations can be minimally and non-trivially embedded without simple repetitions.
With Definition \ref{def:IRAFFS} of the chiral IRAFFS, Georgi's 1979 conjecture~\cite{Georgi:1979md} on the flavor structure in a class of unified ${\rm SU}(N)$ theories was significantly alleviated.
The emergent global symmetries from the chiral fermion sector in Eqs.~\eqref{eq:DRS_SU8} and \eqref{eq:PQ_SU8} play the pivotal roles in several aspects that were not described from previous studies.
Through our previous analysis of the global $\widetilde {\rm U}(1)_{B-L}$ symmetry~\cite{Chen:2023qxi}, which originate from the non-anomalous Abelian component in Eq.~\eqref{eq:DRS_SU8}, we conjectured one unique SM Higgs doublet of $( \rep{1} \,, \repb{2} \,, +\frac{1}{2} )_{\mathbf{H}}^{\prime \prime\prime }\subset \rep{70_H}$ in the spectrum.
It turns out that the top quark is the unique SM fermion that obtains the EW scale mass with natural Yukawa coupling of $Y_\Tc\sim\Oc(1)$ from Eq.~\eqref{eq:top_Yukawa}.
All light SM quark/lepton masses are generated by EWSB VEV from the $( \rep{1} \,, \repb{2} \,, +\frac{1}{2} )_{\mathbf{H}}^{\prime \prime\prime }\subset \rep{70_H}$ through: (i) the $d=5$ bi-linear fermion Yukawa coupling operators directly, (ii) or the $d=5$ non-renormalizable irreducible Higgs mixing operators in Eqs.~\eqref{eqs:d5_Hmixings} indirectly.
Both operators emerge due to the gravitational effects that break the global symmetries of the ${\rm SU}(8)$ chiral fermions in Eq.~\eqref{eq:DRS_SU8} and Tab.~\eqref{tab:U1TU1PQ}.
Three-generational SM quark/lepton masses as well as the CKM mixing pattern can be explained in this framework with: (i) three intermediate symmetry breaking scales as listed in Eq.~\eqref{eq:benchmark} with reasonable hierarchies between the GUT scale and the EW scale;  (ii) the flavor identifications described in Tabs.~\ref{tab:SU8_8barferm}, \ref{tab:SU8_28ferm}, and \ref{tab:SU8_56ferm};  (iii) assumptions of the Higgs VEVs with specific flavor choices in Eqs.~\eqref{eqs:SU8_Higgs_VEVs_mini}.
A typical benchmark point was listed in Tab.~\ref{tab:SU8_benchmark}, where all $d=4/d=5$ Yukawa couplings are $\sim\Oc(0.1)-\Oc(1)$, and all indirect Higgs mixing coefficients are $\sim\Oc(0.01)$.
We wish to highlight that the flavor identifications of all first- and second-generational SM fermions in Tab.~\ref{tab:SU8_56ferm} were not previously reported by Barr in Ref.~\cite{Barr:2008pn}.
Above all, the participation of the gravity in the SM quark/lepton masses makes its role more important than what one could have possibly imagined.

\subsection{Outlook}

\para
The current results of three-generational SM quark/lepton masses and the CKM mixing patterns are tree-level predictions at the GUT scale.
It is straightforward to expect an RG-improved analysis based on the symmetry breaking pattern and the flavor identifications.
One can also expect that the intra-generational mass splittings of the down-type quarks and the charged leptons are due to the RG effects.
Earlier studies of the SM quark/lepton masses based on the ${\rm SU}(5)/{\rm SO}(10)$ groups include Refs.~\cite{Buras:1977yy,Nanopoulos:1978hh,Georgi:1979ga}.
A relevant issue of studying the RG effects is the ${\rm SU}(8)$ symmetry breaking pattern.
Other than the one described in Eq.~\eqref{eq:Pattern} of the current context, two alternative symmetry breaking patterns of
\beqn
&& {\rm SU}(8) \xrightarrow{ v_U } \Gc_{441} \xrightarrow{ v_{441} } \Gc_{431} \xrightarrow{v_{431} } \Gc_{331} \xrightarrow{ v_{331} } \Gc_{\rm SM} \xrightarrow{ v_{\rm EW} } {\rm SU}(3)_{c}  \otimes  {\rm U}(1)_{\rm EM} \,,  \non
&& {\rm SU}(8) \xrightarrow{ v_U } \Gc_{441} \xrightarrow{ v_{441} } \Gc_{431} \xrightarrow{v_{431} } \Gc_{421} \xrightarrow{ v_{421} } \Gc_{\rm SM} \xrightarrow{ v_{\rm EW} } {\rm SU}(3)_{c}  \otimes  {\rm U}(1)_{\rm EM} \,,
\eeqn
are also allowed group-theoretically.
Thus, we defer to compare the three-generational SM quark/lepton masses and the gauge coupling unifications in all three possible patterns in the future work~\cite{Chen:2024deo}.

\para
In addition to explaining the SM flavor puzzle, several BSM ingredients arising from the ${\rm SU}(8)$ spectrum can be sketched as follows:
\begin{itemize}

\item flavor non-universal gauge couplings from the extended the $\Gc_{441}$ and the $\Gc_{341}$ symmetries, which are natural consequence of the non-repetitive flavor structure in the ${\rm SU}(8)$ theory~\cite{Chen:2024deo};

\item vectorial leptoquarks from the extended strong sector of the ${\rm SU}(4)_s$ symmetry, as well as scalar leptoquarks from the Higgs fields;

\item flavor-changing charged currents and neutral currents from the strong/extended weak symmetries;

\item vectorlike mirror quarks and mirror leptons from the $(\rep{10_F} \,, \repb{10_F})$-pair through the decomposition of the $\rep{56_F}$ in Eq.~\eqref{eq:SU8_decompose};

\item light (left-handed) sterile neutrinos through the `t-Hooft anomaly matching of the global $\widetilde{\rm U}(1)_T$ symmetries at different symmetry breaking stages~\cite{Chen:2023qxi}.

\end{itemize}
If the ongoing LHC and one of the projected high-energy $e^+ e^-$ colliders~\cite{ILC:2013jhg,TLEPDesignStudyWorkingGroup:2013myl,CEPCStudyGroup:2018ghi} continue to justify the SM predictions of the Yukawa couplings with unprecedented precisions by measuring one single SM Higgs boson, we conjecture the ${\rm SU}(8)$ theory as a framework to unify both the fundamental symmetries and the elementary particles.
Needless to say,  the intermediate symmetry breaking scales in Eq.~\eqref{eq:benchmark} and the corresponding massless field contents determined in the current discussions will be used for the gauge coupling unifications in the future work~\cite{Chen:2024deo}.

\section*{Acknowledgements}
%
%
\para
We would like to thank Luca Di Luzio, Zhanpeng Hou, Tianjun Li, Yanwen Liu, Kaiwen Sun, Yuan Sun, Yinan Wang, Wenbin Yan, and Ye-Ling Zhou for very enlightening discussions and communications.
N.C. would like to thank Shanxi University, Yantai University, Hangzhou Institute for Advanced Study (HIAS), Nanjing University, and Institute of Theoretical Physics CAS for hospitalities when preparing this work.
N.C. is partially supported by the National Natural Science Foundation of China (under Grants No. 12035008 and No. 12275140) and Nankai University.
Y.N.M. is partially supported by the National Natural Science Foundation of China (under Grant No. 12205227), the Fundamental Research Funds for the Central Universities (WUT: 2022IVA052), and Wuhan University of Technology.

\appendix

\section{Conventions, decomposition rules and charge quantizations in the ${\rm SU}(8)$ theory}
\label{section:Br}

\begin{table}[htp]
\begin{center}
\begin{tabular}{c|ccc}
\hline \hline
Indices   &  group  & irrep &  range    \\
\hline
$\aG\,,\bG\,,\cG$   &  ${\rm SU}(8)$   &  fundamental  & $1\,,...\,,8$   \\
   &      & anti-fundamental  &   \\
 $\AG\,,\BG\,,\CG$   &     ${\rm SU}(8)$  & adjoint  & $1\,,...\,,63$  \\ \hline
$\bar a\,, \bar b\,, \bar c$   &  ${\rm SU}(4)_s$   &  fundamental  & $\clubsuit\,,\diamondsuit\,, \heartsuit\,, \spadesuit$   \\
   &      & anti-fundamental  &   \\
 $\bar A\,,\bar B\,,\bar C$   &     ${\rm SU}(4)_s$  & adjoint  & $1\,,...\,,15$  \\ \hline
$a\,,b\,,c$   &  ${\rm SU}(3)_c$   &  fundamental  & $\clubsuit\,,\diamondsuit\,, \heartsuit$   \\
   &      & anti-fundamental  &   \\
 $A\,,B\,,C$   &     ${\rm SU}(3)_c$  & adjoint  & $1\,,...\,,8$  \\  \hline
 $\bar i\,,\bar j\,, \bar k$   &  ${\rm SU}(4)_W$   &  fundamental  & $1\,,2\,,3\,,4$   \\
   &      & anti-fundamental  &   \\
  $\bar I\,,\bar J\,, \bar K$   &     ${\rm SU}(4)_W$  & adjoint  & $1\,,...\,,15$  \\ \hline
 $\tilde i \,, \tilde j \,, \tilde k $   &  ${\rm SU}(3)_W$   &  fundamental  & $1\,,2\,,3$   \\
   &      & anti-fundamental  &   \\
  $ \tilde I \,, \tilde J \,, \tilde K $   &     ${\rm SU}(3)_W$  & adjoint  & $1\,,...\,,8$  \\ \hline
 $ i\,, j$   &  ${\rm SU}(2)_W$   &  fundamental  & $1\,,2$   \\
   &      & anti-fundamental  &   \\
  $ I\,, J\,, K$   &     ${\rm SU}(2)_W$  & adjoint  & $1\,,2\,,3$  \\
\hline\hline
\end{tabular}
\end{center}
\caption{
Definition of indices for various gauge groups from the maximal ${\rm SU}(8)$ symmetry breaking pattern.
The fundamental and anti-fundamental indices will be distinguished by superscripts and subscripts, respectively.
}
\label{tab:notations}
\end{table}%

%
\para
According to the symmetry breaking pattern in Eq.~\eqref{eq:Pattern}, we define the conventions of gauge group indices in Tab.~\ref{tab:notations}.
The fundamental and anti-fundamental representations will be denoted by superscripts and subscripts, respectively.
We also define the decomposition rules in the ${\rm SU}(8)$ theory.
After the GUT-scale symmetry breaking, we define the ${\rm U}(1)_{X_0}$ charges for the ${\rm SU}(8)$ fundamental representation as follows
\beqn\label{eq:X0charge}
\hat X_0( \rep{8} ) &\equiv& {\rm diag} ( \underbrace{ - \frac{1}{4}  \mathbb{I}_{4 \times 4}  }_{ \rep{4_s} }\,, \underbrace{ +\frac{1}{4} \mathbb{I}_{4 \times 4}   }_{ \rep{4_W} } )\,.
\eeqn
%
%
%
%
%
%
Sequentially, the ${\rm U}(1)_{X_1}$, ${\rm U}(1)_{X_2}$, and ${\rm U}(1)_{Y}$ charges are defined according to the ${\rm SU}(4)_s$ and the ${\rm SU}(4)_W$ fundamental representations as follows
\beqs\label{eq:U1charges_fund}
\beqn
\hat X_1(\rep{4_s}) &\equiv&  {\rm diag} \, \Big( \underbrace{  (- \frac{1}{12}+ \Xc_0 ) \mathbb{I}_{3\times 3} }_{ \rep{3_c} } \,, \frac{1}{4}+ \Xc_0 \Big) \,,\label{eq:X1charge_4sfund}\\[1mm]
\hat X_2 ( \rep{4_W} )&\equiv& {\rm diag} \, \Big( \underbrace{  ( \frac{1}{12} + \Xc_1 ) \mathbb{I}_{3\times 3} }_{ \rep{3_W} } \,, -\frac{1}{4} + \Xc_1 \Big)  \,, \label{eq:X2charge_4Wfund} \\[1mm]
\hat Y ( \rep{4_W} )&\equiv&  {\rm diag} \, \Big(  ( \frac{1}{6}+ \Xc_2 ) \mathbb{I}_{2\times 2} \,,- \frac{1}{3}+ \Xc_2 \,, \Xc_2 \Big) \non
&=& {\rm diag} \, \Big(  \underbrace{ ( \frac{1}{4} + \Xc_1 ) \mathbb{I}_{2\times 2} }_{ \rep{2_W} } \,,  ( - \frac{1}{4} + \Xc_1  ) \mathbb{I}_{2\times 2} \Big) \,, \label{eq:Ycharge_4Wfund} \\[1mm]
\hat Q_e ( \rep{4_W} )&\equiv& T_{ {\rm SU}(4) }^3 +  \hat Y ( \rep{4_W} ) = {\rm diag} \, \Big( \frac{3}{4} + \Xc_1  \,, ( - \frac{1}{4} + \Xc_1  ) \mathbb{I}_{3\times 3}  \Big) \,. \label{eq:Qcharge_4Wfund}
\eeqn
\eeqs
Based on the above definitions for the fundamental representations, the rules for other higher rank anti-symmetric representations can be derived by tensor productions.
For adjoint representations of the ${\rm SU}(4)_s$ and the ${\rm SU}(4)_W$ groups, these charges are defined by setting all ${\rm U}(1)$ charges to zero in Eqs.~\eqref{eq:U1charges_fund}
\beqs\label{eq:U1charges_adj}
\beqn
\hat X_1(\rep{15_s}) &\equiv&  {\rm diag} ( - \frac{1}{12} \mathbb{I}_{3\times 3} \,,  \frac{1}{4} ) \,,\label{eq:X1charge_4sadj}\\[1mm]
\hat X_2 ( \rep{15_W} )&\equiv& {\rm diag} ( \frac{1}{12} \mathbb{I}_{3\times 3} \,, -\frac{1}{4}   )  \,, \label{eq:X2charge_4Wadj} \\[1mm]
\hat Y ( \rep{15_W} )&\equiv&  {\rm diag}(  \frac{1}{4} \mathbb{I}_{2\times 2}  \,,- \frac{1}{4} \mathbb{I}_{2\times 2} )  \,, \label{eq:Ycharge_4Wadj} \\[1mm]
\hat Q_e ( \rep{15_W} )&\equiv&  {\rm diag} \, \Big( \frac{3}{4}   \,, - \frac{1}{4}   \mathbb{I}_{3\times 3}  \Big) \,. \label{eq:Qcharge_4Wadj}
\eeqn
\eeqs


\begin{thebibliography}{10}

\bibitem{Yang:1954ek}
C.-N. Yang and R.~L. Mills, ``{Conservation of Isotopic Spin and Isotopic Gauge
  Invariance},'' \href{http://dx.doi.org/10.1103/PhysRev.96.191}{{\em Phys.
  Rev.} {\bfseries 96} (1954) 191--195}.

\bibitem{Georgi:1974sy}
H.~Georgi and S.~L. Glashow, ``{Unity of All Elementary Particle Forces},''
  \href{http://dx.doi.org/10.1103/PhysRevLett.32.438}{{\em Phys. Rev. Lett.}
  {\bfseries 32} (1974) 438--441}.

\bibitem{Fritzsch:1974nn}
H.~Fritzsch and P.~Minkowski, ``{Unified Interactions of Leptons and
  Hadrons},'' \href{http://dx.doi.org/10.1016/0003-4916(75)90211-0}{{\em Annals
  Phys.} {\bfseries 93} (1975) 193--266}.

\bibitem{Planck:1901oar}
M.~Planck, ``{\"Uber das Gesetz der Energieverteilung im Normalspectrum},''
  \href{http://dx.doi.org/10.1002/andp.19013090310}{{\em Annalen Phys.}
  {\bfseries 309} no.~3, (1901) 553--563}.

\bibitem{Einstein:1905tem}
A.~Einstein, ``{\"Uber einen die Erzeugung und Verwandlung des Lichtes
  betreffenden heuristischen Gesichtspunkt},''
  \href{http://dx.doi.org/10.1002/andp.19053220607}{{\em Annalen Phys.}
  {\bfseries 322} no.~6, (1905) 132--148}.

\bibitem{Bohr:1913zba}
N.~Bohr, ``{On the Constitution of Atoms and Molecules},''
  \href{http://dx.doi.org/10.1080/14786441308634955}{{\em Phil. Mag. Ser. 6}
  {\bfseries 26} (1913) 1--24}.

\bibitem{Georgi:1979md}
H.~Georgi, ``{Towards a Grand Unified Theory of Flavor},''
  \href{http://dx.doi.org/10.1016/0550-3213(79)90497-8}{{\em Nucl. Phys. B}
  {\bfseries 156} (1979) 126--134}.

\bibitem{Frampton:1979cw}
P.~H. Frampton, ``{SU($N$) Grand Unification With Several Quark - Lepton
  Generations},'' \href{http://dx.doi.org/10.1016/0370-2693(79)90472-6}{{\em
  Phys. Lett. B} {\bfseries 88} (1979) 299--301}.

\bibitem{Frampton:1979fd}
P.~Frampton and S.~Nandi, ``{SU(9) Grand Unification of Flavor With Three
  Generations},'' \href{http://dx.doi.org/10.1103/PhysRevLett.43.1460}{{\em
  Phys. Rev. Lett.} {\bfseries 43} (1979) 1460}.

\bibitem{Barr:1979xt}
S.~M. Barr, ``{Light Fermion Mass Hierarchy and Grand Unification},''
  \href{http://dx.doi.org/10.1103/PhysRevD.21.1424}{{\em Phys. Rev. D}
  {\bfseries 21} (1980) 1424}.

\bibitem{Barr:2008pn}
S.~M. Barr, ``{Doubly Lopsided Mass Matrices from Unitary Unification},''
  \href{http://dx.doi.org/10.1103/PhysRevD.78.075001}{{\em Phys. Rev. D}
  {\bfseries 78} (2008) 075001},
  \href{http://arxiv.org/abs/0804.1356}{{\ttfamily arXiv:0804.1356 [hep-ph]}}.

\bibitem{Chen:2023qxi}
N.~Chen, Y.-n. Mao, and Z.~Teng, ``{The global B \ensuremath{-} L symmetry in
  the flavor-unified SU(N) theories},''
  \href{http://dx.doi.org/10.1007/JHEP04(2024)046}{{\em JHEP} {\bfseries 04}
  (2024) 046}, \href{http://arxiv.org/abs/2307.07921}{{\ttfamily
  arXiv:2307.07921 [hep-ph]}}.

\bibitem{Froggatt:1978nt}
C.~D. Froggatt and H.~B. Nielsen, ``{Hierarchy of Quark Masses, Cabibbo Angles
  and CP Violation},''
  \href{http://dx.doi.org/10.1016/0550-3213(79)90316-X}{{\em Nucl. Phys. B}
  {\bfseries 147} (1979) 277--298}.

\bibitem{ATLAS:2012yve}
{\bfseries ATLAS} Collaboration, G.~Aad {\em et~al.}, ``{Observation of a new
  particle in the search for the Standard Model Higgs boson with the ATLAS
  detector at the LHC},''
  \href{http://dx.doi.org/10.1016/j.physletb.2012.08.020}{{\em Phys. Lett. B}
  {\bfseries 716} (2012) 1--29},
  \href{http://arxiv.org/abs/1207.7214}{{\ttfamily arXiv:1207.7214 [hep-ex]}}.

\bibitem{CMS:2012qbp}
{\bfseries CMS} Collaboration, S.~Chatrchyan {\em et~al.}, ``{Observation of a
  New Boson at a Mass of 125 GeV with the CMS Experiment at the LHC},''
  \href{http://dx.doi.org/10.1016/j.physletb.2012.08.021}{{\em Phys. Lett. B}
  {\bfseries 716} (2012) 30--61},
  \href{http://arxiv.org/abs/1207.7235}{{\ttfamily arXiv:1207.7235 [hep-ex]}}.

\bibitem{CMS:2022dwd}
{\bfseries CMS} Collaboration, A.~Tumasyan {\em et~al.}, ``{A portrait of the
  Higgs boson by the CMS experiment ten years after the discovery.},''
  \href{http://dx.doi.org/10.1038/s41586-022-04892-x}{{\em Nature} {\bfseries
  607} no.~7917, (2022) 60--68},
  \href{http://arxiv.org/abs/2207.00043}{{\ttfamily arXiv:2207.00043
  [hep-ex]}}.

\bibitem{ATLAS:2022vkf}
{\bfseries ATLAS} Collaboration, G.~Aad {\em et~al.}, ``{A detailed map of
  Higgs boson interactions by the ATLAS experiment ten years after the
  discovery},'' \href{http://dx.doi.org/10.1038/s41586-022-04893-w}{{\em
  Nature} {\bfseries 607} no.~7917, (2022) 52--59},
  \href{http://arxiv.org/abs/2207.00092}{{\ttfamily arXiv:2207.00092
  [hep-ex]}}. [Erratum: Nature 612, E24 (2022)].

\bibitem{Cabibbo:1963yz}
N.~Cabibbo, ``{Unitary Symmetry and Leptonic Decays},''
  \href{http://dx.doi.org/10.1103/PhysRevLett.10.531}{{\em Phys. Rev. Lett.}
  {\bfseries 10} (1963) 531--533}.

\bibitem{Kobayashi:1973fv}
M.~Kobayashi and T.~Maskawa, ``{CP Violation in the Renormalizable Theory of
  Weak Interaction},'' \href{http://dx.doi.org/10.1143/PTP.49.652}{{\em Prog.
  Theor. Phys.} {\bfseries 49} (1973) 652--657}.

\bibitem{Kallosh:1995hi}
R.~Kallosh, A.~D. Linde, D.~A. Linde, and L.~Susskind, ``{Gravity and global
  symmetries},'' \href{http://dx.doi.org/10.1103/PhysRevD.52.912}{{\em Phys.
  Rev. D} {\bfseries 52} (1995) 912--935},
  \href{http://arxiv.org/abs/hep-th/9502069}{{\ttfamily arXiv:hep-th/9502069}}.

\bibitem{Harlow:2018jwu}
D.~Harlow and H.~Ooguri, ``{Constraints on Symmetries from Holography},''
  \href{http://dx.doi.org/10.1103/PhysRevLett.122.191601}{{\em Phys. Rev.
  Lett.} {\bfseries 122} no.~19, (2019) 191601},
  \href{http://arxiv.org/abs/1810.05337}{{\ttfamily arXiv:1810.05337
  [hep-th]}}.

\bibitem{Harlow:2018tng}
D.~Harlow and H.~Ooguri, ``{Symmetries in quantum field theory and quantum
  gravity},'' \href{http://dx.doi.org/10.1007/s00220-021-04040-y}{{\em Commun.
  Math. Phys.} {\bfseries 383} no.~3, (2021) 1669--1804},
  \href{http://arxiv.org/abs/1810.05338}{{\ttfamily arXiv:1810.05338
  [hep-th]}}.

\bibitem{Dimopoulos:1980hn}
S.~Dimopoulos, S.~Raby, and L.~Susskind, ``{Light Composite Fermions},''
  \href{http://dx.doi.org/10.1016/0550-3213(80)90215-1}{{\em Nucl. Phys. B}
  {\bfseries 173} (1980) 208--228}.

\bibitem{Appelquist:2013oni}
T.~Appelquist and R.~Shrock, ``{Ultraviolet to infrared evolution of chiral
  gauge theories},'' \href{http://dx.doi.org/10.1103/PhysRevD.88.105012}{{\em
  Phys. Rev. D} {\bfseries 88} (2013) 105012},
  \href{http://arxiv.org/abs/1310.6076}{{\ttfamily arXiv:1310.6076 [hep-th]}}.

\bibitem{Shi:2015fna}
Y.-L. Shi and R.~Shrock, ``{$A_k \bar F$ chiral gauge theories},''
  \href{http://dx.doi.org/10.1103/PhysRevD.92.105032}{{\em Phys. Rev. D}
  {\bfseries 92} no.~10, (2015) 105032},
  \href{http://arxiv.org/abs/1510.07663}{{\ttfamily arXiv:1510.07663
  [hep-th]}}.

\bibitem{Peccei:1977hh}
R.~D. Peccei and H.~R. Quinn, ``{CP Conservation in the Presence of
  Instantons},'' \href{http://dx.doi.org/10.1103/PhysRevLett.38.1440}{{\em
  Phys. Rev. Lett.} {\bfseries 38} (1977) 1440--1443}.

\bibitem{Georgi:1981pu}
H.~M. Georgi, L.~J. Hall, and M.~B. Wise, ``{Grand Unified Models With an
  Automatic {Peccei-Quinn} Symmetry},''
  \href{http://dx.doi.org/10.1016/0550-3213(81)90433-8}{{\em Nucl. Phys. B}
  {\bfseries 192} (1981) 409--416}.

\bibitem{Maalampi:1988va}
J.~Maalampi and M.~Roos, ``{Physics of Mirror Fermions},''
  \href{http://dx.doi.org/10.1016/0370-1573(90)90095-J}{{\em Phys. Rept.}
  {\bfseries 186} (1990) 53}.

\bibitem{Li:1973mq}
L.-F. Li, ``{Group Theory of the Spontaneously Broken Gauge Symmetries},''
  \href{http://dx.doi.org/10.1103/PhysRevD.9.1723}{{\em Phys. Rev. D}
  {\bfseries 9} (1974) 1723--1739}.

\bibitem{Li:1981nk}
X.~Li and E.~Ma, ``{Gauge Model of Generation Nonuniversality},''
  \href{http://dx.doi.org/10.1103/PhysRevLett.47.1788}{{\em Phys. Rev. Lett.}
  {\bfseries 47} (1981) 1788}.

\bibitem{Ma:1987ds}
E.~Ma, X.~Li, and S.~F. Tuan, ``{Gauge Model of Generation Nonuniversality
  Revisited},'' \href{http://dx.doi.org/10.1103/PhysRevLett.60.495}{{\em Phys.
  Rev. Lett.} {\bfseries 60} (1988) 495--498}.

\bibitem{Li:1992fi}
X.-y. Li and E.~Ma, ``{Gauge model of generation nonuniversality reexamined},''
  \href{http://dx.doi.org/10.1088/0954-3899/19/9/006}{{\em J. Phys. G}
  {\bfseries 19} (1993) 1265--1278},
  \href{http://arxiv.org/abs/hep-ph/9208210}{{\ttfamily arXiv:hep-ph/9208210}}.

\bibitem{Craig:2011yk}
N.~Craig, D.~Green, and A.~Katz, ``{(De)Constructing a Natural and Flavorful
  Supersymmetric Standard Model},''
  \href{http://dx.doi.org/10.1007/JHEP07(2011)045}{{\em JHEP} {\bfseries 07}
  (2011) 045}, \href{http://arxiv.org/abs/1103.3708}{{\ttfamily arXiv:1103.3708
  [hep-ph]}}.

\bibitem{Bordone:2017bld}
M.~Bordone, C.~Cornella, J.~Fuentes-Martin, and G.~Isidori, ``{A three-site
  gauge model for flavor hierarchies and flavor anomalies},''
  \href{http://dx.doi.org/10.1016/j.physletb.2018.02.011}{{\em Phys. Lett. B}
  {\bfseries 779} (2018) 317--323},
  \href{http://arxiv.org/abs/1712.01368}{{\ttfamily arXiv:1712.01368
  [hep-ph]}}.

\bibitem{Fuentes-Martin:2020pww}
J.~Fuentes-Martin, G.~Isidori, J.~Pag\`es, and B.~A. Stefanek, ``{Flavor
  non-universal Pati-Salam unification and neutrino masses},''
  \href{http://dx.doi.org/10.1016/j.physletb.2021.136484}{{\em Phys. Lett. B}
  {\bfseries 820} (2021) 136484},
  \href{http://arxiv.org/abs/2012.10492}{{\ttfamily arXiv:2012.10492
  [hep-ph]}}.

\bibitem{Davighi:2022fer}
J.~Davighi and J.~Tooby-Smith, ``{Electroweak flavour unification},''
  \href{http://dx.doi.org/10.1007/JHEP09(2022)193}{{\em JHEP} {\bfseries 09}
  (2022) 193}, \href{http://arxiv.org/abs/2201.07245}{{\ttfamily
  arXiv:2201.07245 [hep-ph]}}.

\bibitem{Fuentes-Martin:2022xnb}
J.~Fuentes-Martin, G.~Isidori, J.~M. Lizana, N.~Selimovic, and B.~A. Stefanek,
  ``{Flavor hierarchies, flavor anomalies, and Higgs mass from a warped extra
  dimension},'' \href{http://dx.doi.org/10.1016/j.physletb.2022.137382}{{\em
  Phys. Lett. B} {\bfseries 834} (2022) 137382},
  \href{http://arxiv.org/abs/2203.01952}{{\ttfamily arXiv:2203.01952
  [hep-ph]}}.

\bibitem{Davighi:2022bqf}
J.~Davighi, G.~Isidori, and M.~Pesut, ``{Electroweak-flavour and quark-lepton
  unification: a family non-universal path},''
  \href{http://dx.doi.org/10.1007/JHEP04(2023)030}{{\em JHEP} {\bfseries 04}
  (2023) 030}, \href{http://arxiv.org/abs/2212.06163}{{\ttfamily
  arXiv:2212.06163 [hep-ph]}}.

\bibitem{FernandezNavarro:2023rhv}
M.~Fern\'andez~Navarro and S.~F. King, ``{Tri-hypercharge: a separate gauged
  weak hypercharge for each fermion family as the origin of flavour},''
  \href{http://dx.doi.org/10.1007/JHEP08(2023)020}{{\em JHEP} {\bfseries 08}
  (2023) 020}, \href{http://arxiv.org/abs/2305.07690}{{\ttfamily
  arXiv:2305.07690 [hep-ph]}}.

\bibitem{FernandezNavarro:2023hrf}
M.~Fern\'andez~Navarro, S.~F. King, and A.~Vicente, ``{Tri-unification: a
  separate $SU(5)$ for each fermion family},''
  \href{http://arxiv.org/abs/2311.05683}{{\ttfamily arXiv:2311.05683
  [hep-ph]}}.

\bibitem{Haba:2000be}
N.~Haba and H.~Murayama, ``{Anarchy and hierarchy},''
  \href{http://dx.doi.org/10.1103/PhysRevD.63.053010}{{\em Phys. Rev. D}
  {\bfseries 63} (2001) 053010},
  \href{http://arxiv.org/abs/hep-ph/0009174}{{\ttfamily arXiv:hep-ph/0009174}}.

\bibitem{Glashow:1979nm}
S.~L. Glashow, ``{The Future of Elementary Particle Physics},''
  \href{http://dx.doi.org/10.1007/978-1-4684-7197-7_15}{{\em NATO Sci. Ser. B}
  {\bfseries 61} (1980) 687}.

\bibitem{Barbieri:1979ag}
R.~Barbieri, D.~V. Nanopoulos, G.~Morchio, and F.~Strocchi, ``{Neutrino Masses
  in Grand Unified Theories},''
  \href{http://dx.doi.org/10.1016/0370-2693(80)90058-1}{{\em Phys. Lett. B}
  {\bfseries 90} (1980) 91--97}.

\bibitem{Barbieri:1980vc}
R.~Barbieri and D.~V. Nanopoulos, ``{An Exceptional Model for Grand
  Unification},'' \href{http://dx.doi.org/10.1016/0370-2693(80)90998-3}{{\em
  Phys. Lett. B} {\bfseries 91} (1980) 369--375}.

\bibitem{Barbieri:1980tz}
R.~Barbieri and D.~V. Nanopoulos, ``{Hierarchical Fermion Masses From Grand
  Unification},'' \href{http://dx.doi.org/10.1016/0370-2693(80)90395-0}{{\em
  Phys. Lett. B} {\bfseries 95} (1980) 43--46}.

\bibitem{delAguila:1980qag}
F.~del Aguila and L.~E. Ibanez, ``{Higgs Bosons in SO(10) and Partial
  Unification},'' \href{http://dx.doi.org/10.1016/0550-3213(81)90266-2}{{\em
  Nucl. Phys. B} {\bfseries 177} (1981) 60--86}.

\bibitem{Ellis:1979fg}
J.~R. Ellis and M.~K. Gaillard, ``{Fermion Masses and Higgs Representations in
  SU(5)},'' \href{http://dx.doi.org/10.1016/0370-2693(79)90476-3}{{\em Phys.
  Lett. B} {\bfseries 88} (1979) 315--319}.

\bibitem{Chacko:2020tbu}
Z.~Chacko, P.~S.~B. Dev, R.~N. Mohapatra, and A.~Thapa, ``{Predictive Dirac and
  Majorana Neutrino Mass Textures from $SU(6)$ Grand Unified Theories},''
  \href{http://dx.doi.org/10.1103/PhysRevD.102.035020}{{\em Phys. Rev. D}
  {\bfseries 102} no.~3, (2020) 035020},
  \href{http://arxiv.org/abs/2005.05413}{{\ttfamily arXiv:2005.05413
  [hep-ph]}}.

\bibitem{Wolfenstein:1983yz}
L.~Wolfenstein, ``{Parametrization of the Kobayashi-Maskawa Matrix},''
  \href{http://dx.doi.org/10.1103/PhysRevLett.51.1945}{{\em Phys. Rev. Lett.}
  {\bfseries 51} (1983) 1945}.

\bibitem{Buras:1977yy}
A.~J. Buras, J.~R. Ellis, M.~K. Gaillard, and D.~V. Nanopoulos, ``{Aspects of
  the Grand Unification of Strong, Weak and Electromagnetic Interactions},''
  \href{http://dx.doi.org/10.1016/0550-3213(78)90214-6}{{\em Nucl. Phys. B}
  {\bfseries 135} (1978) 66--92}.

\bibitem{Nanopoulos:1978hh}
D.~V. Nanopoulos and D.~A. Ross, ``{Limits on the Number of Flavors in Grand
  Unified Theories from Higher Order Corrections to Fermion Masses},''
  \href{http://dx.doi.org/10.1016/0550-3213(79)90507-8}{{\em Nucl. Phys. B}
  {\bfseries 157} (1979) 273--284}.

\bibitem{Georgi:1979ga}
H.~Georgi and D.~V. Nanopoulos, ``{Masses and Mixing in Unified Theories},''
  \href{http://dx.doi.org/10.1016/0550-3213(79)90323-7}{{\em Nucl. Phys. B}
  {\bfseries 159} (1979) 16--28}.

\bibitem{Chen:2024deo}
N.~Chen, Z.~Hou, Y.-n. Mao, and Z.~Teng, ``{The gauge coupling evolutions of an
  ${\rm SU}(8)$ theory with the maximally symmetry breaking pattern},''
  \href{http://arxiv.org/abs/2406.09970}{{\ttfamily arXiv:2406.09970
  [hep-ph]}}.

\bibitem{ILC:2013jhg}
{\bfseries ILC} Collaboration, ``{The International Linear Collider Technical
  Design Report - Volume 2: Physics},''
  \href{http://arxiv.org/abs/1306.6352}{{\ttfamily arXiv:1306.6352 [hep-ph]}}.

\bibitem{TLEPDesignStudyWorkingGroup:2013myl}
{\bfseries TLEP Design Study Working Group} Collaboration, M.~Bicer {\em
  et~al.}, ``{First Look at the Physics Case of TLEP},''
  \href{http://dx.doi.org/10.1007/JHEP01(2014)164}{{\em JHEP} {\bfseries 01}
  (2014) 164}, \href{http://arxiv.org/abs/1308.6176}{{\ttfamily arXiv:1308.6176
  [hep-ex]}}.

\bibitem{CEPCStudyGroup:2018ghi}
{\bfseries CEPC Study Group} Collaboration, M.~Dong {\em et~al.}, ``{CEPC
  Conceptual Design Report: Volume 2 - Physics \& Detector},''
  \href{http://arxiv.org/abs/1811.10545}{{\ttfamily arXiv:1811.10545
  [hep-ex]}}.

\end{thebibliography}

\providecommand{\href}[2]{#2}\begingroup\raggedright\endgroup

\end{document}